\documentclass{emulateapj}
\usepackage{longtable}
\usepackage{lscape}
\usepackage{graphicx}

\slugcomment{ApJ in press}
\shorttitle{The light of MgII absorbing galaxies}
\shortauthors{Stefano Zibetti et al.}

\newcommand{\MgII}{MgII}
\newcommand{\be}{\begin{equation}}
\newcommand{\ee}{\end{equation}}
\newcommand{\bea}{\begin{eqnarray}}
\newcommand{\eea}{\end{eqnarray}}

\begin{document}
\title{Optical properties and spatial distribution\\ of MgII absorbers
from SDSS image stacking}

\author{Stefano Zibetti\altaffilmark{1}, Brice
M\'enard\altaffilmark{2,3},\\ Daniel B. Nestor\altaffilmark{4}, Anna
M. Quider\altaffilmark{5}, Sandhya M. Rao\altaffilmark{5} and David
A. Turnshek\altaffilmark{5} }
 
\altaffiltext{1}{Max-Planck-Institut f\"ur Extraterrestrische Physik,
Postfach 1312, D-85741, Garching bei M\"unchen, Germany, e-mail
szibetti@mpe.mpg.de} 
\altaffiltext{2}{Institute for Advanced Study, Einstein Drive,
 Princeton NJ 08540, USA, e-mail menard@ias.edu}
\altaffiltext{3}{Present address: Canadian Institute for Theoretical
 Astrophysics (CITA), University of Toronto, 60 St. George Street,
 Toronto, Ontario, M5S 3H8, email: menard@cita.utoronto.ca }
\altaffiltext{4} {Institute of Astronomy,
University of Cambridge, Madingley Road, Cambridge. CB3 0HA, UK,
e-mail: dbn@ast.cam.ac.uk } 
\altaffiltext{5}{Dept. of Physics and Astronomy, University of
Pittsburgh, Pittsburgh, PA 15260, USA, e-mail:
amq3,srao,turnshek@pitt.edu}

\begin{abstract}
We present a statistical analysis of the photometric properties and
spatial distribution of more than 2,800 MgII absorbers with
$0.37<z<1$ and rest equivalent width
$W_0(\lambda2796)>0.8$\AA~detected in SDSS quasar spectra. Using an
improved image stacking technique, we measure the cross-correlation
between MgII gas and light (in the $g$, $r$, $i$ and $z$-bands) from
10 to 200 kpc and infer the light-weighted impact parameter
distribution of MgII absorbers. Such a quantity is well described by a
power-law with an index that strongly depends on absorption rest
equivalent width $W_0$, ranging from $\sim-1$ for $W_0\lesssim 1$\AA~
to $\sim -2$ for $W_0\gtrsim 1.5$\AA.

At redshift $0.37<z_{abs}\leq 0.55$, we find the average luminosity
enclosed within 100 kpc around MgII absorbers to be
$M_g=-20.65\pm0.11$ mag, which is $\sim0.5~L_g^\star$. The global
luminosity-weighted colors are typical of present-day intermediate
type galaxies.  We then investigate these colors as a function of MgII
rest equivalent width and find that they follow the track between
spiral and elliptical galaxies in color space: while the light of
weaker absorbers originates mostly from red passive galaxies, stronger
systems display the colors of blue star-forming galaxies.
Based on these observations, we argue that the origin of strong MgII
absorber systems might be better explained by models of metal-enriched
gas outflows from star-forming/bursting galaxies.

Our analysis does not show any redshift dependence for both impact
parameter and rest-frame colors up to $z=1$.  However, we do observe a
brightening of the absorbers related light at high redshift ($\sim
50\%$ from $z_{abs}\sim 0.4$ to 1). We argue that MgII absorbers are a
phenomenon typical of a given evolutionary phase that more massive
galaxies experience earlier than less massive ones, in a
\emph{downsizing} fashion.

This analysis provides us with robust and quantitative constraints of
interest for further modeling of the gas distribution around galaxies.
As a side product we also show that the stacking technique allows us
to detect the light of quasar hosts and their environment.
\end{abstract}


\section{Introduction}\label{intro}

Quasar absorption lines provide us with a unique tool to probe the gas
content in the Universe with a sensitivity that does not depend on
redshift.  In order to use this information for investigating galaxy
formation and evolution, characterizing the underlying population of
galaxies seen in absorption is an important task to achieve.  

In 1969, Bahcall \& Spitzer suggested that metal
absorption lines seen in quasar spectra are induced by large halos of
gas around galaxies extending up to a hundred kpc.  Since then, the
study of quasar absorption lines has become an active field of
investigation, but fundamental questions regarding the nature of this
gas and its distribution around galaxies still remain.

Studies of absorber-galaxy connections began with the
investigation of MgII systems. Such a choice results from a practical
constraint: among the dominant ions in HI gas, MgII has the
longest-wavelength resonance transition at the doublet $\lambda\lambda
2796.35,2803.53$.  Several studies showed that MgII arises in gas
spanning more than five decades of neutral hydrogen column density
\citep{bergeron_stasinska_86,steidel_sargent_92,churchill+00}, and at a
similar time the first galaxies responsible for some metal absorption
features were identified \citep{bergeron_86,bergeron+87,
cristiani_87,bergeron_boisse_91}. \cite{SDP94} then built a sample of
58 identified MgII systems and found that various types of galaxies
with $\sim L^\star$ luminosities exhibit MgII absorption up to a
radius of $R\sim 50 h^{-1}$ kpc.
For a review of results regarding the connection between MgII
absorbers and galaxies see \citet{churchill+05}. It is also
interesting to mention that \cite{bowen+06} recently reported the
detection of MgII absorption around quasars.

While the connection between galaxies and metal absorbers has been
clearly established, the origin of the absorbing gas and its
properties remains unclear.  Whether these absorption lines trace gas
being accreted by a galaxy or outflowing from it is still a matter of
debate.
Recently the SDSS database has provided us with significantly larger
samples of absorbers (by two orders of magnitude), allowing us to
accurately measure a number of statistical properties such as the
redshift distribution, $dN/dz$, and the distribution of absorber rest
equivalent widths, $dN/dW_0$, which is now measured over more than a
decade in $W_0$ \citep{nestor+05,prochter+06}. Unfortunately, physical
models or numerical simulations reproducing these observations are
still lacking. While efforts on the theoretical side are needed,
additional \emph{statistical} observational constraints are required,
especially to characterize the spatial distribution of the gas around
galaxies and their corresponding optical properties.

Previous attempts to constrain such properties of MgII absorbers have
relied on deep imaging and spectroscopic follow-up of (arbitrarily
faint) galaxies in the quasar field in order to identify a potential
galaxy responsible for the absorption seen in the quasar spectrum.
Such expensive studies have been limited to samples of a few tens of
quasars. Moreover recent results have shown discrepancies in the
inferred distributions of absorber impact parameters.  Whereas
\cite{SDP94} claimed that $\sim L^\star$ galaxies are surrounded by
MgII halos extending up to $R\sim 50 h^{-1}$ kpc, \cite{churchill+05}
recently showed that a more systematic analysis reveals the same type
of absorption up to $R\sim 100 h^{-1}$ kpc in a number of cases. The
different results apparently arise from different ways of preselecting
the candidate galaxies for the spectroscopic follow-up.  Furthermore,
it is interesting to note that the recent detections of MgII around
quasars \citep{bowen+06} reach similar distances. Estimating the
distribution of absorber impact parameters \emph{without the use of
any assumption} is therefore needed.

It should be emphasized that the identification of a galaxy
responsible for a given absorption feature is in general not a well
defined process. Indeed it is always possible that an object fainter
than the limiting magnitude, such as a dwarf galaxy, gives rise to the
absorbing gas, or that more than one galaxy has a redshift consistent
with that of the absorber system which is likely to happen in group
environments.  A well-defined quantity that can be obtained from
observations and modeled theoretically is the cross-correlation
between absorbers and the measured light within a given radius:
$\langle W_0 \, L\rangle(\theta)$.

In \cite{abs_letter05} we proposed a new approach to measure the
systematic photometric properties of large samples of absorbing
systems, which combines statistical analysis of both spectroscopic and
imaging datasets of the Sloan Digital Sky Survey \citep[hereafter
SDSS][]{york+00}. As MgII lines can only be detected in SDSS spectra
at $z\gtrsim0.4$, the absorbing galaxies are only marginally detected
in individual SDSS images at the lowest redshifts, and below the noise
level otherwise.
However, they do produce systematic light excesses around the
background QSOs. Such excesses can be detected with a statistical
analysis consisting of stacking a large number of absorbed QSO
images. After having demonstrated the feasibility of this technique
with EDR data \citep{abs_letter05}, we now extend our analysis to the
SDSS DR4 sample, i.e., a dataset about twenty times larger.

Stacking analysis has the intrinsic limitation of only probing the
mean or median value of a quantity for which the underlying
distribution is not necessarily known. However, it can lead to
detections of low-level signals not detectable otherwise. Moreover,
one can also investigate the dependence of an observed quantity $O$ as
a function of a set of parameters that describe a given sample. Some
properties of the underlying distribution giving rise to $<O>$ can
therefore be inferred.

Stacking approaches have provided us with valuable results in a number
of areas. For example, \cite{bartelmann_white_03} have demonstrated
that stacking of ROSAT All-Sky Survey X-ray images of high redshift
clusters detected in the SDSS can be used to derive their mean X-ray
properties. \cite{zibetti+05} detected low levels of intracluster
light by stacking SDSS cluster images, while \cite{lin_mohr_04}
measured their infrared (IR) luminosity by coadding  2MASS data. In
the context of galaxies, \cite{hogg+97} constrained the IR signal from
faint galaxies using stacked Keck data. Similarly, \cite{brandt+01}
measured the mean X-ray flux of Lyman Break galaxies,
\cite{zibetti+04} characterized the very low-surface brightness
\emph{diffuse} light in galaxy halos and, more recently,
\cite{white+06} constrained the radio properties of SDSS quasars down
to the nanoJansky level.  In spectroscopic studies, stacking
techniques have been extensively used to look for weak signals.
Composite spectra of SDSS quasars were recently  used to detect weak
absorption lines as well as dust reddening effects that are are well
below the noise level in individual spectra
\citep{nestor+03,menard05,york+06}.

In order to emphasize the power of the stacking technique using the
SDSS, it is instructive to compare the sensitivity that can be reached
with respect to that of classical pointed observations. In the present
study we make use of SDSS imaging data which is obtained from a 2.5
meter telescope, using a drift-scan technique giving rise to an
exposure time of about one minute. The samples of quasars with and
without absorbers used for the stacking analysis include approximately
2,500 and 10,000 objects respectively. The careful analysis of the
corresponding 0.5 Terabyte of imaging data allows us to reach
equivalent exposure times of about 40 and 160 hours respectively,
i.e., 4 and 16 nights of observations with the SDSS telescope. Such
numbers imply that, as long as systematic effects are small, detecting
faint levels of light from high-redshift becomes feasible.

In this paper we present the results of a stacking analysis aimed at
characterizing the light associated with MgII absorbers and
investigating its properties as a function of absorber rest equivalent
width and redshift. As we will show, the sensitivity achieved with
this technique even allows us to detect the light from the
high-redshift quasar hosts. The paper is organized as follows: in
\S\ref{data_sec} we present our sample of MgII absorbers; in
\S\ref{improc_sec} we describe the image processing and the technique
adopted to extract photometric information from the stack images;
in \S\ref{spatial_sec} we analyze the surface brightness profiles of the
light in excess around absorbed QSOs and derive the luminosity
weighted impact parameter distribution of the MgII systems around
galaxies; \S\ref{SED_subsec} is devoted to the interpretation of the
integral photometry in terms of rest-frame SEDs and luminosity;
in \S\ref{discussion_sect} we briefly discuss a possible unified scenario
for the interpretation of the results; finally, in \S\ref{summary}
we present a schematic summary of the paper.

``Standard'' cosmological parameters are assumed throughout the paper,
i.e., $\Omega_{tot}=1.0$, $\Omega_{\Lambda}=0.7$, and $h=0.7$.

\section{The data}\label{data_sec}

\subsection{QSOs with MgII absorbers}

We use the sample of \MgII\ absorber systems compiled with the method
presented in \cite{nestor+05} and based on SDSS DR4 data \citep{DR4}.
In this section we briefly summarize the main steps involved in the
absorption line detection procedure.  For more details we refer the
reader to \cite{nestor+05}.
\begin{figure*}
\plottwo{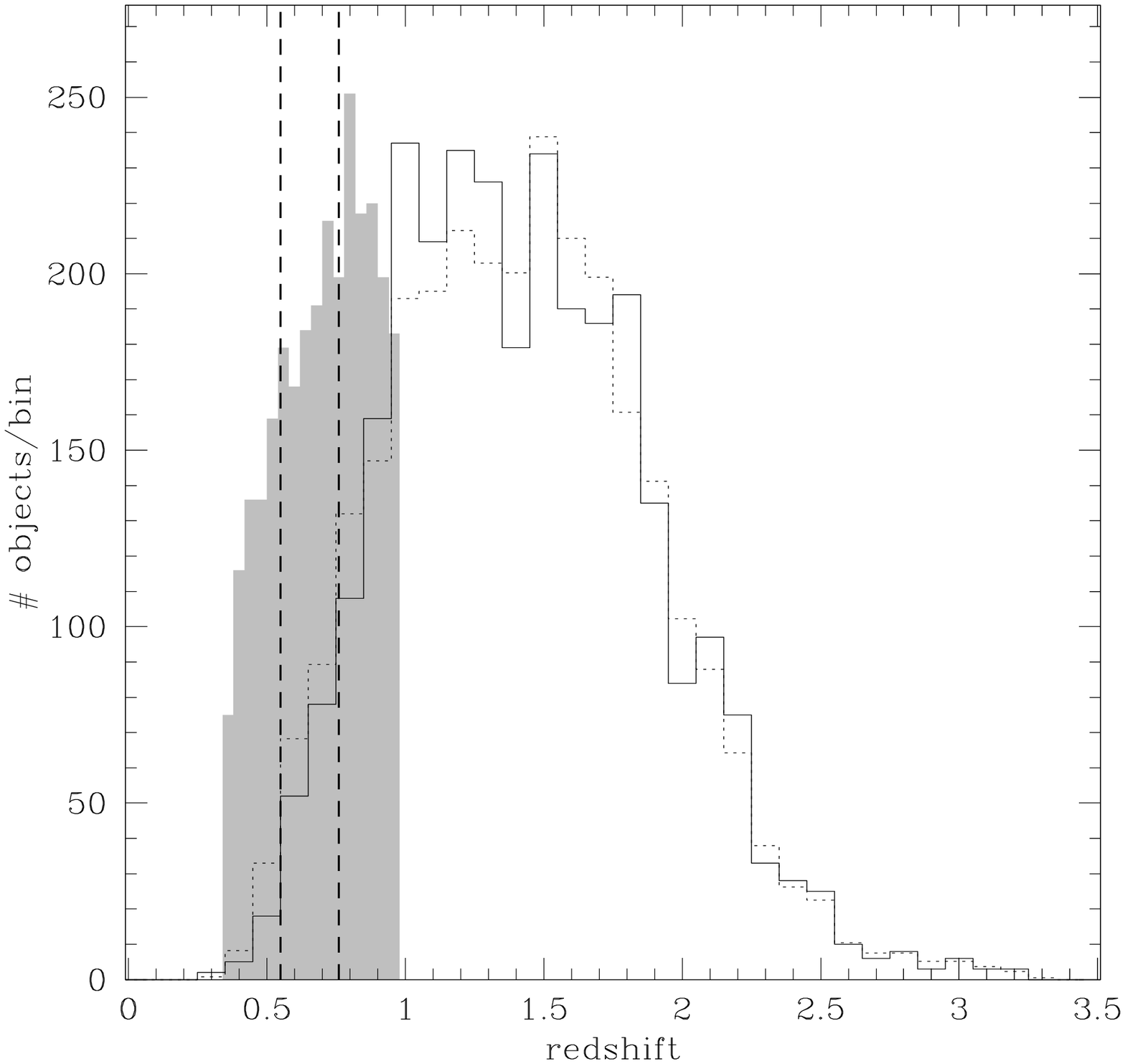}{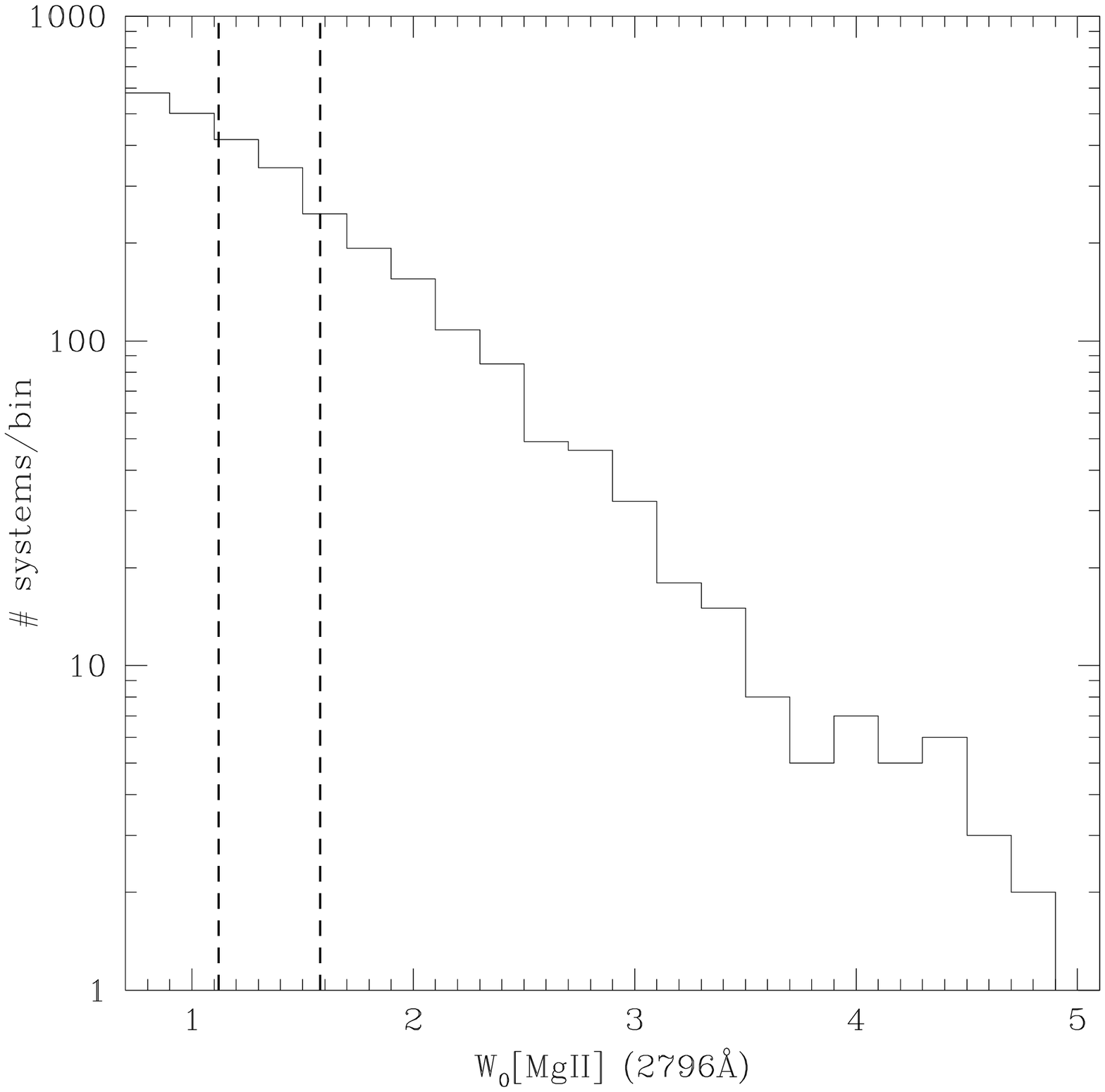}
  \caption{ \emph{Left panel}: redshift distributions of MgII
  absorbers with $W_0>0.8 \mathrm{\AA}$ used in this analysis (shaded
  histogram), of their background quasars (solid line) and of the
  complete population of reference quasars (dashed line). The latter
  is rescaled by a factor 4 to allow a direct comparison with the
  population of absorbed quasars. The vertical dashed lines mark the
  boundaries between the three bins in absorbers'
  redshifts. \emph{Right panel}: distribution of absorber rest
  equivalent widths. The vertical dashed lines separate the three bins
  in absorption strength.}
\label{plot_absorber_parameters}
\end{figure*}
All QSO spectra from the SDSS DR4 database (see Richards et al. 2002
for a definition of the SDSS QSO sample, also Schneider et al. 2005)
were analyzed, regardless of QSO magnitude. The continuum-normalized
SDSS QSO spectra were searched for \MgII\ $\lambda\lambda2796,2803$
doublets using an optimal extraction method employing a Gaussian
line-profile to measure each rest equivalent width $W_0$. All
candidates were interactively checked for false detections, a
satisfactory continuum fit, blends with absorption lines from other
systems, and special cases.  The identification of Mg II doublets
required the detection of the $\lambda2796$ line and at least one
additional line, the most convenient being the $\lambda2803$ doublet
partner.  A 5$\sigma$ significance level was required for all
$\lambda2796$ lines, as well as a $3\sigma$ significance level for the
corresponding $\lambda2803$ lines.  Only systems 0.1c blueward of the
quasar redshift and redward of Lyman $\alpha$ emission were accepted.
For simplicity, systems with separations of less than 500 km/s were
considered as a single system. We note that \MgII\ absorption lines
are in general saturated, and in these cases no column density
information can be directly extracted from $W_0(\lambda2796)$.  Given
the typical S/N of SDSS quasar spectra, most of the detected MgII
absorption lines have $W_0>0.8\;\mathrm{\AA}$.  Weaker systems can
occasionally be detected in the spectra of bright quasars but are
significantly less numerous. As our technique requires a large number
of absorbers we will only consider the stronger ones,
i.e., $W_0>0.8\;\mathrm{\AA}$.

In this sample, multiple absorbers detected in the same quasar
spectrum are found in $\sim 15\%$ of the systems. In order to ease the
interpretation of our analysis, we select quasars with only one strong
absorber detected in their spectrum. Finally, we restrict our analysis
to \MgII\ systems with $z<1$. After excluding a few tens of QSOs for
which poor imaging data is available (mainly because of very bright
saturated stars close by), the sample of absorbed QSOs includes 2844
objects. Their properties are summarized in Figure
\ref{plot_absorber_parameters}: the left panel shows the redshift
distribution of the selected sample of MgII absorbers (shaded
histogram) and their background quasars (solid line).  The right panel
presents the distribution of absorber rest equivalent widths.

\subsection{Reference QSOs}\label{ref_sample_sec}

In order to use a control sample in our analysis we need to select a
population of quasars without absorbers. A number of criteria must be
applied to ensure the selection of an unbiased sample: the two quasar 
populations must have the same redshift and apparent magnitude 
distributions, and must present the same absorption-line detectability. The
selection of the reference quasar population is therefore defined as
follows: for each quasar with a detected absorber with rest equivalent
width $W_0$ and redshift $z_{abs}$, we randomly look for a quasar
without any detected absorber above $W_0(\lambda2796)=0.8\,$\AA\ such
that:

\newcounter{Lcount}
\begin{list}{\emph{(\roman{Lcount})}}{}
{\usecounter{Lcount}
\item they have similar redshifts: $\Delta z<0.1$;
\item their magnitudes do not differ significantly: $\Delta m_i<0.5$;
\item the absorption line is detectable in the reference spectrum,
i.e., the S/N of the reference spectrum at
the corresponding wavelength is high enough to detect the 
MgII $\lambda 2796$ line at the required level of significance. 
Such a requirement
translates into $W_{min}(QSO_{ref},z_{abs})<W_0$, where $W_{min}$ is
the minimum equivalent width that can be detected by the line finder
at the corresponding wavelength.}  
\end{list}
The quasars selected in this way are
called \emph{reference} quasars in the following.  Our selection
procedure ensures that if similar absorbers were present in front of
the selected reference quasars, we would have been able to detect
them.
In order to improve the noise properties of our analysis we select
four \emph{independent} reference quasars for each quasar with an
absorber, resulting in a total of 11,376 objects.

\subsection{Definition of absorbers sub-samples}\label{def_subsample_sec}
The complete sample of MgII absorbers used in this study spans
a large range in $z_{abs}$ (from 0.37 to 1.00), where substantial
wavelength shift occurs. In fact, at the lowest redshift, $z_{abs}=0.37$, the
4000\AA~break feature lies in the observed $r$-band, whereas at
$z_{abs}=1.00$ it occurs in the observed $z$-band. Moreover, the
distance modulus between $z_{abs}=0.37$ and $z_{abs}=1.00$ increases
by 2.6 mag, thus making low-$z_{abs}$ absorbing galaxies the dominant
sources of signal in the overall stack. Therefore, in the following
analysis we will consider three $z_{abs}$ bins: $0.37\leq
z_{abs}<0.55$ (``low-redshift'' sub-sample), $0.55\leq z_{abs}<0.76$
(``intermediate-redshift''), and $0.76\leq z_{abs}<1.00$
(``high-redshift''). The boundaries of this binning in $z_{abs}$ are
indicated by the vertical dashed lines in the left panel of
Figure \ref{plot_absorber_parameters}. Possible dependence of the
photometric and spatial properties on the absorption strength will be
investigated by further splitting each $z_{abs}$ subsample into three
different $W_0(\lambda2796)$ subsamples: $0.8\mathrm{\AA} \leq W_0 <
1.12\mathrm{\AA}$ (``low-$W_0$'' systems),
$1.12\mathrm{\AA} \leq W_0 < 1.58\mathrm{\AA}$ (``intermediate-$W_0$''
systems), and $W_0 \geq 1.58\mathrm{\AA}$ (``high-$W_0$''
systems). The boundaries of these bins are also shown as vertical
dashed lines in the right panel of
fig. \ref{plot_absorber_parameters}.
Finally, it is important to note that very strong systems
($W_0\gtrsim 3 $\AA~) are not expected to contribute significantly to
our results as they are substantially less numerous than weaker ones.

\section{Image processing and stack photometry}\label{improc_sec}

As indicated above \citep[see also][]{abs_letter05}, our approach
makes use of the fact that the galaxies linked to an absorbing system
\emph{statistically} produce an excess of surface brightness (SB)
around the absorbed QSOs with respect to unabsorbed ones.  Such a SB
excess can be measured to obtain both global photometric quantities
for the absorbing galaxies in different bands and the spatial
distribution of the light of the absorbing galaxies, from which the
impact parameter distribution of absorbing gas clouds can be derived.

In this section we describe the techniques that allow us to optimally
integrate and measure the flux distribution of all and only the
galaxies
that cross-correlate with the presence of an absorber. By
``optimally'' we mean that \emph{any source of noise is minimized}.
In this measurement we are faced with three main sources of noise:
\emph{(i)} the intrinsic photon noise, which is fixed by the number of
stacked images and their depth, \emph{(ii)} the ``background'' signal
produced by stars and galaxies which are not correlated with the
absorbers, and \emph{(iii)} the light from the QSO itself. While the
intrinsic photon noise appears to be sufficiently low for a few
hundred SDSS images and absorbers up to $z=1$ or more, the signal from
background sources and the brightness mismatch which is allowed
between absorbed and reference QSOs, although small, produce a noise
which is orders of magnitude larger than the signal to be
detected. These two sources of noise must be drastically reduced by
applying accurate masking and QSO/Point Spread Function (PSF)
subtraction algorithms on each image as described in the next two
sections \S\S \ref{mask_subsec} and \ref{PSFsub_subsec}.

The stacking and subsequent photometric analysis is conducted
simultaneously in the four bands for which most of the flux is
expected, i.e. $g$, $r$, $i$, and $z$. We use the so-called SDSS
``corrected frame'' images (fpC's), which are bias subtracted and
flat-field corrected, but preserve the original background signal.
The error computation takes into account not only formal photometric
uncertainties, but also sample variance and error correlation between
different quantities in different bands. This is described in detail
in \S \ref{stackphotocode_subsec}.

\subsection{Masking algorithm}\label{mask_subsec}

The goal of the masking algorithm is to mask out any region of the
image frame where flux is contributed by sources that have negligible
probability of being a galaxy at the redshift of the absorber,
$z_{abs}$.  As a preliminary step, the light of the QSO is removed
from a working copy of the image. This is done by subtracting the
so-called postage-stamp image of the QSO output by the SDSS
\texttt{PHOTO} pipeline \citep{lupton+01}, in which the QSO is
optimally de-blended from all surrounding sources. \textsc{SExtractor}
\citep[][V.2.3.2]{sex} is then run on the QSO-subtracted image to
obtain a segmentation image and a catalogue of all sources detected $1
\sigma$ above the local background over a minimum area of 10 pixels.
The \textsc{SExtractor}'s segmentation image is used as base for the
final mask. Only \emph{segments} (i.e., groups of pixels) attributed to
sources that are too bright to be a galaxy at $z_{abs}$ are kept in
the mask.

\begin{figure}
\plotone{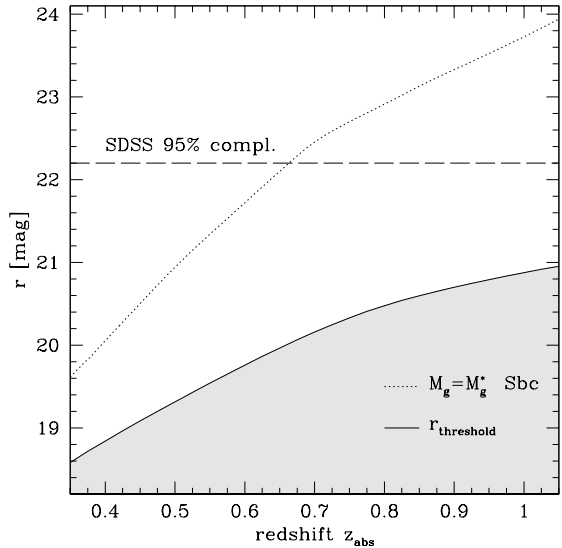}\caption{The threshold magnitude for
the masks in $r$-band as a function of the absorber's redshift
$z_{abs}$ (solid line), compared to the apparent magnitude of a
``normal'' (Sbc SED template) $M_g=M_g^\star$ galaxy (dotted
line). The horizontal long dashed line is the magnitude where
completeness of the SDSS photometric catalog drops below 95\% (point
sources).}\label{r_thresholds}
\end{figure}

For the mask thresholds, we adopt the fluxes derived for an
unobscured, metal poor stellar population produced in a 10 Myr long
burst, observed right after the end of the burst itself, with
rest-frame $g$-band absolute magnitude
$M_{g,\mathrm{threshold}}=-22.4~\mathrm{mag}$ \citep[as computed by
the code of][]{BC03}, observed at $z_{abs}$. The flux-limited masks
are computed separately in $g$-, $r$- and $i$-band, that roughly
correspond to the rest-frame $g$- and $u$-bands in the redshift range
spanned by the absorbers in our sample. The choice of this very
``blue'', UV-bright SED is motivated by the need to avoid the
exclusion of UV-bright galaxies, while keeping the flux thresholds
relatively low in order to minimize the contamination by bright
interlopers.  Compared to the field luminosity function (LF) in a
similar redshift range \citep[cf.][]{gabasch+04}, our flux thresholds
are $\sim1$ mag brighter than $M^\star$ in the rest-frame $g$-band, and
$\sim1.8$ mag brighter in $u$. We make sure that the exclusion of
galaxies at the very bright end of the LF does not affect our results
by performing the stacking using a more conservative value of
$M_{g,\mathrm{threshold}}=-23.9~\mathrm{mag}$ (i.e. 1.5 mag brighter
than previously assumed). While the noise from bright interlopers
increases significantly, no systematic effect results from this change
in the masks. Therefore, all the analysis of this paper is conducted
on images masked with
$M_{g,\mathrm{threshold}}=-22.4~\mathrm{mag}$. 
For illustration, 
we plot the apparent $r$-band threshold magnitude as a function of the
absorber's redshift (solid line) in Figure \ref{r_thresholds}.  The
dotted line displays the magnitude of a $M_g=M_g^\star$ galaxy for an
intermediate (Sbc) SED template. The choice of a UV-bright template
yields flux thresholds that become increasingly brighter than
``normal'' $M_g^\star$ galaxies at increasingly higher redshift.

It is worth noting that at low redshift some absorbing galaxies can
possibly be detected in the SDSS images: in fact, the SDSS photometric
sample has a completeness larger than 95\% (for point sources) at
$r<22.2$ mag and thus would allow to detect objects as faint as
$M_g=M_g^\star$ up to $z=0.65$, and 10 times fainter than that at
$z=0.37$ (an analysis of galaxy number counts will be presented in a
forthcoming paper). However, at $z_{abs}\gtrsim0.7$ the detection of
individual galaxies is completely precluded and only the stacking
technique is able to reach the required sensitivity.
In the second step, we build a mask where all stars brighter than 21.0
mag in $r$ are covered with a circular patch of radius $R=1.2\times
a_{25}+3$ (in pixels), where $a_{25}$ is the median of the isophotal
semi-major axis at 25 mag arcsec$^{-2}$ in $g$, $r$, and $i$. For
saturated stars, $a_{25}$ is replaced by three times its value, in
order to avoid the extended halos of scattered light that surround
these stars. For this step we entirely rely on the SDSS-DR4
photometric catalogs (PhotoPrimary table) in order to ensure optimal
star/galaxy separation \citep[see ][]{ivezic+02}.

Finally, the flux-limited masks in the three bands and the star mask
are combined into a final mask, where a pixel is masked if and only if
it is masked in any of the input masks (logical {\sc or}).

\subsection{QSO's PSF subtraction}\label{PSFsub_subsec}

In very high S/N images, like the stacked images, the spread light of
a QSO can be detected out to angular distances of a few tens of
arcsecond and still contribute a non-negligible fraction of the
azimuthally averaged SB linked to the absorbers. In order to suppress
this contribution with the best possible accuracy, the SB distribution
of the QSO is estimated from the high S/N PSF of bright, unsaturated
stars. More specifically, for each QSO we look for a star that
fulfills the following requirements. \emph{(i)} The star must not be
saturated, nor fainter than 17 mag ($r$ band); stars with interpolated
pixels, cosmic rays, or flagged for any photometry defect are
rejected, as well as those stars whose centers happen to lie less than
5 pixels apart from any mask in the frame (see below how PSF star
frames are masked). \emph{(ii)} The star must have been observed
during the same run as the QSO, in the same SDSS camera column, and
within 10 arcmin angular separation from the QSO. These two criteria,
\emph{(i)} and \emph{(ii)}, ensure that the PSF is photometrically
accurate and it is taken in observing conditions that are consistent
with those of the QSO. Among the stars selected in this way, the one
that best matches \emph{(iii)} the size (as determined by the second
moment of the light distribution) and colors ($g-r$ and $r-i$) of the
QSO is chosen as representative of the QSO's PSF. This last selection
aims at minimizing the effects of the space/time\footnote{As the SDSS
images are acquired in drift scan mode, a position offset in the scan
direction implies a time delay as well.} and color dependence of the
PSF. In fact, it must be noted that QSOs are on average bluer than
stars and therefore, have broader PSFs at any given position, on
average.  The selection at point \emph{(iii)} minimizes this effect
and, at the same time, compensates for it by exploiting the spatial
variation of the PSF.  The stars selected to model the PSF have median $r\sim
15.7$ mag, and span the range between 13.4 and 17 mag. The median
difference in color between QSOs and the corresponding stars is
$-0.28$ mag (rms 0.21) in $g-r$ and $-0.06$ mag (rms 0.16) in
$r-i$. The distribution of difference in size is centered at 0, with
0.16 pixels rms.

The masking algorithm described above (\S \ref{mask_subsec}) is then
applied to each selected reference star; however, no lower flux
threshold is applied and all SExtracted sources are retained in the
masks. The subtraction of the PSF frame from the QSO frame is only
possible for pixels that are unmasked in both frames. Thus, in order
to limit the loss of effective detection area around the QSOs, we
``restore'' as many as possible masked pixels in the PSF frame in the
following way. The PSF frame is virtually divided into four quadrants
centered on the reference star. Each masked pixel is then replaced by
the average of all (if any) unmasked pixels at axisymmetric
positions in the other three quadrants. If none of the symmetric
pixels is unmasked, the pixel is kept masked. This ``symmetrization''
process is very effective and results in no more than 0.05\% of the
area masked in 95\% of the PSF frames.

The normalization of the PSF to the QSO, necessary before the subtraction
can be done, is computed using the PSF magnitudes from the SDSS-DR4.

\subsection{The stacking and photometry procedures}\label{stackphotocode_subsec}

Before all images are actually stacked, we rescale them to the same
physical scale at the redshift of the absorber, $z_{abs}$. This is
determined such that at the lowest redshift, $z_{abs}=0.37$, the original SDSS
pixel scale is preserved (0.396 arcsec pixel$^{-1}$); the resulting common
pixel scale is thus 2.016 kpc pixel$^{-1}$. Pixel intensity is
accordingly rescaled in order to conserve the total flux in any
object. The images of reference QSOs are rescaled according to the
redshift of the absorber in the corresponding absorbed QSO. PSF stars
are rescaled with the same geometry as their corresponding QSO.

As individual images contribute to the final stacked image only with
their unmasked regions, it is mandatory that all images are accurately
background subtracted to avoid spatial background fluctuation in the
final stacked image that might strongly reduce the photometric
accuracy. A constant background level in each image (QSOs and stars)
is computed as the median of the unmasked pixel values within an
annulus between 400 and 500 kpc. Two $\sigma$-clipping iterations with
$\pm 3\sigma$ cut are performed to exclude the faint sources that are
left unmasked by the masking algorithm (see above,
\S \ref{mask_subsec}).

All images are also intensity rescaled in order to match a common
photometric calibration \citep[according to the SDSS photometric
standard defined by][]{fukugita+96,smith+02}, that includes the
correction for Galactic extinction. More specifically, the calibrated
pixel intensity $I_{calib}$ is derived from the raw pixel intensity in
the fpC $I_{raw}$ according to the following equation:
\begin{equation}
I_{calib} = I_{raw}\cdot \frac{f_{20,\mathrm{ref}}}{f_{20}}\cdot
10^{0.4\cdot A_\lambda}\label{phototrans}
\end{equation}
where $f_{20}$ and $f_{20,\mathrm{ref}}$ are the pixel counts
corresponding to 20 mag (observed) in the fpC original frame and in
the final reference photometric system, respectively. $A_\lambda$ (in
mag) is the galactic attenuation in the working pass-band, computed
from the \cite{schlegel_dust} dust distribution map.

The final stacked image in each band, for any given sample of QSOs
(either absorbed or reference), is computed such that the resulting
intensity at any pixel is the simple average of the net PSF-subtracted
counts over all corresponding pixels which are unmasked in the
individual images:
\begin{equation}
I_{stacked}=\frac{\sum_{i=1}^{N} mask(i) \times
(I_{QSO,i}-I_{PSF,i})}{\sum_{i=1}^{N} mask(i)}
\end{equation}\label{eqtn_stack}
where $I_{stacked}$ is the intensity of a given stacked pixel,
$I_{QSO,i}$ and $I_{PSF,i}$ are the calibrated, background-subtracted
intensities of the QSO and PSF frames at the corresponding pixel in
the $i^{th}$ image, and $mask(i)$ equals 1 if the pixel is unmasked in
both the QSO and the PSF image, 0 otherwise.  It is worth mentioning
that, at a given position on a stack image, only 6 to 11\% of the
individual images are masked in the low-$z_{abs}$ bin and even less at
higher redshift. The final intensity, $I_{stacked}$, is therefore a
reliable estimate of the average value.
A careful re-determination of the (residual) background level is done
on the final stacked image in an annulus between 350 and 500
kpc. Circular aperture photometry is then performed, by simply
averaging the pixel intensities in a series of circular annuli.

For the purpose of determining uncertainties that take into account
not only formal photometric errors, but also the effects of sample
variance and the correlation between different apertures and bands,
the full covariance matrix is computed by means of a jackknife
algorithm. For any given subsample of $N$ QSOs, the stacking and
photometric measurements in the four bands are repeated $N$ times,
leaving out a different QSO each time. The covariance matrix of the
aperture photometry fluxes obtained from these $N$ realizations is
then multiplied by $N-1$ to give the sample covariance matrix. All the
errors relative to the photometric quantities derived from the primary
aperture fluxes, such as integrated fluxes, colors, and profile shape
parameterizations, are computed using this covariance matrix.

Absorbed and reference QSOs are stacked and analyzed separately as far
as primary aperture photometry is concerned. However, all the results
that will be presented and discussed in the following sections are
obtained from \emph{net} (i.e., absorbed$-$reference QSOs) aperture
photometry. For the error estimates of the derived quantities,
absorbed and reference QSOs are treated as independent datasets, thus
the combined covariance matrices have null cross terms. We note that
this assumption is correct only as a first order approximation; as we
will discuss below in \S\ref{stack_SB_subsec} and Appendix
\ref{refsys_appendix}, there are systematic effects in the stacked
reference QSOs that depend on the redshift and magnitude of the QSO,
thus making the absorbed and reference samples correlated. Neglecting
such second-order error correlation results in slightly
over-estimating errors on the final \emph{net} photometry, especially
within the seeing radius (i.e., the radius that corresponds to the FWHM
of the PSF image).

All the stacking and photometric measurements described in this
subsection and in the following sections are performed using a
dedicated package of programs and utilities, which are developed in C
language by S.Z. The package relies on the \texttt{cfitsio} libraries
\citep{cfitsio} for input/output interface and makes use of a number
of Numerical Recipes routines \citep{NR}. The analysis presented in
this paper results from the processing of 0.5 Terabyte of imaging
data.

\section{The spatial distribution of MgII absorbing galaxies}\label{spatial_sec}
In this Section we present an analysis of the surface brightness
distribution of the excess light around absorbed QSOs and infer from
that the spatial distribution of absorbing galaxies with respect to
the MgII systems probed by the QSOs sight-lines. In particular, in \S
\ref{stack_SB_subsec} we present the SB profiles extracted from the
stack images, and parameterize their shape as a function of the
absorption rest equivalent width, $W_0(\lambda2796)$, and absorption redshift,
$z_{abs}$. In \S \ref{SB_impact_subsect} we will show that the SB
profile is directly related to the impact parameter distribution
weighted by the luminosity of the absorbing galaxy. Such a
distribution will be explicitly derived and discussed in \S
\ref{impactpars_sec}.

\subsection{Stack images and surface brightness profiles}\label{stack_SB_subsec}

\begin{figure*}
\plotone{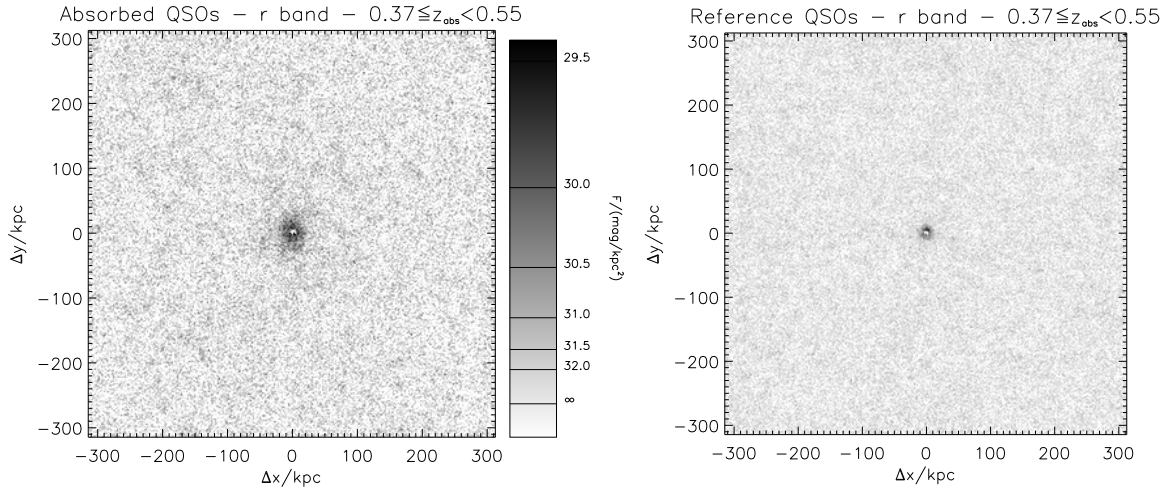}
\caption{The $r$-band stack images of absorbed (\emph{left panel}) and
corresponding reference QSOs (\emph{right panel}) in the redshift
range $0.37\leq z_{abs}<0.55$. The sample of reference quasars is four
times larger which results in a lower noise level. Both images are
PSF-subtracted and are visualized with the same color scale, as
indicated by the color bar marked in units of magnitudes per kpc$^2$
($\infty$ corresponds to the background level). The excess of light around the
absorbed QSOs is apparent.\label{map_gridz0_r}}
\end{figure*}

In Figure \ref{map_gridz0_r} we report an example of typical final stack
images for the absorbed QSOs (left panel) and for the corresponding
control sample of reference QSOs (right panel). More specifically,
this is the $r$-band stack for the low $z_{abs}$ sample that includes all
$W_0$. The two panels reproduce a region of 600$\times$600 kpc$^2$
projected at $z_{abs}$, centered on the QSOs. The grey level intensity
maps the surface brightness linearly. The corresponding levels,
expressed in physical units of apparent magnitudes per projected
kpc$^2$, are indicated by the vertical color bar between the two
panels. The infinity level ($\infty$) indicates the background signal.
In both panels, we note that the intensity of the very central pixels
is zero, within the typical background noise uncertainties. This is
obtained \emph{by construction}, by subtracting the PSF from each
individual stacked image. The fact that these central pixels are
zeroed to such high accuracy is indicative of the good quality of
our PSF subtraction algorithm.  What is most important to notice in
Figure \ref{map_gridz0_r} is the SB excess around the QSOs, beyond the
very central pixels. Although an excess is seen in both panels, it is
apparent that the SB excess around absorbed QSOs is much more intense
and extended than around the reference QSOs; it can be seen, even by
eye, that the excess around the absorbed QSOs extends well
beyond 50 kpc, whereas nothing is apparent around the reference QSOs
beyond 10-15 kpc. This effect is even more significant if one
considers that the reference QSO image results from 4 times as many
QSOs as in the absorbed QSO image (see \S\ref{ref_sample_sec}), and
therefore has twice as low background noise.

The SB excess observed around the reference QSOs is a systematic
effect which is mainly caused by the emission of the QSO's host
galaxy/environment. We analyze the origin of this effect in
greater detail in appendix \ref{refsys_appendix}. Regarding
the measurement of the light of absorbing galaxies, we can treat
the SB excess found around reference QSOs as a pure
systematic \emph{additive} contribution, which is inherent to any
subsample of QSOs with a given redshift and magnitude
distribution. \emph{Therefore, for each subsample of absorbed QSOs, the
final stack image (SB profile) of the corresponding reference QSOs is
subtracted from the absorbers' one to obtain a \emph{net} image (SB
profile).}

Next we analyze the SB distribution of the galaxies responsible for
the MgII absorption by extracting one-dimensional SB profiles from
circular annuli, within which the average SB is computed. The spacing
between subsequent apertures is chosen to ensure adequate S/N in all
apertures; this results in a progression that increases slightly
faster than a geometric one.
\begin{figure*}
\plotone{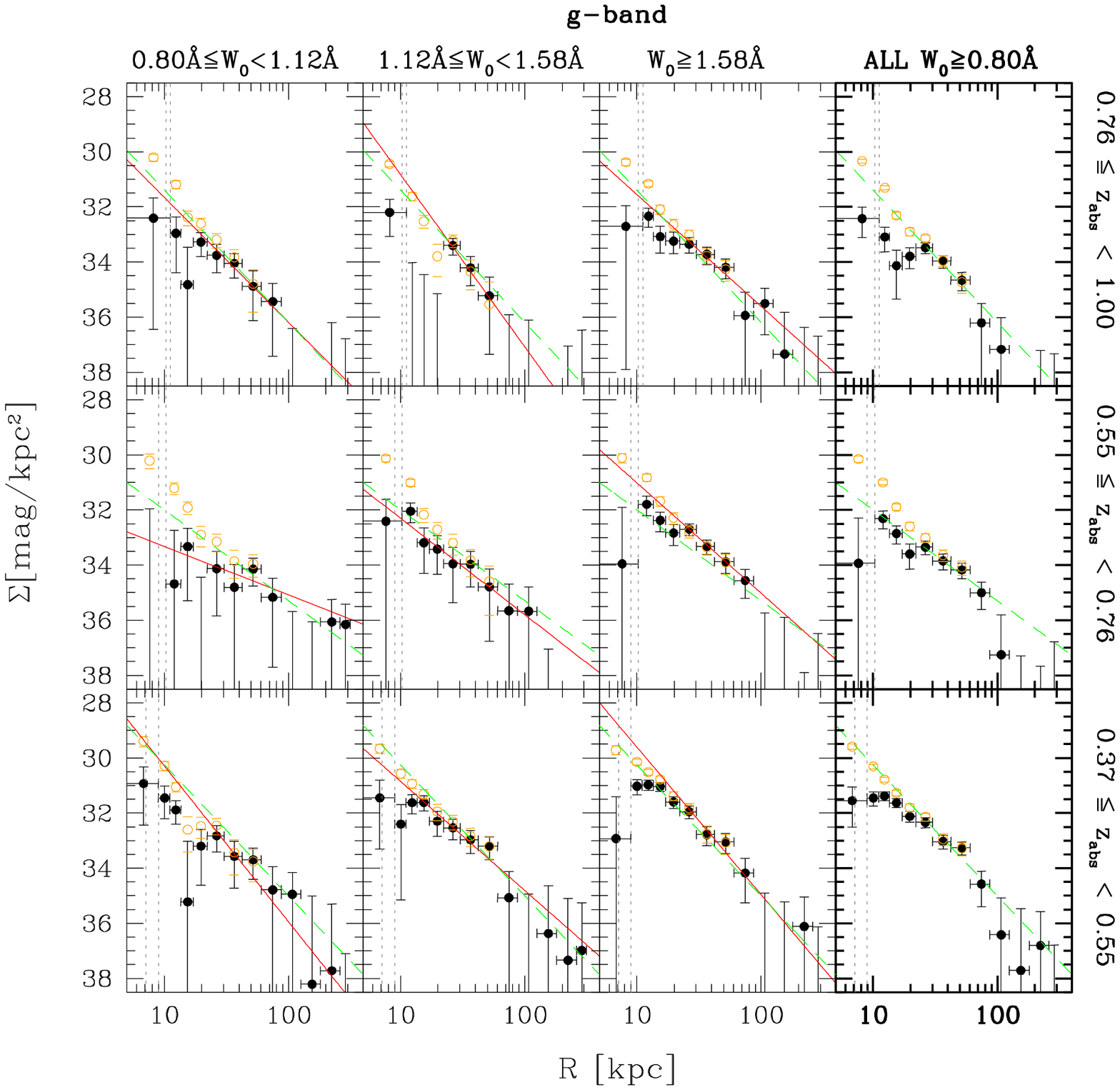}
\caption{The $g$-band surface brightness profiles of the light around
absorbed QSOs, in different bins of redshift $z_{abs}$ (indicated by
the labels on the right-hand side) and rest-frame equivalent width of
the 2796\AA~line (labels along the top of the figure). Black filled
dots with error bars represent the \emph{net} SB, i.e., after
subtracting the residual systematic contribution measured from the
reference QSOs inside 50 kpc. The orange open circles with error
bars show the SB prior to the subtraction of the SB of the
reference sample. The dashed green line is the same in all four panels
of each row, and displays the best power-law fit for the sample
including all $W_0(\lambda2796)$ in that $z_{abs}$ bin. The power-law fits for
the three $W_0(\lambda2796)$ subsamples in each redshift interval are shown
by the red solid lines. In each panel, vertical dotted lines mark the
distance corresponding to the 1.4\arcsec\ average seeing (FWHM) at the
minimum and maximum redshift in the subsample.
\label{profiles_all_g}}
\end{figure*}
\begin{figure*}
\plotone{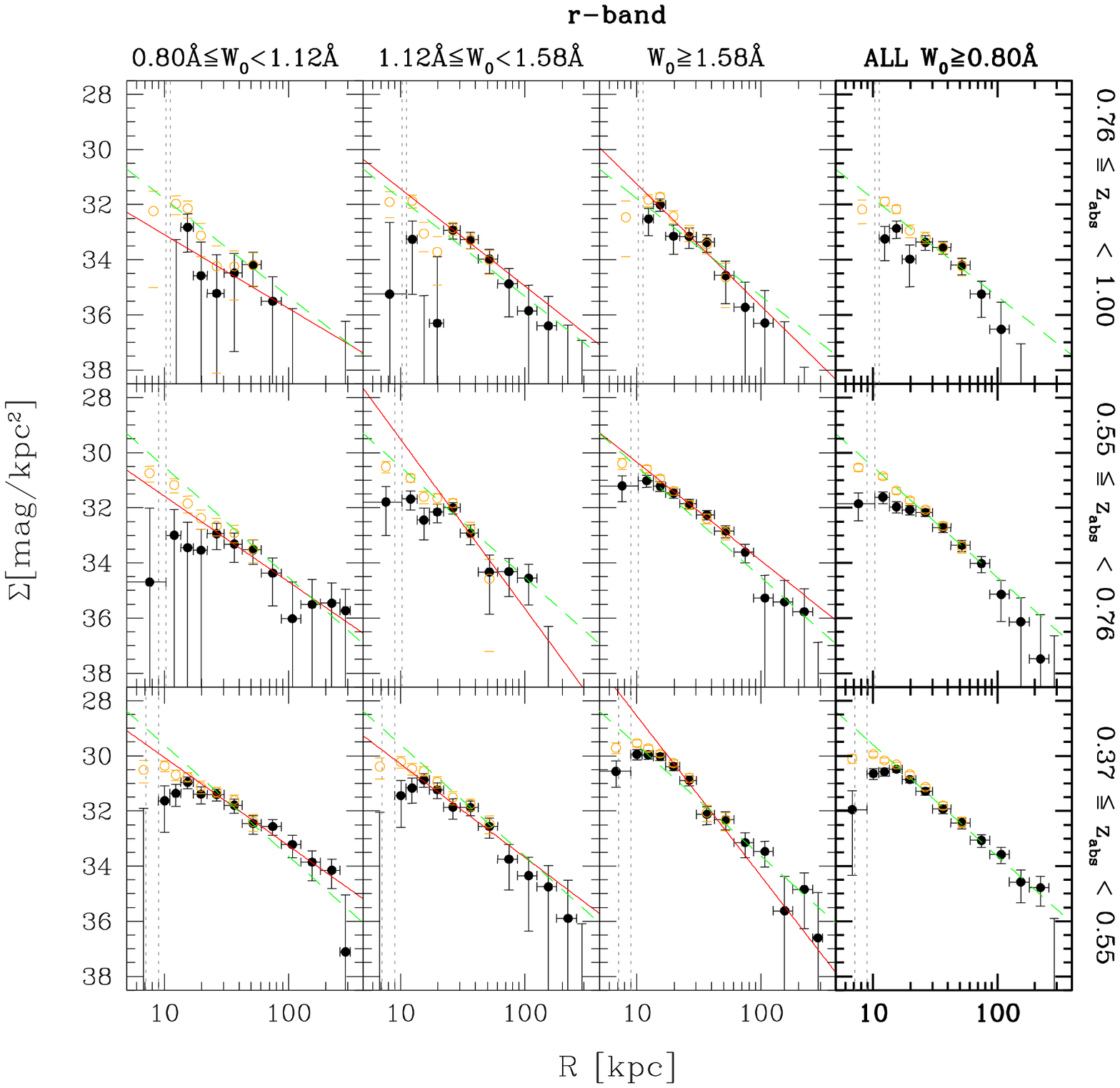}
\caption{Same as Figure \ref{profiles_all_g}, but for the
$r$-band.\label{profiles_all_r}}
\end{figure*}
\begin{figure*}
\plotone{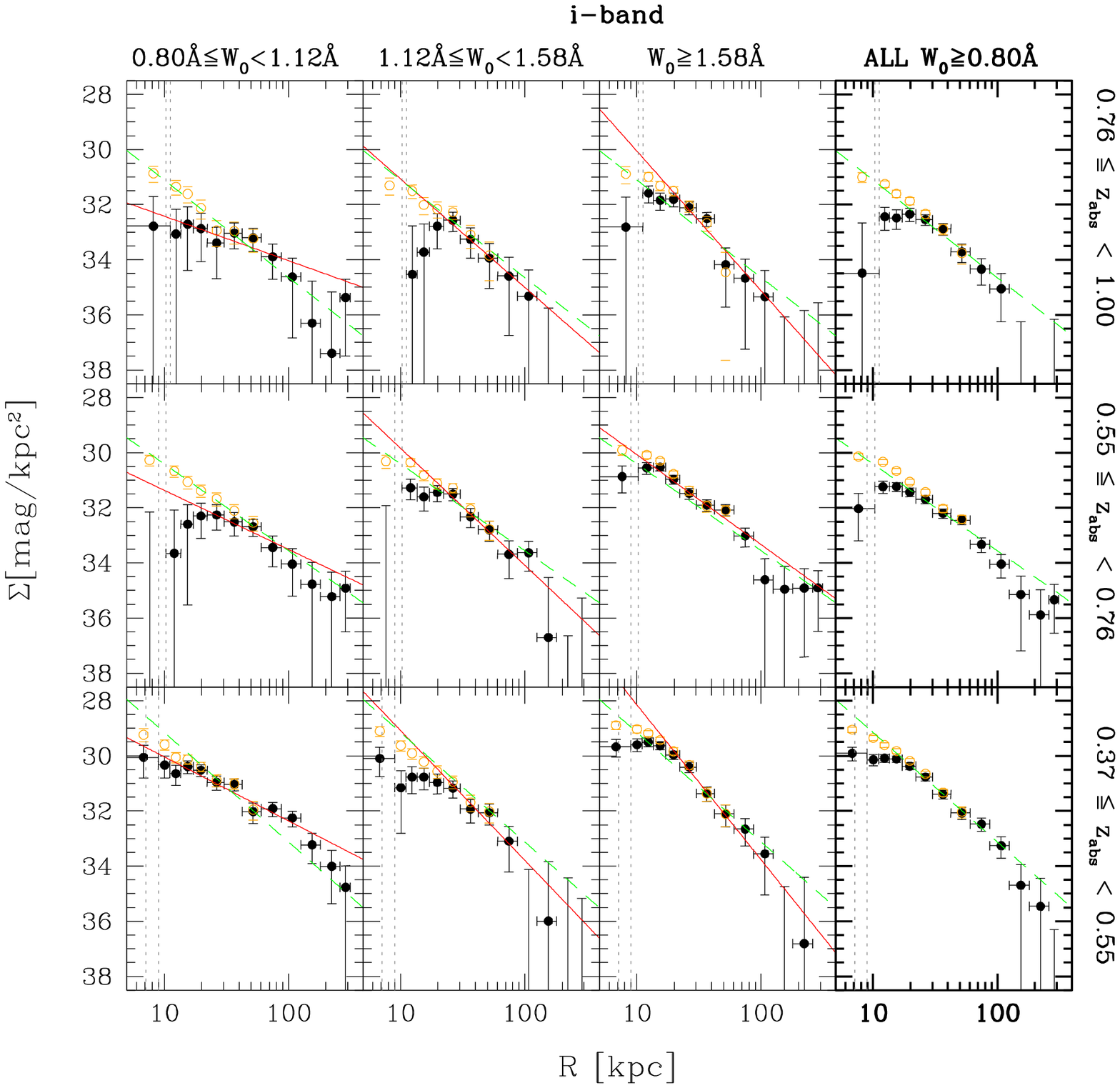}
\caption{Same as Figure \ref{profiles_all_g}, but for the
$i$-band.\label{profiles_all_i}}
\end{figure*}
\begin{figure*}
\plotone{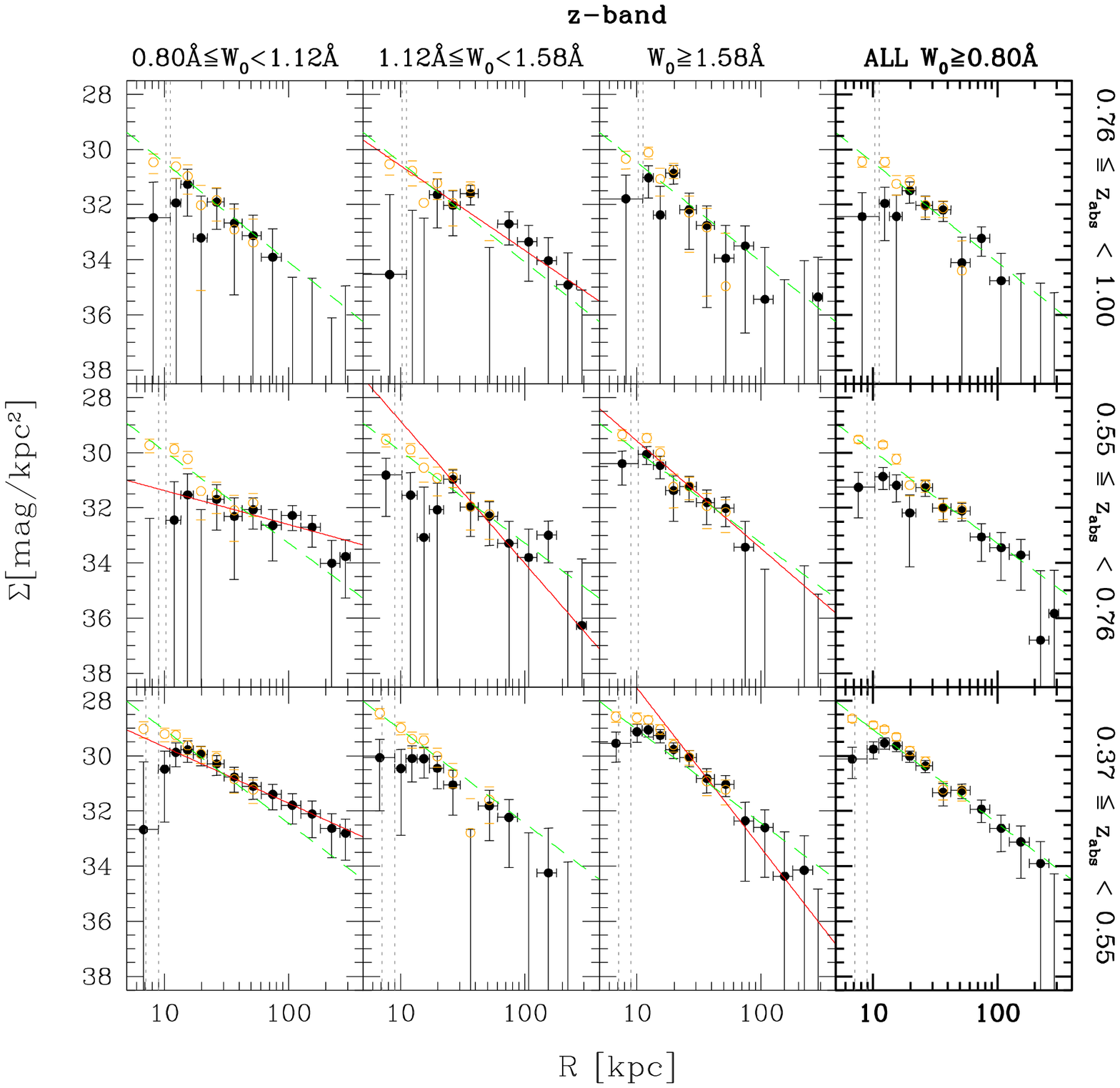}
\caption{Same as Figure \ref{profiles_all_g}, but for the $z$-band. Note
that for three out of the nine $W_0$ subsamples no power-law fits (red
lines) are reported, because the large uncertainties make the fit
meaningless.\label{profiles_all_z}}
\end{figure*}
The SB profiles of the absorbers in all our subsamples are presented
in Figures \ref{profiles_all_g} to \ref{profiles_all_z}, for the $g$,
$r$, $i$, and $z$ band, respectively. Each of these figures is
organized in twelve panels: three rows corresponding to the three
$z_{abs}$-bins (from the lowest in the bottom row to the highest in
the top one), and four columns for the low-, intermediate- and
high-$W_0$ systems, plus the total of the $z_{abs}$-bin
(``\textsc{all}''), from left to the right. The impact parameter $R$
is reported on the x-axis in logarithmic scale, while the surface
brightness is given on the y-axis in units of apparent magnitudes per
projected kpc$^2$.  The \emph{net} SB profile is shown by the black
filled dots with error bars. Vertical error bars represent the
jackknife estimate of the SB uncertainty computed as above
(\S\ref{stackphotocode_subsec}), while the horizontal error bars just
mark the width of the radial bin. The orange open circles with upper
and lower ticks indicate the \emph{uncorrected} SB determination and
relative jackknife uncertainty, prior to the subtraction of the
systematic excess measured from the reference QSOs sample. It must be
noted that beyond 50 kpc no subtraction is performed (hence no orange
points) as the reference stack images are completely consistent with
zero at this impact parameter, and the subtraction would just increase
the noise in the \emph{net} profile.

As a preliminary sanity check, we note that the signal decreases from
the lowest to the highest $z_{abs}$ bin by roughly 2 mag, which is
consistent with the cosmological dimming expected for the covered
redshift range.  The first important result that we can extract from
Figures \ref{profiles_all_g} to \ref{profiles_all_z} is that there is
light that cross-correlates with the MgII absorbing clouds out to
100-200 kpc projected distance. This clearly indicates that MgII
absorption systems probe the intergalactic medium out to quite large
galactocentric distances, and may possibly arise in complex galaxy
environments.

It must be noted that the PSF subtraction artificially introduces the
constraint $\mathrm{SB}=0$ within the first resolution element
(corresponding to the seeing). In Figure \ref{profiles_all_g}, the two
vertical short-dashed lines in grey reported in each panel show, at
the lowest and highest $z_{abs}$ in the bin, the value of 1.4\arcsec\
which corresponds to the median seeing (FWHM) in SDSS images. Due to
the above normalization we expect the signal to be strongly decreased
within this radius ($\sim10$kpc), and significant deviations from the
``true'' SB profile to occur out to twice this distance from the QSO
($\sim 20$kpc) as a result of the breadth of the seeing distribution
in our images.

The SB profiles between $R=20$ kpc and $R=100$ kpc in all bands are
generally well approximated by a single power-law with index ranging
from $-1$ to $-2.5$. Between 10 and 20 kpc (roughly corresponding to
one to two times the median seeing value at all redshifts) we observe
a break and a deficit with respect to the powerlaw, which can be
mostly ascribed to the central normalization. However, real departures
from a powerlaw profile in the inner 20 kpc cannot be excluded based
on the present analysis. Also, it is worth noting that previous
studies show that very few galaxies, if any, are associated with MgII
absorbers at impact parameters $<10$ kpc
\citep{SDP94,churchill+05}. Thus the signal that we miss from the
central part of the profile is probably negligible. Beyond 100 kpc the
photometric uncertainties are too large to derive any reliable
information about the profile shape, although significant detections
are obtained in several bands and bins.

Power-law fits of the form
\begin{equation}
S(R)=S(\overline{R}){\left(\frac{R}{\overline{R}}\right)}^{\alpha}
\end{equation}
are performed in the impact parameter range
$20~\mathrm{kpc}<R<100~\mathrm{kpc}$ by minimizing the $\chi^2$
computed using the full data covariance matrix. The slopes, $\alpha$,
for each absorber bin are reported in Table \ref{shape_pars_tab}
(columns 2-5, for each band respectively), along with their upper and
lower confidence limits (1-$\sigma$), which are obtained from the
2-dimensional $\chi^2$ distribution in the $S(\overline{R})-\alpha$
parameter space. Column 6 reports the weighted average of the slopes
in the four bands, with relative error\footnote{We note that
the error that is quoted in this case is slightly under-estimated
because we neglect that the slope determinations in the four bands are
not completely uncorrelated.\label{w_ave_note}}. The best fitting
power-laws are also shown in the panels of Figure \ref{profiles_all_g}
to \ref{profiles_all_z}.  They are shown in red for the three EW 
bins\footnote{No red line is reported in three panels for the $z$ band
(Figure \ref{profiles_all_z}) as the very low S/N of the profiles
resulted in 1-$\sigma$ uncertainty range on the power-law slope larger
than 3.} and with a dashed green line for the ``\textsc{All}''-EW
panels.
The fit to the ``\textsc{All}''-EW sample is also reported as a dashed
green line in all other panels to facilitate a prompt comparison
between different subsamples.

Looking at the ``\textsc{All}''-EW samples we observe that, within the
quoted uncertainties, all profiles in the four observed bands and in
the three redshift bins are consistent with a common slope, $\alpha$,
between $-1.4$ and $-1.6$. There is no evidence for any redshift
evolution of the power-law slope for the average SB distribution of
absorbers related galaxies between $z_{abs}=0.37$ and 1. This is
confirmed and better constrained by the weighted average of the slopes
in the four bands reported in column 6 of Table \ref{shape_pars_tab}:
their values are consistent within 1-$\sigma~(\leq 0.17)$ among the
three $z_{abs}$ bins.
\begin{figure*}
\plotone{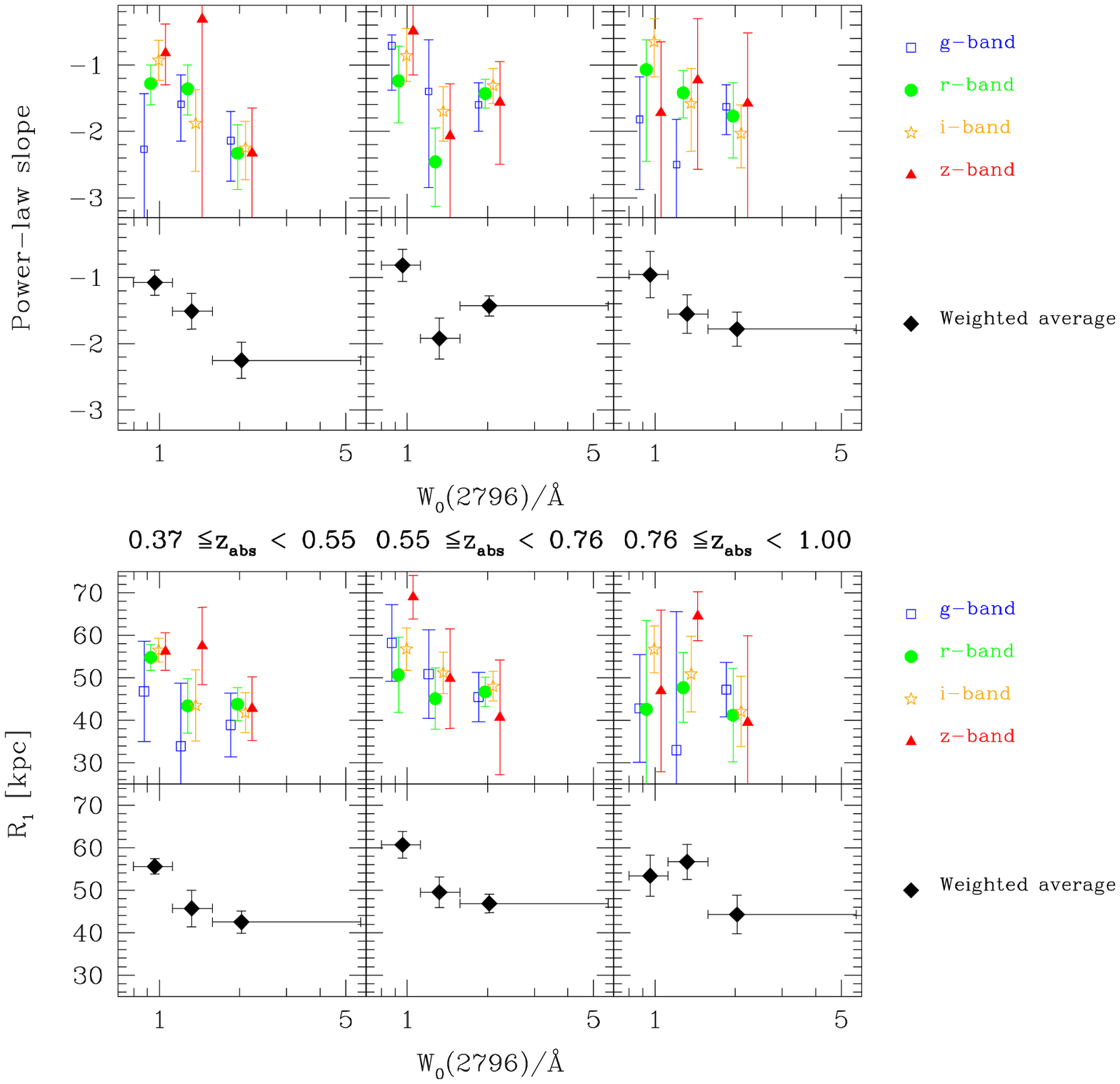}
\caption{The correlation between the shape of the surface brightness
profiles and the absorption strength. In the \emph{top} panels, the
slope of the best fit powerlaw slope
($20~\mathrm{kpc}<R<100~\mathrm{kpc}$) is plotted against the
absorption strength, for the three $z_{abs}$ bins, as indicate by the
labels in the middle. Blue open squares, green filled circles, orange
empty stars, and red filled triangles are shown in the upper row for
the $g$-, $r$-, $i$-, and $z$-band respectively.  The black filled
diamond in the lower row is the weighted average of the four
bands. The horizontal error bars show the extent of each EW bin. The
\emph{bottom} panels are the same as the top ones, but for the first
moment of the SB distribution $R_1$ (see text).\label{shapes}}
\end{figure*}
A second interesting result that emerges from
Figure \ref{profiles_all_g} to \ref{profiles_all_z} can be derived by
comparing different EW bins, especially in the most sensitive $r$ and
$i$ bands for the low-$z_{abs}$ systems. It is apparent that the
high-$W_0$ MgII systems have systematically steeper power-law profiles
than the weaker systems. Despite the relatively large uncertainty of
our slope measurements, this trend is clear in the low-$z_{abs}$ bin
from the values reported in Table \ref{shape_pars_tab}. We also plot
these values in the top panels of Figure \ref{shapes}, where different
symbols represent the measurements in the four bands as a function of
the absorption strength, binned as defined in
\S\ref{def_subsample_sec}. The three columns of panels are for the
three $z_{abs}$ bins. The large filled diamonds in the lower row
represent the weighted average of the four bands, while the values for
the different bands are shown in the upper row. At the lowest
$z_{abs}$ the trend of steepening of the profile with absorption
strength is unambiguously detected in $r$ and $i$ and statistically
confirmed by the weighted averages. The power-law slope steepens from
$-1.1$ in the weakest systems, to $-2.3$ in the strongest. Similar
trends (in particular for the weighted averages) are seen at higher
redshift too.

We further investigate the systematic variation of the surface
brightness profiles as a function of the absorption strength in a
model independent, non-parametric way, by means of the first moment of
the SB distribution in $R$. This is defined as:
\begin{equation}
R_1=\frac{\int\limits_{10~\mathrm{kpc}}^{100~\mathrm{kpc}}S(R)~R^2~dR}
{\int\limits_{10~\mathrm{kpc}}^{100~\mathrm{kpc}}S(R)~R~dR}
\end{equation}
$R_1$ is thus given in kpc and can take any value between 10 and 100
(see above in this section why 10 kpc is chosen as lower limit).  As
we will demonstrate in the next \S\ref{SB_impact_subsect}, $R_1$ can
be interpreted as the luminosity-weighted average impact parameter of
MgII absorbing galaxies. Small values of $R_1$ characterize centrally
concentrated light profiles, while less concentrated profiles have
larger $R_1$. The values of $R_1$ computed for the four bands and
their statistical errors are reported in columns 7 to 10 of table
\ref{shape_pars_tab}, for all bins; the weighted average among the
four bands\footnote{No correlation between the four bands is taken
into account in this case, possibly resulting in underestimating the
associated error by a few per cents.} is reported in column 11. We
note that $R_1$ is remarkably constant among different redshift. For
``\textsc{All}'' EW $<R_1>$ is 48 kpc. However, for strong systems
this value decreases to 43 kpc (at low-$z_{abs}$, similar at higher
redshift) while it approaches 60 kpc for the low-$W_0$ systems. As for
the power-law index, we also plot these values in fig. \ref{shapes}
(bottom panels). Again, these plots confirm that stronger systems have
more centrally concentrated profiles. The effect appears statistically
significant at low and intermediate $z_{abs}$, especially when the
weighted average of the four bands is considered. For the highest
redshift bin the trend is much more uncertain because of the lower
S/N, but is still consistent with the lower redshift samples.

\subsection{From surface brightness profiles to the impact parameter 
distribution of MgII systems}\label{SB_impact_subsect}

In this section we provide a derivation of the link between the
surface brightness profiles obtained from the stacked images and the
impact parameter distribution of the MgII absorbing clouds. First, we
demonstrate that if each MgII absorber is linked to only one galaxy,
then the SB at any radius $R$ is proportional to the luminosity
weighted impact parameter distribution.

Let us consider initially the case of galaxies with fixed luminosity
$\bar L$, surrounded by MgII absorbing clouds resulting in a
probability of producing an absorption along the sight-line
\begin{equation}
dP(R,\bar L)=p(R,\bar L)~dA,\label{dP_eq}
\end{equation}
where $p(R,\bar L)$ is the differential probability per unit area at
given luminosity $\bar L$, which we assume to be independent of 
position angle. Although this assumption might not be verified for
single galaxies, it is always automatically realized in the case of 
stacked images. Let us now consider a set of QSOs homogeneously distributed
in sight-lines around $\bar L$ galaxies. The number of QSOs that are
actually absorbed at given impact parameter $R$ is thus given by
\begin{eqnarray}
dN_{abs}(R,\bar L)&=&N_{abs}(\bar L)~dP(R,\bar L)\\
&=&2\pi N_{abs}(\bar L)p(R,\bar L)~R~dR,
\label{dn_eq}
\end{eqnarray}
where $N_{abs}(\bar L)$ is the total number of QSOs with detected
absorption. If we now stack all the absorbed lines-of-sight, we find
the azimuthally averaged surface brightness:
\begin{equation}
SB(R,\bar L)= \frac{dN_{abs}(R,\bar L)\cdot\bar L}{2\pi~R~dR\cdot
N_{abs}(\bar L)}.\label{eq_SB_long}
\end{equation}
This is valid in the limit that the effective extent of each galaxy is
significantly smaller than $R$.  As galaxies with $L\sim L^\star$ have
typical effective radii of the order of 3 kpc, at $R\gtrsim 10$~kpc
Equation \ref{eq_SB_long} is a fair approximation. Substituting
Equation \ref{dn_eq} into this expression we see that the area from
the probability cancels out with the normalization of the azimuthal
average and we have
\begin{equation}
SB(R,\bar L)= \bar L \cdot p(R,\bar L)\,.
\label{SBsingle_eq}
\end{equation}
Therefore, the SB profiles presented in the previous figures are
directly proportional to the probability of producing absorption at a
given distance from a galaxy.
If we now consider contributions from galaxies of different
luminosities, we can write
\begin{equation}
SB(R)=\int L~\frac{dp(R,L)}{dL}~dL,\label{SBallL_eq}
\end{equation}
that is to say that the SB at any radius $R$ is proportional to the
luminosity-weighted impact parameter distribution.

In order to take into account the effect of satellite galaxies, or
environment if a MgII absorber is associated with more than one galaxy
(for example in a group), we can extend the argument that leads to
Equation \ref{SBsingle_eq}. Let now $\bar L$ be the luminosity of the
brightest associated galaxy and $R$ measure the impact parameter from
it. The presence of secondary galaxies increases the total luminosity
per absorber and produces a spatially extended contribution to the
average SB.
By defining $K_{\bar L}(R)$ as the average surface brightness around
the location of an $\bar L$ galaxy, including all its satellites and
possible companions, and normalized dividing by $\bar L$, we can then
rewrite Equation \ref{SBsingle_eq} as
\begin{equation}
SB(R,\bar L)= C_{\bar L}\cdot\bar L \cdot (p(R,\bar L)\otimes K_{\bar
L}(R))
\end{equation}
where the $C_{\bar L}$ accounts for the increased total luminosity per
absorber and $p(\bar L, R)$ is convolved with $K_{\bar L}(R)$.
Finally, integrating over luminosity we obtain the analogous of
Equation \ref{SBallL_eq} for multiple associated galaxies
\begin{equation}
SB(R)=\int C_{L}\cdot L\frac{d(p(R,L)\otimes K_{L}(R))}{dL}~dL
\end{equation}
It is immediately seen that this last equation simplifies to Equation
\ref{SBallL_eq} when $C_L=1$ and $K_{L}(R)=\delta_{\mathrm{Dirac}}(R)$
for every $L$, i.e., when there is only one associated galaxy per
absorber, with negligible spatial extent.

\subsubsection{How many galaxies per absorber?}

Although the association of an individual galaxy with an absorber
is a somewhat ill-defined procedure, in order to
interpret the SB profiles and especially the SEDs, it is highly
relevant to understand how many (bright) galaxies per absorber
contribute to the cross-correlating light.
Preliminary results by M\'enard et al. (in preparation) indicate that
at $z_{abs}<0.6$ there is on
average one bright galaxy detected in excess within $\sim100$ kpc
around absorbed QSOs in comparison to reference QSOs. 
As we will show in \S\ref{SED_subsec}, the average total luminosity
$<C_{L}\cdot L>$ (within 100 kpc, in the low-$z_{abs}$ bin) is
$m_r = -21.5~(\pm0.2)$ or $\approx 2.8\times 10^{10}L_{\odot,r}$,
corresponding to the $L^\star_r$ of the local field luminosity
function in the SDSS by \cite{blanton+03} \citep[we adopt ][to convert
from magnitudes in the z=0.1 system to the rest-frame]{blanton+05}. If
we compare this value to the LF at $z\sim0.5$, then this luminosity
corresponds to $0.5~L^\star$ \citep[e.g. in COMBO-17,
][]{wolf+03}. \emph{Such a relatively low luminosity inside a 100 kpc
circle shows that, on average, strong MgII absorbers with
$0.8$\AA\ $<W_0\lesssim3$\AA\ do not reside in dense galaxy environments.}
This statement however cannot be proved for stronger systems (with
$W_0\gtrsim3$\AA) alone as their relative contribution to the signal
is substantially lower.

Therefore, in the following we will interpret the SB profile $SB(R)$
to first order approximation as the luminosity weighted average
probability of intercepting an absorber at a given impact parameter $R$
from a bright galaxy. Regarding the corresponding spectral energy
distribution (SED), the inclusion of small companions and satellite
galaxies in the integrated light is expected to have a minor influence
on our estimates of total luminosity and spectral properties. In
practice we will treat such properties as the properties of a single
absorbing galaxy.

\subsection{The impact parameter distribution of MgII absorbers}
\label{impactpars_sec}
\begin{figure*}
\plottwo{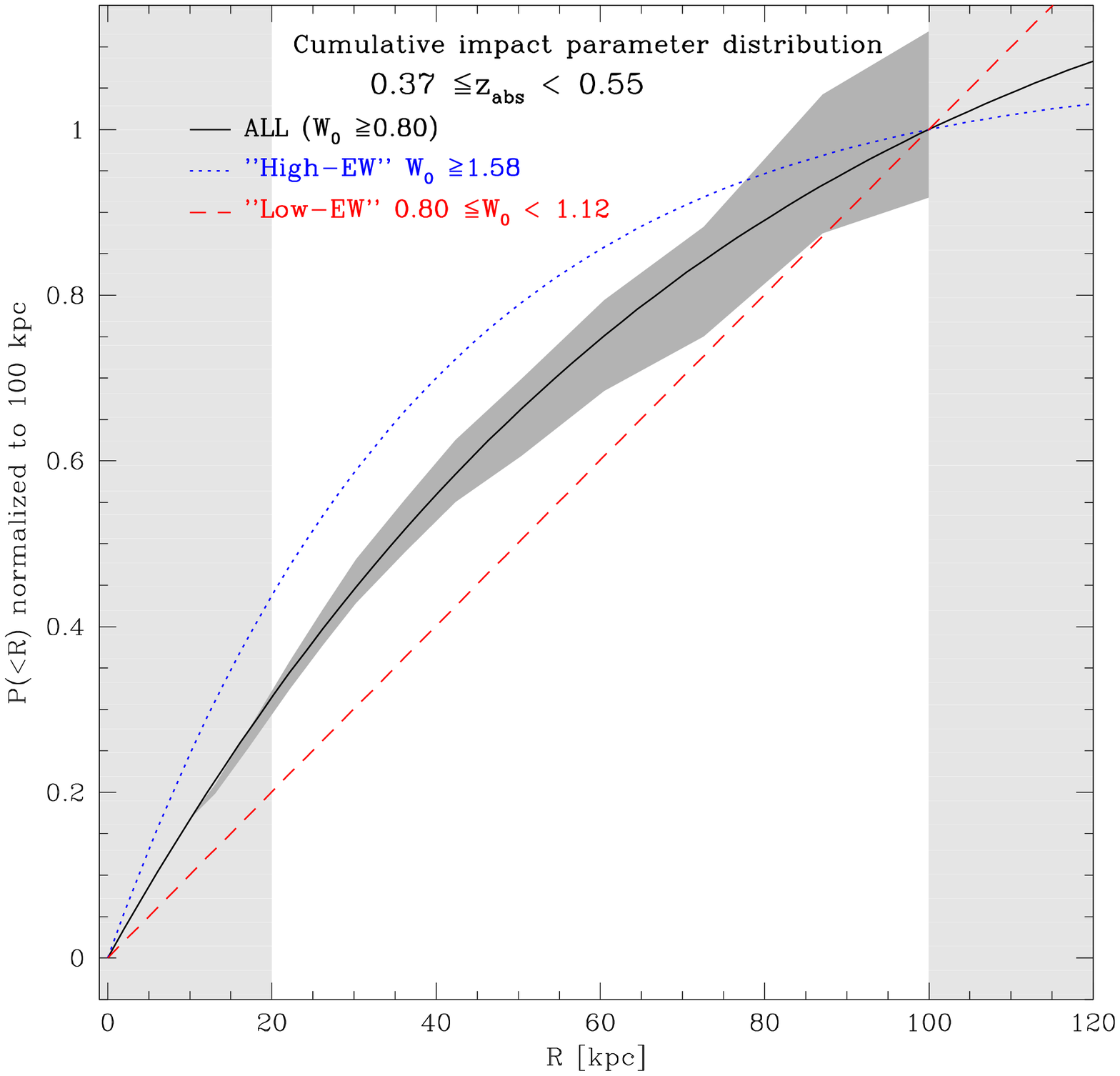}{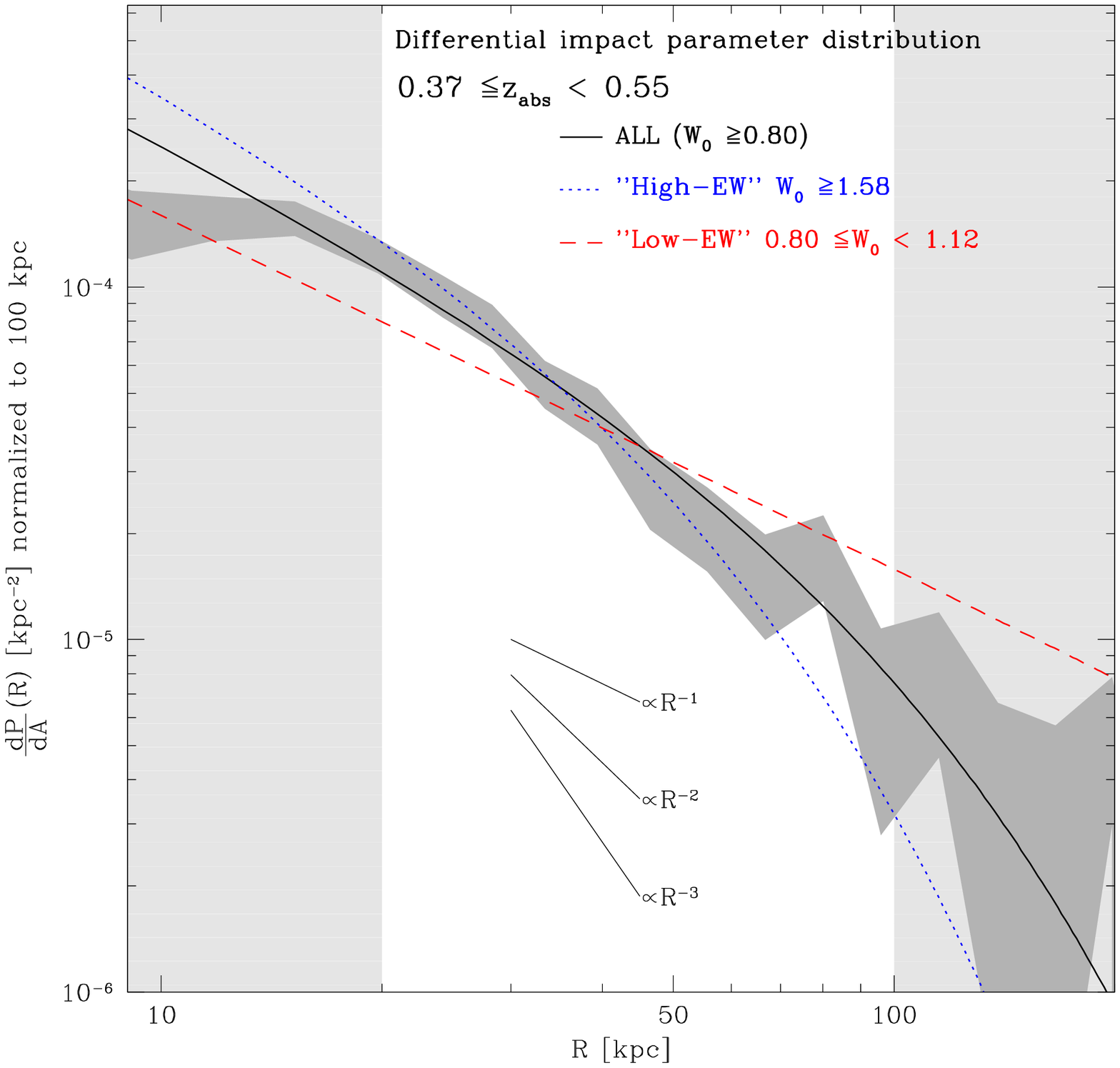}
\caption{Luminosity-weighted impact parameter distributions for the
low $z_{abs}$ sample, normalized between 0 and 100 kpc. The \emph{left
panel} shows the cumulative probability of observing a galaxy at
impact parameter $<R$ from an MgII-absorbed QSO (see
Equation \ref{cumulative_eq}). The three lines are the $\tanh$ analytic
fits to the distributions for all (black solid line), high-$W_0$ (blue
dotted line), and low-$W_0$ (red dashed line) systems.  The dark grey
shaded area represents the 1-$\sigma$ confidence interval empirically
derived from the photometry for the whole low-$z_{abs}$ sample. Data
points in the regions shaded in light grey were excluded from the
fit. The \emph{right panel} reports the differential probability
$\frac{dP}{dA}(R)$, as analytically derived from the cumulative
one. Different line styles and shadings represent the same quantities
as in the left panel. Three lines with different logarithmic slopes
are also plotted for easy comparison.}\label{distr_functions}
\end{figure*}
Based on the arguments presented in the previous paragraphs, we
translate the observed SB distributions around the absorbed QSOs into a
luminosity weighted probability of intercepting an MgII-absorbing
system at a given impact parameter from a galaxy. In differential form
this is directly given by the SB profiles (see
\S\ref{stack_SB_subsec}), modulo an arbitrary normalization factor. As
already noted, simple power laws are able reproduce the SB profiles,
and hence also the differential probability profiles, reasonably
well. Here we introduce the integral form
\begin{equation}
P(<R)=\int\limits_{0}^R 2\pi~R^\prime~dR^\prime\int L
\frac{dp(R^\prime,L)}{dL}dL\label{cumulative_eq}
\end{equation}
where $P(<R)$ is the luminosity weighted probability of measuring an
impact parameter between 0 and $R$. This form of cumulative
probability distribution is particularly useful for defining
characteristic scale lengths.  According to Equations \ref{dP_eq} and
\ref{SBallL_eq}, this is simply proportional to the apparent
luminosity integrated between $0$ and $R$, $F(<R)$. The
proportionality factor between the two distributions is just given by
the total integral flux.

Observationally, however, we are faced to two limitations. At large
radii small background uncertainties yield huge uncertainties in the
estimate of the total integral flux. Therefore we are forced to limit
the flux integral to an arbitrary large radius where the background
level uncertainty is negligible. This is chosen as 100 kpc, which
implies that the impact parameter distribution will be normalized to 1
between 0 and 100 kpc.
The second limitation is that at small radii ($R\lesssim~10$-20 kpc) the
finite size of the PSF does not allow a reliable estimate of the flux
(see \S\ref{stack_SB_subsec}), and therefore we can produce flux
growth curves only from 10-20 kpc outwards.

We overcome this problem by extrapolating the growth curve to $R=0$
using an analytic fit to the empirical growth curve. Only the
conservative range between 20 and 100 kpc is used. As a convenient
fitting function we choose an hyperbolic tangent generalized as
follows:
\begin{equation}
F(R)=\alpha\tanh[\beta\cdot(R-\delta)]+\gamma
\end{equation}
From this fit we can derive the (extrapolated) total flux between 0
and 100 kpc $F_{100}\equiv F(100)-F(0)$, and hence express
\begin{equation}
P(<R)=c_1\tanh[c_2\cdot(R-c_3)]+c_4,\label{tanhfit}
\end{equation}
where
\begin{eqnarray}
c_1=\alpha/F_{100}\\
c_2=\beta\\
c_3=\delta\\
c_4\equiv c_1\cdot\tanh(c_2\cdot c_3).
\end{eqnarray}

The $i$-band photometry in the low-redshift sample is used to derive
the cumulative impact parameter distribution $P(<R)$. Thus what we
present here is weighted on the rest-frame $g$-band light. In table
\ref{tanh_fit_coef} we report the four coefficients of the analytic
formula given in Equation \ref{tanhfit} for several bins of $W_0$.
These analytic expressions are meant to be handy representations for
the $P(<R)$ extracted from a robust statistical analysis. No physical
model is implied. However, the reader should be aware that below 20
kpc these formulae are extrapolated and that the normalization within
100 kpc is arbitrary.

In the left panel of Figure \ref{distr_functions}, $P(<R)$ is plotted
for all EWs as a solid line. The dark grey shaded area is the
1-$\sigma$ confidence region for $P(<R)$ as it is directly derived
from the $i$-band (roughly rest-frame $g$-band) flux growth curve of
the entire sample at low redshift. The integral impact parameter
distributions for the low- and high-$W_0$ systems are overplotted in
the same panel with  dashed red  and dotted blue lines,
respectively. Again, the higher concentration of stronger systems is
evident.  This is also quantified by the values of $R_{50}$ and
$R_{90}$ reported in Table \ref{tanh_fit_coef} that represent the
distances from the ``parent'' galaxy within which the probability of
finding an absorber system is 50 and 90\% respectively (as computed
from Equation \ref{tanhfit}).
\setcounter{table}{1}
\begin{deluxetable*}{lrrrrrr}
\tablewidth{0pt} \tablecolumns{6} \tabletypesize{\small}
\tablecaption{Parameters of the impact parameter distribution
$P(<R)$ ($0.37\leq z_{abs}<0.55$)\label{tanh_fit_coef}}

\tablehead{
\colhead{$W_0(\lambda2796)$} & \colhead{$c_1$} &\colhead{$c_2$} &\colhead{$c_3$} &\colhead{$c_4$} &
\colhead{$R_{50}$} & \colhead{$R_{90}$} \\
                          &                 & \colhead{kpc$^{-1}$} &\colhead{kpc}   &
                & \colhead{kpc} & \colhead{kpc}
}
\startdata
\textsc{All}    & 7.900  & $7.275\cdot10^{-3}$  & $-163.91$ & $-6.567$  & 34.6 & 81.4  \\
\textsc{low-$W_0$}   & 13.555 & $7.406\cdot10^{-4}$  & $-29.98$  & $-0.301$  & 49.9 & 90.0  \\
\textsc{high-$W_0$} & 11.177 & $1.359\cdot10^{-2}$  & $-109.89$ & $-10.103$  & 23.9 & 68.5  \\

\enddata
\end{deluxetable*}

In the right panel of Figure \ref{distr_functions}, we differentiate
$P(<R)$ to derive the differential impact parameter distribution
$\frac{dP}{dA}(R)$. As in the left panel, the different lines
represent the three EW samples. The dark grey shaded area in this plot
shows the normalized SB profile (1-$\sigma$ confidence level) for the
``All-EW'' sample. The agreement with the corresponding analytic curve
(black solid line) is remarkable and extends even beyond the fitted
range; this testifies to the goodness of the fit.

Three lines of different logarithmic slopes are also plotted in
figure \ref{distr_functions} (right panel); they can be compared with
the lines representing the analytic $\tanh$ fit, and with the
power-law best fit slopes reported in Table \ref{shape_pars_tab},
column 4.

It is instructive to compare our results with previous studies on MgII
impact parameters. For this, the main sample available in the
literature is that of \cite{SDP94} with about 50
``identified'' MgII absorbing galaxies of which 70\% have
spectroscopic redshifts. As described in \cite{steidel_proc_95}, the
global picture that arose from their study implies that bright
galaxies are surrounded by a halo of gas detectable through strong
MgII absorption lines up to about 40 kpc. More specifically, using
several quasar sightlines without any strong MgII absorption
feature, \cite{steidel_proc_95} derived that the size of MgII halos
follows $R(L_K)\simeq\,38\,h^{-1}\,(L_K/L_K^*)^{0.15}$ kpc.  It is
interesting to note that the mean impact parameters inferred from his
study are similar (although systematically lower) than the ones
obtained in our analysis.

However, whereas Steidel et al. did not report detections of galaxy
impact parameters larger than 40 $h^{-1}$ kpc, we find a significant
SB signal up to about 200 kpc. More specifically we find a similar
amount of correlated light on scales $r<50$ kpc and $50<r<100$ kpc.
As recently pointed out by \cite{churchill+05} the study
performed by Steidel et al.  suffered from several selection
biases. For example, once a galaxy was identified at the absorber
redshift, no further redshifts were determined in the field thus
discarding the possible detection of several galaxies contributing to
the absorption.  Moreover galaxies were targeted depending on their
colors and magnitudes, and  spectroscopy was performed beginning
with the smallest impact parameters.  Churchill et al. have started to
reinvestigate some of these absorber systems more systematically.
They have reported several mis-identifications.
Their preliminary
results show the existence of more than one galaxy at the absorber
redshift in some cases, as well as a number of impact parameters as
large as 80 kpc.

In light of these results, it is worth emphasizing the power of
our stacking approach and its ability in producing unbiased results
over a range of impact parameters that are hardly accessible with any
other ``classical'' method.  Our analysis provides us with a
well-defined cross-correlation between gas and light: $\langle
W_0(\phi)\,L(\phi+\theta)\rangle$. Trying to identify a galaxy 
responsible for the absorption is not awell-defined process. 
On the one hand, there can very well be galaxies at smaller impact 
parameters that are too faint to be detected. On the other, MgII 
clouds could be at such large distances from the associated galaxy 
that confusion can arise due to galaxy clustering.
Correlation functions (using a stacking
technique or galaxy number counts) are needed in order to probe the
statistical properties of connections between absorption and emission
properties.

\section{Integral photometric properties of MgII absorbing galaxies}
\label{SED_subsec}

In this section we focus on the integrated absorbers' light, with the
aim of characterizing the absorber-related galaxies in terms of
spectral energy distribution (SED) and total optical luminosity. Our
approach consists of comparing the observed fluxes with suitably
transformed galaxy templates observed in the local Universe. 
Our approach and the subsequent interpretation are based on the assumption
that there are
galaxies in the local Universe whose SED closely resembles the average
SED of the galaxies connected to the MgII absorbers at $z\gtrsim
0.4$. Moreover we assume that evolution within a redshift interval 
$\Delta z=0.2$ is
negligible (which will be verified as shown below).  We begin
by selecting 39 nearby galaxies
spanning the whole range of morphology, star formation history and
current activity, for which good quality spectral coverage from the UV
to the near IR is provided by \cite{calzetti+94}, \cite{mcquade+95},
and \cite{storchibergmann+95}\footnote{The observed spectra are
shifted back to their rest frame and corrected for foreground Galactic
extinction using the \cite{cardelli+89} extinction curve and the
\cite{schlegel_dust} dust distribution.}. Let us now consider a given
sample of $N$ absorbers and its redshift distribution
$\{z_{abs,i}\}_{i=1,N}$. For each template SED, we compute the
effective SED which would be observed if all absorbing galaxies had
this same template spectrum and the same absolute magnitude in the
rest-frame.

The expected effective flux density at the observed wavelength
$\lambda$ is then given by:
\begin{equation}
f_{\mathrm{eff}}(\lambda)=\frac{10^{-0.4\Delta
M}}{N}\sum\limits_{i=1}^{N}
\frac{f_{0}(\lambda/(1+z_{abs,i}))}{1+z_{abs,i}}\times
10^{-0.4\mu(z_{abs,i})}
\end{equation}\label{obs_to_rest_eq}
where $f_0$ is the flux density of the template observed at the
canonical distance of 10 pc with a fixed absolute magnitude $M_0$, $N$
is the number of absorbers in the sample, $\mu(z_{abs,i})$ is the
distance modulus corresponding the given absorber redshift, and
$\Delta M$ is the difference between the actual absolute magnitude of
the absorbing galaxies (which is assumed to be the same for the all
absorbers in the sub-sample) and $M_0$. $\Delta M$ is treated as a
free normalization parameter in the following.  From the effective
SED, synthetic observed-frame flux densities in the four bands are
computed and compared to the observed fluxes from the stacking, using
the standard definition of $\chi^2$ for correlated errors:
\begin{equation}
\chi^2=\sum\limits_{b=g,r,i,z}\sum\limits_{b^\prime=g,r,i,z}\Delta f_b
\hat\sigma_{b,b^\prime} \Delta f_{b^\prime} 
\end{equation}
where $\Delta f_b$ are the differences between the observed and the
effective template fluxes, and $\hat\sigma_{b,b^\prime}$ is the
inverse covariance matrix of the observed fluxes in the four
bands. For each template, we determine the normalization factor
$\Delta M$ that minimizes the $\chi^2$, and consider this minimum
$\chi^2$ as a measure of the goodness of the fit attainable with that
given template.
\begin{figure*}
\includegraphics{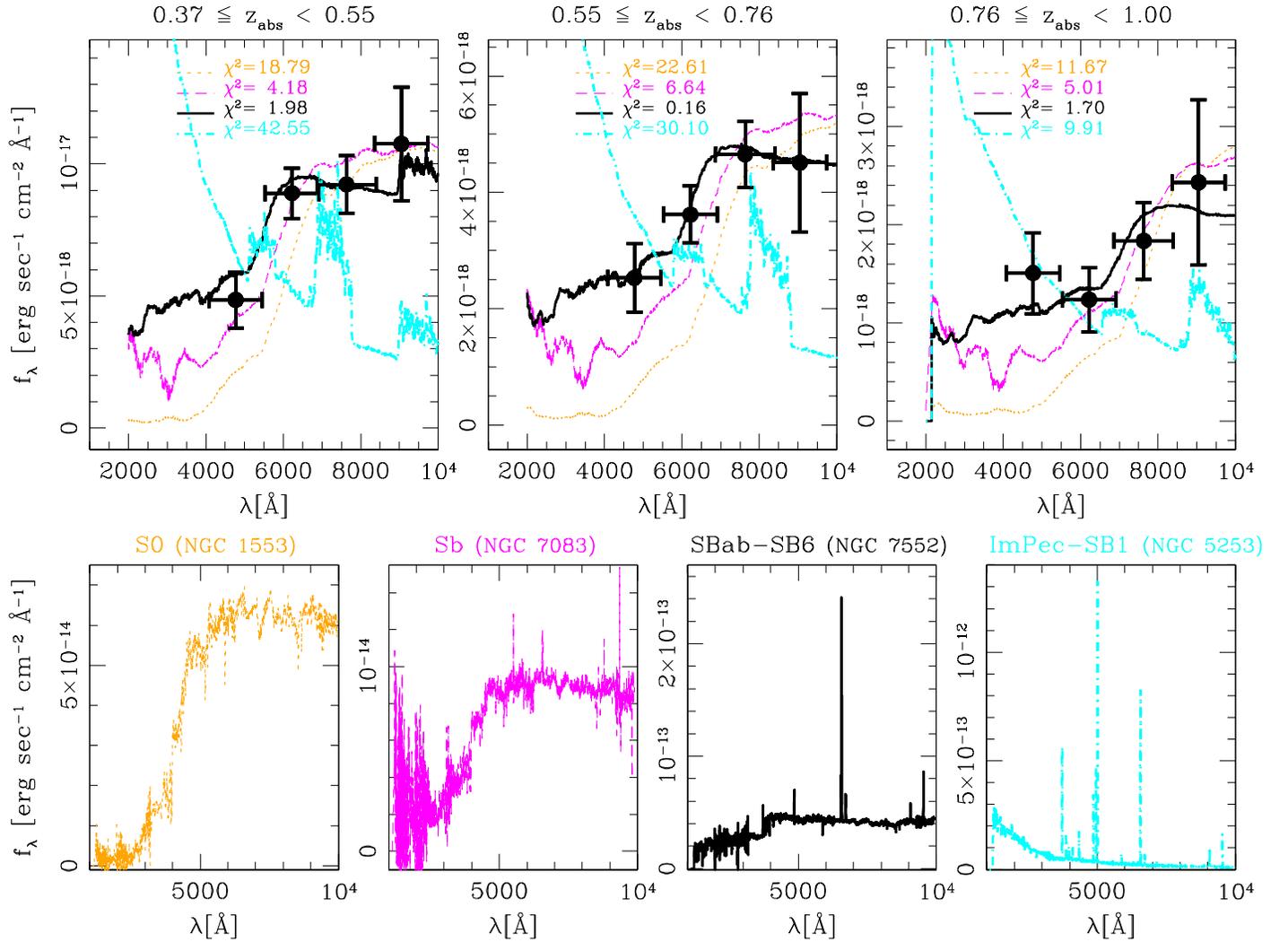}
\caption{Illustration of the SED fitting technique. In the
\emph{bottom} row we plot the rest-frame spectra (UV to near-IR) of
four galaxies which are representative of the whole range of galaxy
SEDs, from the early types (to the \emph{left}) to the late and
star-bursting types (to the \emph{right}). The flux normalization is
arbitrary. In the \emph{top} panels the filled dots with error bars
display the flux density in the four SDSS bands (observed frame) for
the three $z_{abs}$ bins (low is to the \emph{left} and high to the
\emph{right}). The horizontal error bars span the filter
band-width. The lines represent the best-fit renormalized spectra
computed by convolving the redshifted templates of the bottom row with
the redshift distribution of the absorbers (see text for details).
The line styles and colors are the same as in the bottom panels. The
$\chi^2$ of each model is reported as well. Note how much better the
intermediate type templates (NGC 7552 in particular) match the
observational data at all redshifts.\label{SEDcompare_fig}}
\end{figure*}
\begin{deluxetable*}{rcccccccc}
\tablewidth{0pt}
\tablecolumns{9}
\tabletypesize{\tiny}
\tablecaption{Integral photometry and colors\label{photo_par_tab}}
\tablehead{
\colhead{$W_0(2796)$ bin} & \colhead{Radial range} & \multicolumn{4}{c}{Integral flux (mag)} & \multicolumn{3}{c}{Color (mag)} \\
& &\colhead{g} &\colhead{r} &\colhead{i} &\colhead{z} &\colhead{g-r}&\colhead{r-i} &\colhead{i-z}}
\startdata
\cutinhead{$0.37\leq z_{abs}<0.55$}
\textsc{Low-$W_0$} & 10-- 50 kpc & $23.50^{+ 0.49}_{-0.34}$ & $22.00^{+ 0.21}_{-0.17}$ & $21.40^{+ 0.21}_{-0.17}$ & $20.82^{+ 0.29}_{-0.23}$ & $  1.50 \pm 0.34$ & $ 0.60 \pm 0.12$ & $ 0.57 \pm 0.23$ \\
         & 50--100 kpc & $23.88^{+ \inf}_{-0.77}$ & $21.81^{+ 0.30}_{-0.24}$ & $21.12^{+ 0.29}_{-0.23}$ & $20.52^{+ 0.53}_{-0.36}$ & $  2.08 \pm 0.99$ & $ 0.69 \pm 0.18$ & $ 0.60 \pm 0.39$ \\
         & 10--100 kpc & $22.92^{+ 0.88}_{-0.48}$ & $21.14^{+ 0.21}_{-0.18}$ & $20.50^{+ 0.22}_{-0.18}$ & $19.91^{+ 0.36}_{-0.27}$ & $  1.78 \pm 0.53$ & $ 0.65 \pm 0.13$ & $ 0.59 \pm 0.29$ \\
\tableline
\textsc{Intermediate-$W_0$} & 10-- 50 kpc & $22.95^{+ 0.28}_{-0.22}$ & $22.18^{+ 0.24}_{-0.19}$ & $21.88^{+ 0.30}_{-0.24}$ & $22.02^{+ 1.26}_{-0.57}$ & $  0.76 \pm 0.22$ & $ 0.31 \pm 0.19$ & $-0.14 \pm 0.67$ \\ 
                      & 50--100 kpc & $24.38^{+ \inf}_{-1.04}$ & $22.71^{+ 0.73}_{-0.43}$ & $22.27^{+ 0.96}_{-0.50}$ & $21.25^{+ 1.15}_{-0.55}$ & $  1.67 \pm 1.60$ & $ 0.44 \pm 0.48$ & $ 1.02 \pm 0.78$ \\ 
                      & 10--100 kpc & $22.69^{+ 0.64}_{-0.40}$ & $21.66^{+ 0.29}_{-0.23}$ & $21.30^{+ 0.41}_{-0.30}$ & $20.81^{+ 0.95}_{-0.50}$ & $  1.03 \pm 0.44$ & $ 0.36 \pm 0.26$ & $ 0.49 \pm 0.61$ \\ 
\tableline
\textsc{High-$W_0$} & 10-- 50 kpc & $22.54^{+ 0.19}_{-0.16}$ & $21.58^{+ 0.14}_{-0.13}$ & $21.09^{+ 0.14}_{-0.13}$ & $20.59^{+ 0.22}_{-0.18}$ & $  0.97 \pm 0.14$ & $ 0.48 \pm 0.08$ & $ 0.50 \pm 0.18$ \\
                & 50--100 kpc & $23.34^{+ 1.08}_{-0.53}$ & $22.13^{+ 0.41}_{-0.30}$ & $21.85^{+ 0.54}_{-0.36}$ & $21.35^{+ 1.24}_{-0.56}$ & $  1.21 \pm 0.59$ & $ 0.28 \pm 0.31$ & $ 0.50 \pm 0.70$ \\
                & 10--100 kpc & $22.12^{+ 0.35}_{-0.26}$ & $21.07^{+ 0.19}_{-0.16}$ & $20.65^{+ 0.21}_{-0.17}$ & $20.15^{+ 0.39}_{-0.29}$ & $  1.05 \pm 0.25$ & $ 0.41 \pm 0.14$ & $ 0.50 \pm 0.31$ \\ 
\tableline
\textsc{All} & 10-- 50 kpc & $22.89^{+ 0.15}_{-0.13}$ & $21.86^{+ 0.10}_{-0.09}$ & $21.39^{+ 0.11}_{-0.10}$ & $20.97^{+ 0.18}_{-0.15}$ & $  1.03 \pm 0.11$ & $ 0.47 \pm 0.07$ & $ 0.42 \pm 0.15$ \\ 
             & 50--100 kpc & $23.75^{+ 0.82}_{-0.46}$ & $22.16^{+ 0.22}_{-0.18}$ & $21.67^{+ 0.25}_{-0.20}$ & $21.00^{+ 0.40}_{-0.29}$ & $  1.59 \pm 0.50$ & $ 0.49 \pm 0.16$ & $ 0.67 \pm 0.32$ \\ 
        & 10--100 kpc & $22.49^{+ 0.27}_{-0.21}$ & $21.25^{+ 0.12}_{-0.11}$ & $20.77^{+ 0.14}_{-0.12}$ & $20.23^{+ 0.24}_{-0.20}$ & $  1.24 \pm 0.21$ & $ 0.48 \pm 0.09$ & $ 0.54 \pm 0.21$ \\ 
\cutinhead{$0.55\leq z_{abs}<0.76$}
\textsc{low-$W_0$} & 10-- 50 kpc & $24.64^{+ 1.15}_{-0.54}$ & $23.48^{+ 0.45}_{-0.32}$ & $22.83^{+ 0.41}_{-0.30}$ & $22.60^{+ 1.18}_{-0.55}$ & $  1.16 \pm 0.60$ & $ 0.65 \pm 0.29$ & $ 0.23 \pm 0.70$ \\
              & 50--100 kpc & $24.07^{+ 1.31}_{-0.58}$ & $23.61^{+ 1.27}_{-0.57}$ & $22.43^{+ 0.53}_{-0.35}$ & $21.19^{+ 0.52}_{-0.35}$ & $  0.45 \pm 0.80$ & $ 1.18 \pm 0.59$ & $ 1.24 \pm 0.51$ \\
              & 10--100 kpc & $23.56^{+ 0.98}_{-0.51}$ & $22.79^{+ 0.59}_{-0.38}$ & $21.86^{+ 0.38}_{-0.28}$ & $20.93^{+ 0.54}_{-0.36}$ & $  0.77 \pm 0.59$ & $ 0.93 \pm 0.35$ & $ 0.93 \pm 0.47$ \\
\tableline
\textsc{Intermediate-$W_0$}  & 10-- 50 kpc & $24.14^{+ 0.54}_{-0.36}$ & $22.90^{+ 0.23}_{-0.19}$ & $22.28^{+ 0.23}_{-0.19}$ & $22.02^{+ 0.51}_{-0.35}$ & $  1.23 \pm 0.38$ & $ 0.62 \pm 0.17$ & $ 0.26 \pm 0.39$ \\ 
                       & 50--100 kpc & $24.30^{+ 2.40}_{-0.69}$ & $23.73^{+ 1.26}_{-0.57}$ & $22.51^{+ 0.56}_{-0.37}$ & $22.34^{+ 4.20}_{-0.74}$ & $  0.56 \pm 1.01$ & $ 1.22 \pm 0.59$ & $ 0.17 \pm 1.09$ \\
                       & 10--100 kpc & $23.46^{+ 0.85}_{-0.47}$ & $22.49^{+ 0.36}_{-0.27}$ & $21.64^{+ 0.29}_{-0.23}$ & $21.42^{+ 0.81}_{-0.46}$ & $  0.97 \pm 0.57$ & $ 0.85 \pm 0.25$ & $ 0.22 \pm 0.60$ \\ 
\tableline
\textsc{High-$W_0$} & 10-- 50 kpc & $23.35^{+ 0.21}_{-0.18}$ & $22.33^{+ 0.12}_{-0.11}$ & $21.81^{+ 0.13}_{-0.12}$ & $21.74^{+ 0.37}_{-0.28}$ & $  1.02 \pm 0.17$ & $ 0.51 \pm 0.10$ & $ 0.07 \pm 0.30$ \\ 
                & 50--100 kpc & $23.78^{+ 0.67}_{-0.41}$ & $22.71^{+ 0.35}_{-0.26}$ & $22.08^{+ 0.35}_{-0.27}$ & $22.39^{+ \inf}_{-0.79}$ & $  1.08 \pm 0.47$ & $ 0.63 \pm 0.26$ & $-0.31 \pm 1.11$ \\ 
                & 10--100 kpc & $22.79^{+ 0.31}_{-0.24}$ & $21.75^{+ 0.17}_{-0.15}$ & $21.19^{+ 0.18}_{-0.16}$ & $21.27^{+ 0.73}_{-0.43}$ & $  1.05 \pm 0.25$ & $ 0.56 \pm 0.14$ & $-0.08 \pm 0.52$ \\ 
\tableline
\textsc{All}  & 10-- 50 kpc & $23.88^{+ 0.21}_{-0.18}$ & $22.79^{+ 0.11}_{-0.10}$ & $22.22^{+ 0.12}_{-0.10}$ & $22.06^{+ 0.28}_{-0.22}$ & $  1.10 \pm 0.17$ & $ 0.57 \pm 0.09$ & $ 0.16 \pm 0.24$ \\ 
              & 50--100 kpc & $24.01^{+ 0.50}_{-0.34}$ & $23.21^{+ 0.32}_{-0.25}$ & $22.31^{+ 0.24}_{-0.19}$ & $21.81^{+ 0.50}_{-0.34}$ & $  0.80 \pm 0.39$ & $ 0.90 \pm 0.22$ & $ 0.50 \pm 0.41$ \\ 
              & 10--100 kpc & $23.19^{+ 0.29}_{-0.23}$ & $22.23^{+ 0.16}_{-0.14}$ & $21.51^{+ 0.14}_{-0.13}$ & $21.17^{+ 0.33}_{-0.25}$ & $  0.97 \pm 0.23$ & $ 0.72 \pm 0.12$ & $ 0.34 \pm 0.29$ \\ 
\cutinhead{$0.76\leq z_{abs}<1.00$}
\textsc{low-$W_0$} & 10-- 50 kpc & $24.26^{+ 0.43}_{-0.31}$ & $24.53^{+ 1.10}_{-0.54}$ & $23.23^{+ 0.41}_{-0.30}$ & $22.65^{+ 0.77}_{-0.45}$ & $ -0.27 \pm 0.66$ & $ 1.30 \pm 0.57$ & $ 0.58 \pm 0.56$ \\ 
              & 50--100 kpc & $24.87^{+ \inf}_{-0.78}$ & $24.95^{+ \inf}_{-1.03}$ & $23.03^{+ 0.77}_{-0.44}$ & $23.11^{+ \inf}_{-1.02}$ & $ -0.08 \pm 1.80$ & $ 1.91 \pm 1.47$ & $-0.08 \pm 1.63$ \\ 
              & 10--100 kpc & $23.77^{+ 0.80}_{-0.45}$ & $23.97^{+ 2.30}_{-0.69}$ & $22.38^{+ 0.50}_{-0.34}$ & $22.10^{+ 1.76}_{-0.64}$ & $ -0.20 \pm 0.95$ & $ 1.59 \pm 0.78$ & $ 0.27 \pm 0.85$ \\ 
\tableline
\textsc{Intermediate-$W_0$} & 10-- 50 kpc & $24.52^{+ 0.50}_{-0.34}$ & $23.82^{+ 0.33}_{-0.25}$ & $23.59^{+ 0.51}_{-0.35}$ & $22.65^{+ 0.67}_{-0.41}$ & $  0.70 \pm 0.39$ & $ 0.23 \pm 0.37$ & $ 0.94 \pm 0.58$ \\ 
                      & 50--100 kpc & $...   ^{+   ...}_{-  ...}$ & $24.13^{+ 1.27}_{-0.57}$ & $23.55^{+ 1.30}_{-0.57}$ & $21.94^{+ 0.78}_{-0.45}$ & $    ... \pm   ...$ & $ 0.58 \pm 0.73$ & $ 1.61 \pm 0.85$ \\ 
                      & 10--100 kpc & $24.57^{+ 9.19}_{-0.75}$ & $23.21^{+ 0.53}_{-0.35}$ & $22.82^{+ 0.67}_{-0.41}$ & $21.48^{+ 0.61}_{-0.39}$ & $  1.37 \pm 1.01$ & $ 0.39 \pm 0.44$ & $ 1.33 \pm 0.62$ \\ 
\tableline
\textsc{High-$W_0$} & 10-- 50 kpc & $23.90^{+ 0.25}_{-0.20}$ & $23.74^{+ 0.29}_{-0.23}$ & $22.75^{+ 0.21}_{-0.17}$ & $22.52^{+ 0.58}_{-0.38}$ & $  0.16 \pm 0.27$ & $ 0.99 \pm 0.24$ & $ 0.23 \pm 0.45$ \\
                & 50--100 kpc & $24.24^{+ 0.80}_{-0.46}$ & $24.47^{+ 2.39}_{-0.69}$ & $23.87^{+ 2.71}_{-0.71}$ & $23.43^{+ \inf}_{-1.14}$ & $ -0.23 \pm 0.90$ & $ 0.59 \pm 0.99$ & $ 0.44 \pm 2.17$ \\ 
                & 10--100 kpc & $23.30^{+ 0.38}_{-0.28}$ & $23.29^{+ 0.51}_{-0.35}$ & $22.42^{+ 0.41}_{-0.30}$ & $22.13^{+ 1.46}_{-0.60}$ & $  0.01 \pm 0.40$ & $ 0.87 \pm 0.37$ & $ 0.29 \pm 0.85$ \\ 
\tableline
\textsc{All}  & 10-- 50 kpc & $24.19^{+ 0.19}_{-0.16}$ & $23.94^{+ 0.22}_{-0.18}$ & $23.12^{+ 0.17}_{-0.15}$ & $22.59^{+ 0.33}_{-0.25}$ & $  0.24 \pm 0.21$ & $ 0.82 \pm 0.18$ & $ 0.53 \pm 0.29$ \\ 
              & 50--100 kpc & $24.96^{+ 1.04}_{-0.52}$ & $24.39^{+ 0.77}_{-0.45}$ & $23.44^{+ 0.52}_{-0.35}$ & $22.60^{+ 0.83}_{-0.47}$ & $  0.57 \pm 0.71$ & $ 0.94 \pm 0.48$ & $ 0.84 \pm 0.65$ \\ 
              & 10--100 kpc & $23.75^{+ 0.35}_{-0.26}$ & $23.39^{+ 0.33}_{-0.25}$ & $22.52^{+ 0.26}_{-0.21}$ & $21.84^{+ 0.46}_{-0.32}$ & $  0.36 \pm 0.33$ & $ 0.87 \pm 0.25$ & $ 0.68 \pm 0.40$
\enddata
\end{deluxetable*}
We illustrate the main results of this analysis in Figure
\ref{SEDcompare_fig}. Here we consider the three $z_{abs}$ subsamples.
In the three panels in the top row the observed flux densities,
derived from the integrated magnitudes between 10 and 100 kpc (see
table \ref{photo_par_tab}) in the three $z_{abs}$ bins, are reported
as the orange filled dots with error-bars (the horizontal error-bars
represent the effective width of each passband). For the sake of
illustration, we now consider only four SED templates that are
representive of the whole range of galaxies. Their rest-frame
spectra are shown in the four panels of the bottom row. From left
to the right, we consider:\\
\indent \emph{a)} a typical early-type spectrum
(from the S0 galaxy NGC 1553), characterized by a very suppressed UV
flux, strong 4000 \AA~ break and metal lines, and flat shape at
$\lambda>6000$~\AA. \\
\indent \emph{b)} the spectrum of an intermediate spiral
galaxy (from the Sb galaxy NGC 7083), which is characterized by stronger 
UV flux, a weaker 4000 \AA~ break, weak emission lines, and very moderate blue
slope in the continuum at $\lambda>6000$~\AA.\\
\indent \emph{c)} an early type spiral (NGC7552, SBab) which is
undergoing a starburst, partially attenuated by dust \citep[StarBurst
class 6 in the atlas of][]{calzetti+94}; the unattenuated blue
continuum longward of the 4000 \AA~ break and the strong emission
lines coming from the star-burst coexist with a strong 4000 \AA~ break
and strongly suppressed UV flux, typical of the unextincted old
stellar population.\\
\indent  \emph{d)} Finally, the fourth panel shows the spectrum
of an unextincted starburst galaxy \citep[NGC 5253, starburst class
1 in the atlas of][]{calzetti+94}, which is characterized by blue
optical continuum, strong emission lines, and highly enhanced UV flux.

These four spectra are redshifted and convolved with the distribution
of $z_{abs}$ and normalized to the observed data points as explained
above. The results are reported in the three top panels of Figure
\ref{SEDcompare_fig}, where different line styles and colors code the
four original spectra. We also report the $\chi^2$ relative to each
model. It is apparent that the two extreme templates (passive early
type and pure star burst) completely fail to reproduce the
observations at any redshift. Good fits are obtained only with the
intermediate types, with the best results coming from the intermediate
spiral with partially obscured starburst. This latter young component
appears to be particularly required in order to reproduce the observed
``jump'' due to the 4000 \AA~ break, which in addition, is required to
be weaker than in a normal Sb galaxy.  A striking result from Figure
\ref{SEDcompare_fig} is also the fact that the same template is the
best fit to the observed fluxes at all redshifts. In other words, we
do not observe any significant evolution in the average SED of
galaxies associated with MgII absorbers from $z=0.4$ to $z=1$, which
validates our assumption of negligible redshift evolution.

From FIgure \ref{SEDcompare_fig} we also note that the SDSS bands mainly
constrain the rest-frame $u$ and $g$ flux. In the highest redshift bin
the situation is even more extreme, as the $g$-, $r$-, and $i$-bands
sample the flux blue-wards of the 4000 \AA~ break, while the $z$-band
is not very sensitive; this results in a very poor constraint to the
global properties of the stellar populations, and in an
over-sensitivity to the effects of dust attenuation in the rest-frame
UV at $z_{abs}\gtrsim 0.6$.

\subsection{SED fitting analysis}\label{SEDfitting_par}

\begin{figure*}
\plottwo{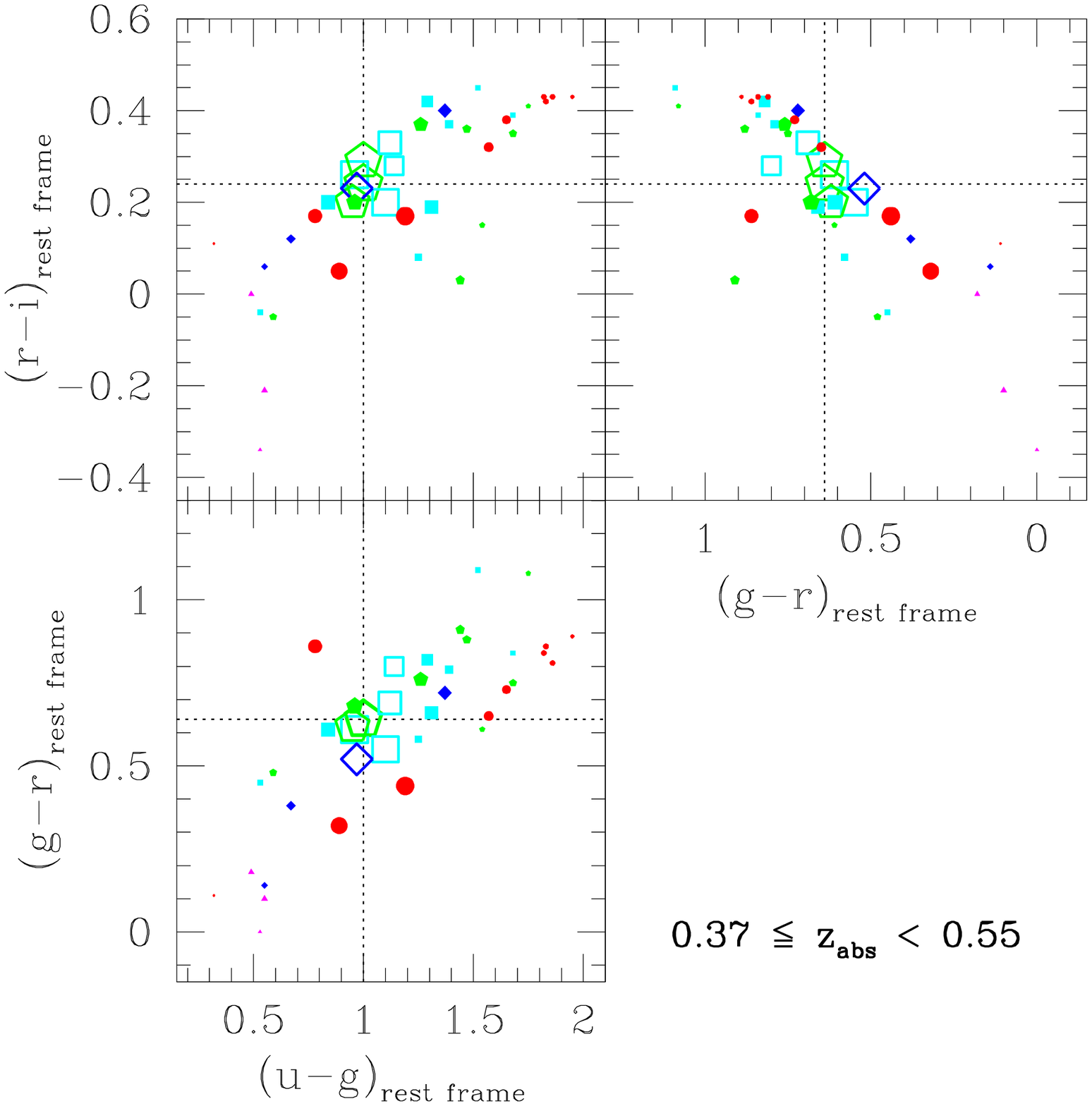}{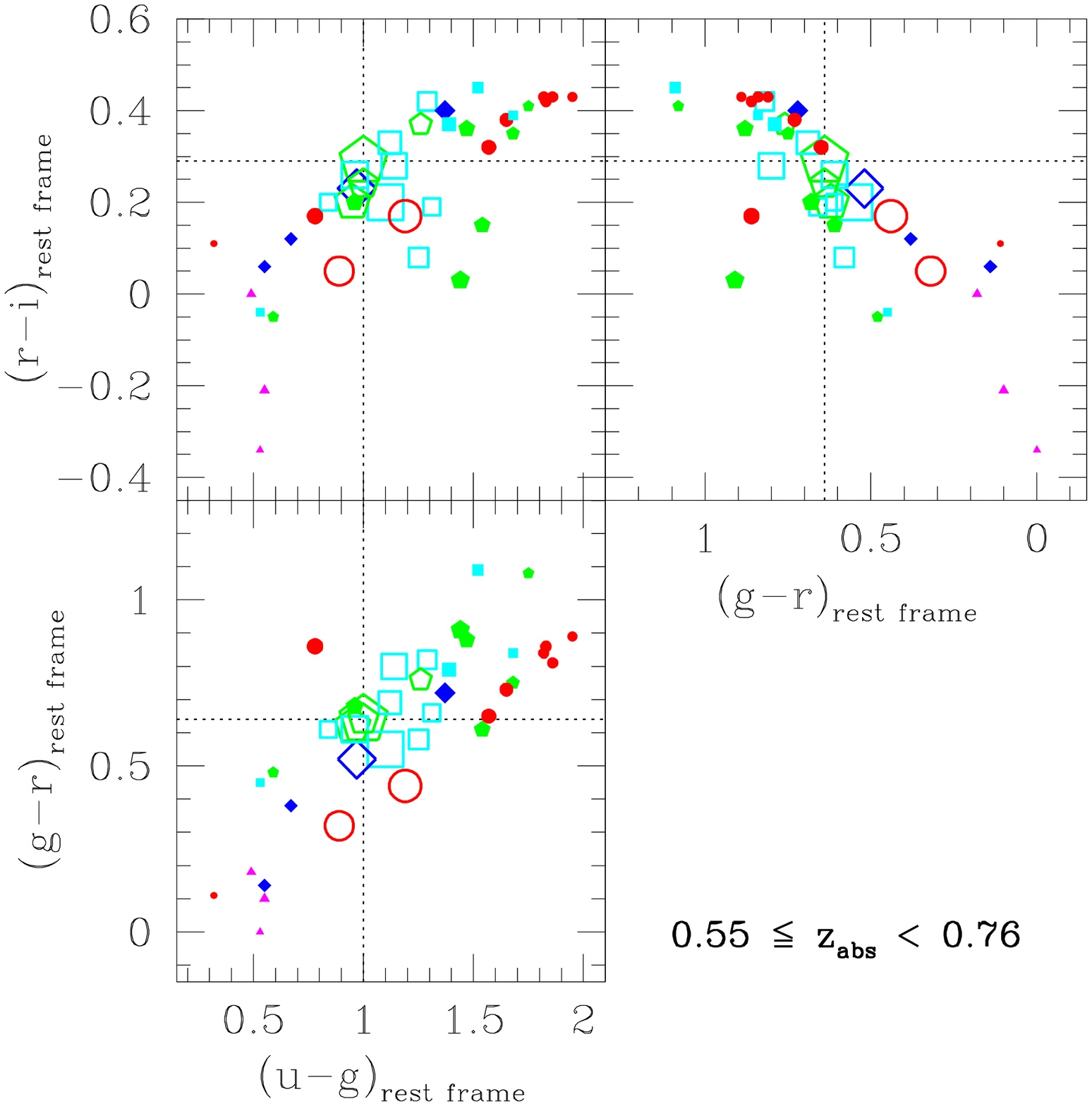}
\plottwo{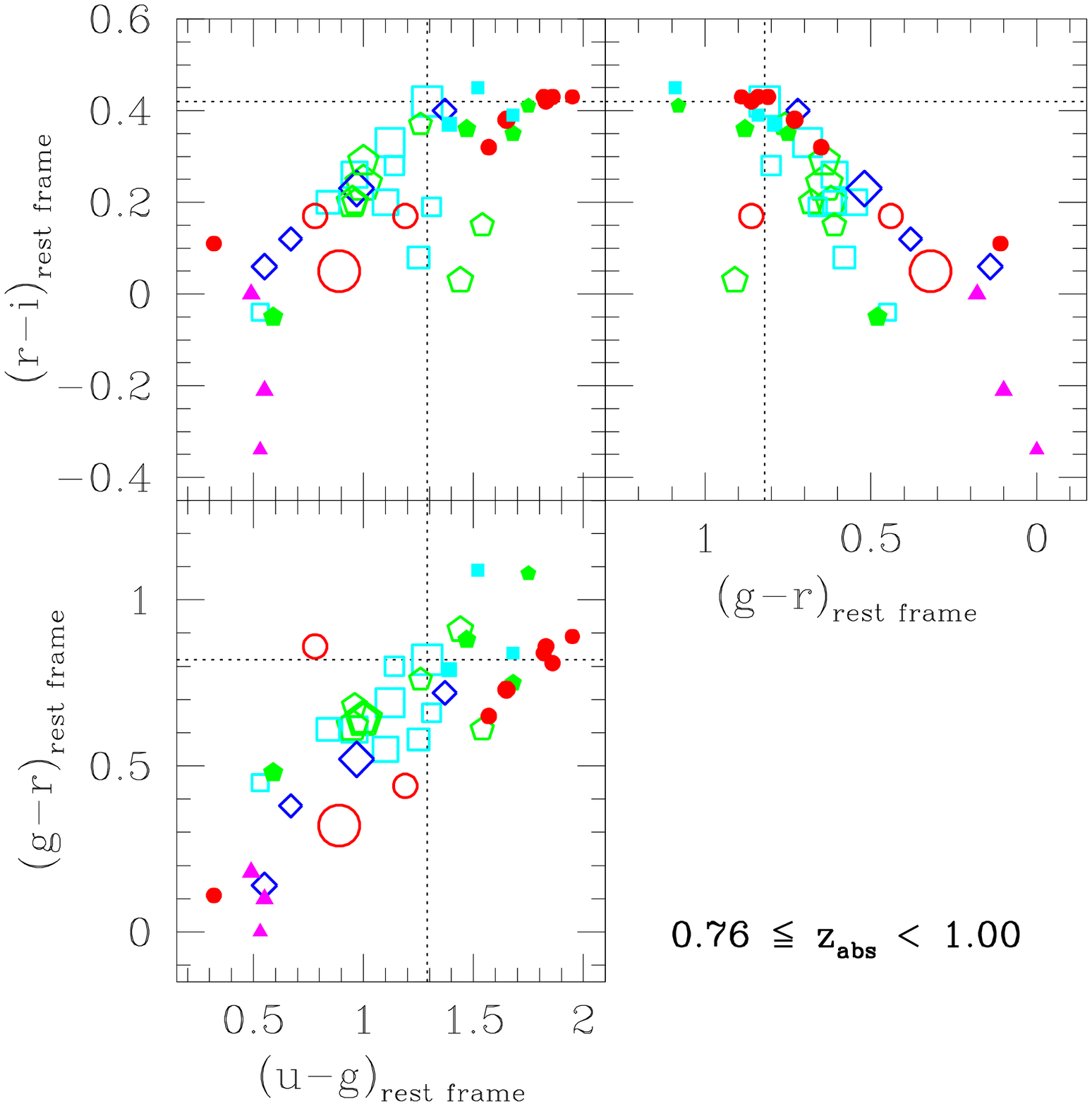}{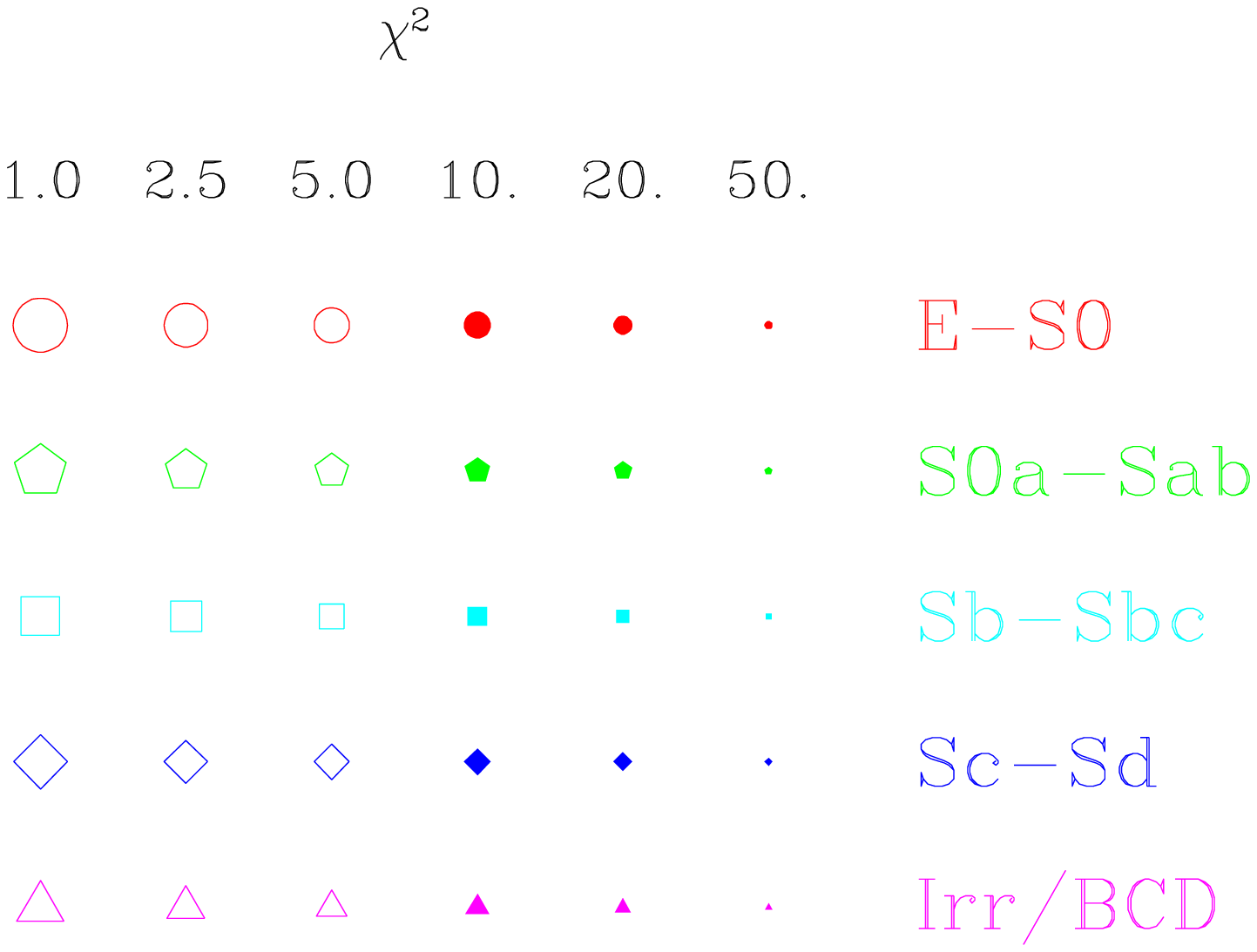}
\caption{``Goodness of fit'' distribution in the rest-frame color
space for the three $z_{abs}$ samples. Each point represents a
template galaxy spectrum \citep[taken
from][]{calzetti+94,mcquade+95,storchibergmann+95}, whose
corresponding morphology is coded by its shape and color, as indicated
in the legend. The size of the symbol is inversely related to the
$\chi^2$ of the template, as from legend. The best fitting template is
marked by the cross-hairs, while all templates yielding $\chi^2<6$
(our confidence limit, see in the text for details) are coded with
open symbols. All other templates are shown as filled symbols.
\label{CC_gridz}}
\end{figure*}

\begin{figure*}
\plottwo{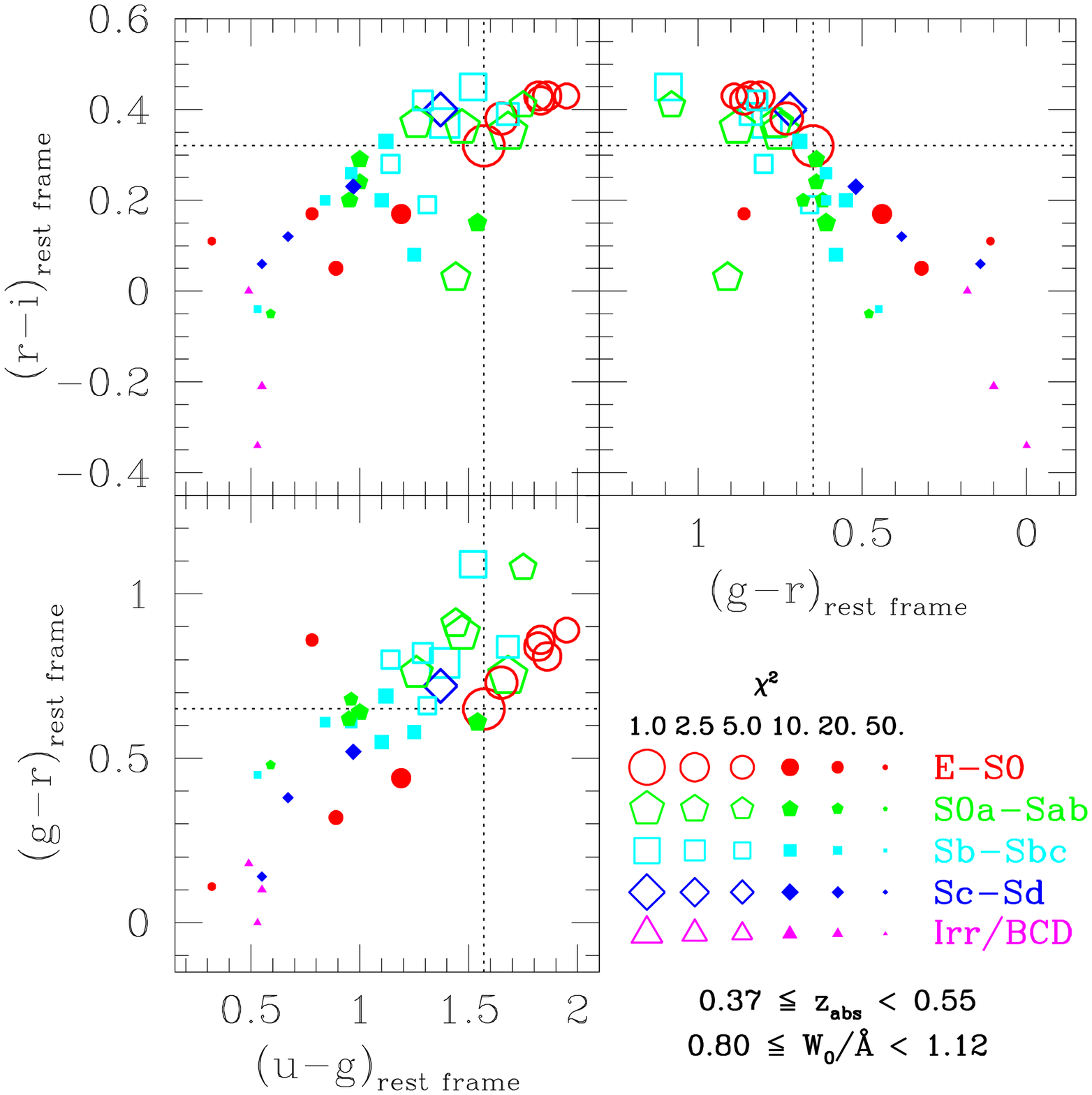}{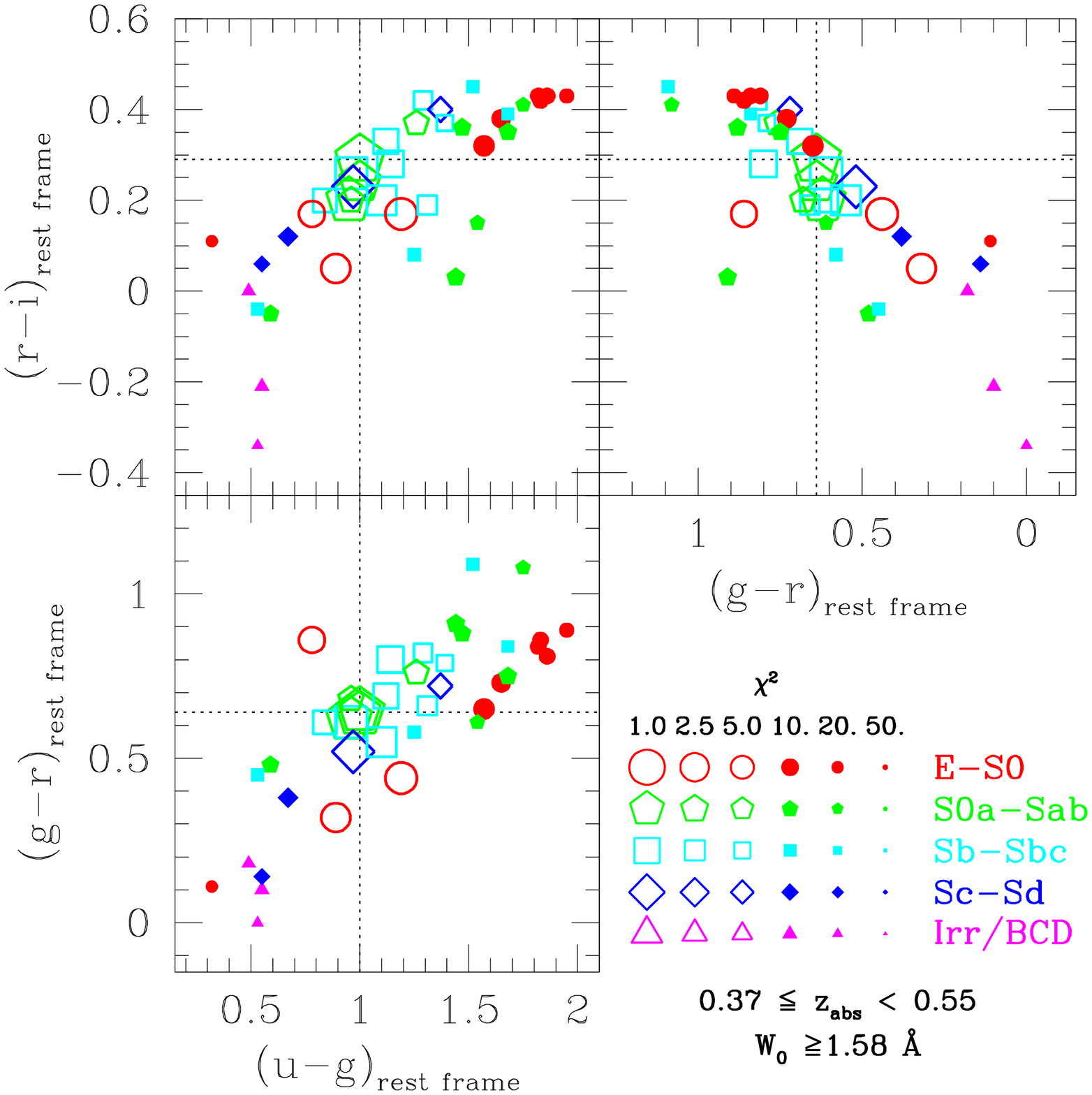}
\caption{``Goodness of fit'' distribution in the rest-frame color
space for the two extreme bin of EW at low $z_{abs}$: the results for
the `low-$W_0$ systems are shown in the \emph{left} panel, high-$W_0$
systems are shown on the \emph{right}. Plotted quantities are the same
as Figure \ref{CC_gridz}.\label{CC_gridEW}}
\end{figure*}

We now extend our analysis, using the complete library of the 39
UV-optical spectra selected as reported above, to assess the previous
result statistically, and in terms of rest-frame colors and
luminosities. All templates are classified by morphology (according to
NED\footnote{The NASA/IPAC Extragalactic Database,
http://nedwww.ipac.caltech.edu}) and their rest-frame colors, $u-g$,
$g-r$, $r-i$, are synthesized from the spectrum. The distribution of
the templates in this 3-dimensional color space is plotted in Figures
\ref{CC_gridz} and \ref{CC_gridEW}, for different absorber subsamples,
respectively. In each of these figures, the morphological type of the
SED template, from Elliptical-S0 to Irregular/Blue Compact Dwarf
galaxies, is coded according to the symbol shape (from circles to
triangles) and color, whereas the size scales inversely to the
$\chi^2$, as indicated in the legend. The best-fit model is identified
by the cross-hairs; all models that result in $\chi^2<6$ (or a
$\chi^2$ per degree of freedom less than 2) are identified by the open
symbols, while other models are represented by the filled ones.
We immediately observe that the rest-frame $u-g$ color, and, to a lesser
extent, the $g-r$ are well constrained by our observations, and
indicate that the average SED of the absorbing galaxies is consistent
with intermediate-type spiral galaxies (Sab to Sbc) of the local
Universe at all studied redshifts (Figure \ref{CC_gridz}). We consider
the colors of the best-fit model as best estimates of the rest-frame
colors and $\chi^2=6$ as a confidence limit, and report these values
in  columns 2, 3, and 4  of Table \ref{lum_col_tab}\footnote{More
specifically, the highest value corresponds to the reddest rest-frame
color of any model for which $\chi^2<6$. Vice versa, the lowest value
is for the bluest color of any model with $\chi^2<6$. Note that the
highest (lowest) values for two different colors do not need to be
taken from the same template.}. We conclude that \emph{in the entire
redshift range 0.36$<z<$1.0} the average SED of the absorbing galaxies
is characterized by rest-frame colors $u-g=1.1\pm0.2$,
$g-r=0.8\pm0.2$, and $r-i=0.3\pm0.1$.

\begin{deluxetable*}{lcccccccc}
\tablewidth{0pt} \tablecolumns{9} \tabletypesize{\small}
\tablecaption{Rest-frame colors and luminosities\label{lum_col_tab}}
\tablehead{
\colhead{$W_0(2796)$ bin} & \colhead{$u-g$} & \colhead{$g-r$} & \colhead{$r-i$} &
\colhead{$u$} & \colhead{$g$} & \colhead{$r$} & \colhead{$i$} &\colhead{$\delta M$}\\
 &\colhead{mag} &\colhead{mag} &\colhead{mag} &\colhead{mag} 
&\colhead{mag} &\colhead{mag} &\colhead{mag} &\colhead{mag} \\
\colhead{(1)} &\colhead{(2)} &\colhead{(3)} &\colhead{(4)} &\colhead{(5)} 
&\colhead{(6)} &\colhead{(7)} &\colhead{(8)} &\colhead{(9)} 
}
\startdata
\cutinhead{$0.37\leq z_{abs}<0.55$}
\textsc{All}    ($<100$ kpc)  &  $1.14^{+.25}_{-.19}$ & $0.80^{+.02}_{-.36}$ & $0.28^{+.14}_{-.11}$ & $-19.57$ & $-20.65$ & $-21.45$ &$-21.73$ & $0.11$\\
\textsc{All}    ($<50$ kpc)   &  $1.00^{+.14}_{-.05}$ & $0.64^{+.16}_{-.12}$ & $0.24^{+.09}_{-.04}$ & $-19.02$ & $-20.02$ & $-20.66$ &$-20.90$ & $0.09$\\
\textsc{low-$W_0$}   ($<100$ kpc)  &  $1.57^{+.38}_{-.43}$ & $0.65^{+.44}_{-.00}$ & $0.32^{+.13}_{-.29}$ & $-19.37$ & $-20.94$ & $-21.59$ &$-21.91$ & $0.17$\\
\textsc{high-$W_0$} ($<100$ kpc)  &  $1.00^{+.39}_{-.22}$ & $0.64^{+.22}_{-.32}$ & $0.29^{+.13}_{-.24}$ & $-19.71$ & $-20.71$ & $-21.35$ &$-21.64$ & $0.17$\\
\cutinhead{$0.55\leq z_{abs}<0.76$}
\textsc{All}    ($<100$ kpc)  &  $1.00^{+.31}_{-.16}$ & $0.64^{+.18}_{-.32}$ & $0.29^{+.13}_{-.14}$ & $-19.92$ & $-20.92$ & $-21.56$ &$-21.85$ & $0.12$\\
\cutinhead{$0.76\leq z_{abs}<1.00$}
\textsc{All}    ($<100$ kpc)  &  $1.29^{+.25}_{-.29}$ & $0.82^{+.09}_{-.68}$ & $0.42^{+.00}_{-.46}$ & $-19.90$ & $-21.19$ & $-22.01$ &$-22.43$ & $0.20$\\
\enddata
\end{deluxetable*}
By applying the same kind of analysis to the two subsamples of
low-$W_0$ (Figure \ref{CC_gridEW} \emph{left panel}) and high-$W_0$
systems (Figure \ref{CC_gridEW} \emph{right panel}) at low redshift, we
observe significant differences between the two classes; i.e., the colors of
low-$W_0$ systems are best fitted by the template of an elliptical
galaxy and the high-$W_0$ systems appear to be significantly bluer.
This is shown quantitatively in Table \ref{lum_col_tab} where we can
see strong differences in $u-g$ between the two subsamples. At low
redshift high-$W_0$ systems are about 0.6 mag bluer than low-$W_0$
ones. At higher redshift the signal is too weak to perform a similar
analysis.

\begin{figure*}
\centerline{
\includegraphics[width=.33\hsize]{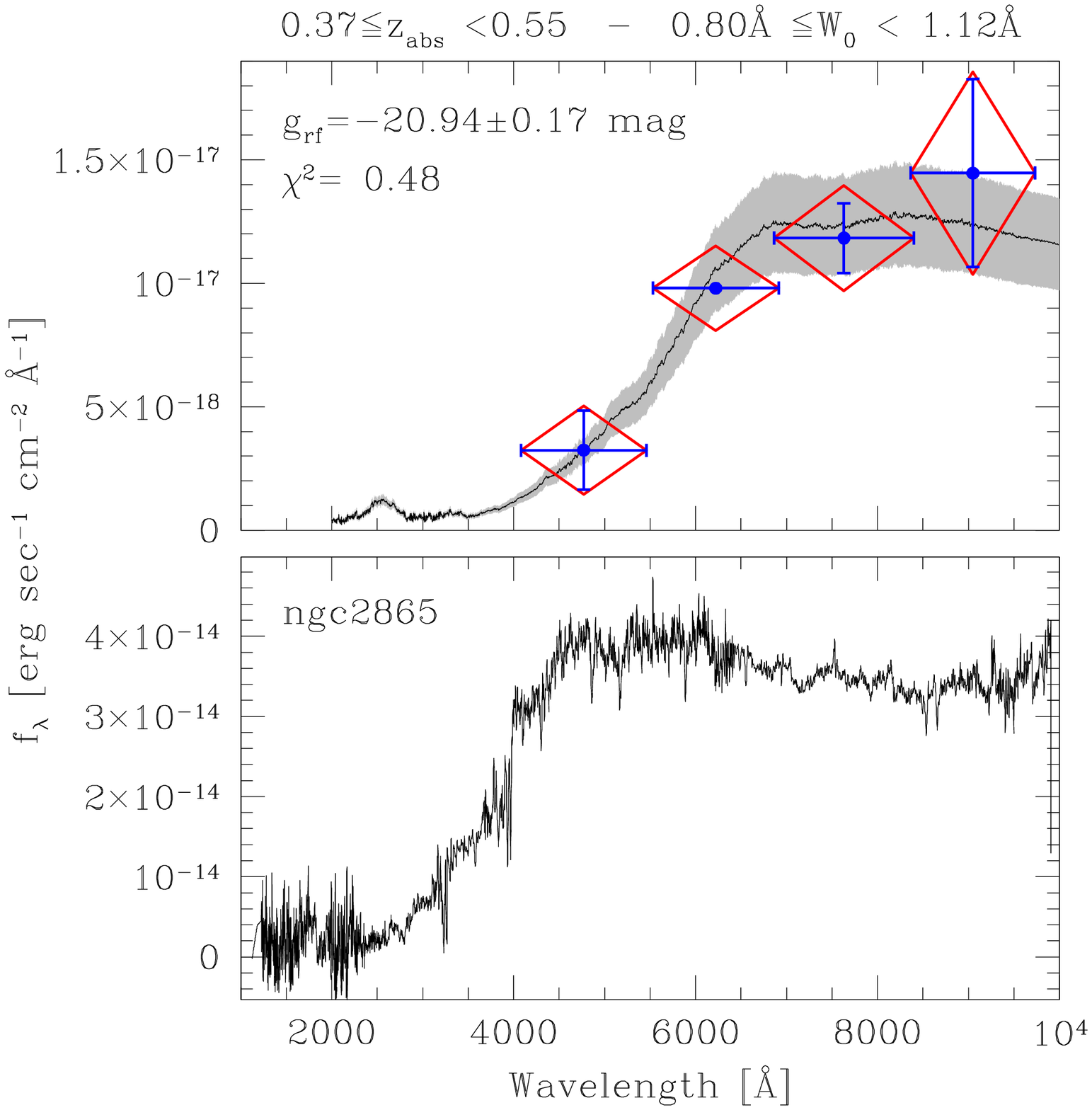}
\includegraphics[width=.33\hsize]{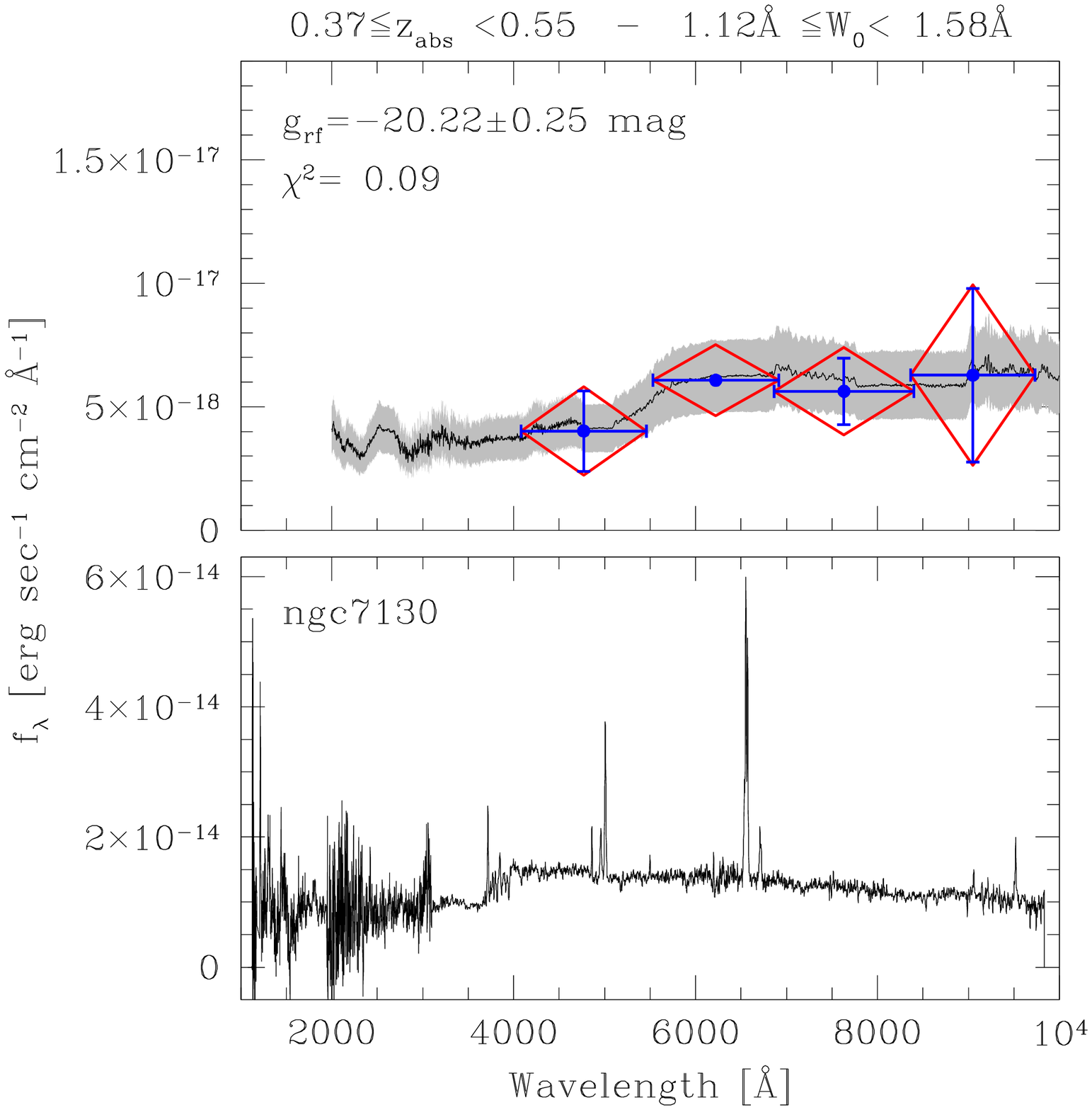}
\includegraphics[width=.33\hsize]{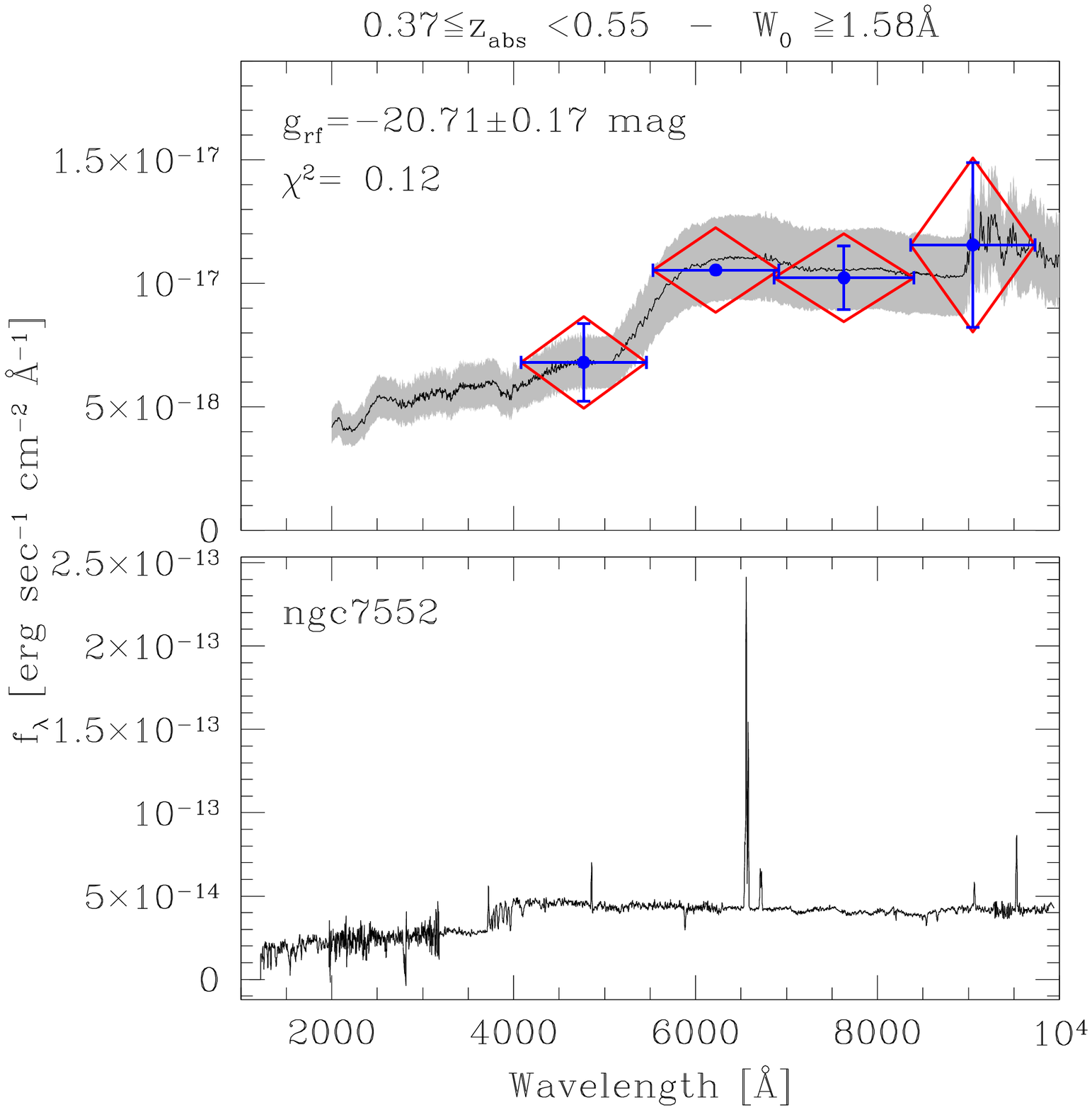}}
\caption{The best fitting templates for the three $W_0$ bins at low
redshift, from the weakest systems (\emph{to the left}) to the
strongest ones (\emph{to the right}). In each graph, the rest-frame
template spectrum is plotted in the bottom panel, with some arbitrary
intensity normalization, and identified by the NGC name of the
galaxy. The top panels show the measured photometric points (red
diamonds, whose vertical extent indicates the absolute photometric
error in each band individually) overlaid on the convolved and
re-normalized template spectrum (see text for details). The blue
points with error bars show the error \emph{relative} to the $r$-band
data points. The shaded area is the 1-$\sigma$ confidence range allowed for
the normalization. The best fitting rest-frame $g$ magnitude is also
reported, along with the (non-reduced) $\chi^2$.\label{best_fits_w0}}
\end{figure*}

The difference between the SEDs of low- and high-$W_0$ systems is
particularly well illustrated in Figure \ref{best_fits_w0}, where we
plot the best fitting SEDs for the three bins of $W_0$ at low
redshift. The top half of each panel reproduces the observed
photometric data points overlaid on the convolved and normalized
template spectrum, while the bottom half displays the original
template spectrum. It is apparent that the weakest systems are best
reproduced by a red passive template, while at $W_0\gtrsim1.1$\AA~the
SED closely resembles those of local actively star-forming galaxies.

From the fitting of the SED we also derive the normalization factor,
which enables us to compute rest-frame luminosities in different
bands. As mentioned above, the best constraints are obtained for the 
rest-frame $u-$ and $g-$ band fluxes, which correspond to the central and most 
sensitive $r$- and $i$-band in the observed frame. The
uncertainty in these fluxes is driven mainly by the photometric
errors.  On the other hand, the rest-frame $r-$ and $i-$ band fluxes are
extrapolated with the aid of the template SED, and therefore are
subject to the additional uncertainty of the choice of the SED
template. This is clearly the major source of error in the reddest
rest-frame bands at high redshift. In Table \ref{lum_col_tab}, we
report the rest-frame absolute magnitudes in $u$, $g$, $r$, and $i$ 
in columns 5, 6, 7, and 8, respectively, for different absorber
subsamples. The uncertainty in the normalization factor, which is given 
purely by the photometric error, is given in column 9. This is computed as the
variation of the normalization factor which is required to increase 
the value of $\chi^2$ by 1 with respect to the minimum while adopting the best
fit template.  All the following considerations are made using
photometric quantities integrated between 10 and 100 kpc. Looking at
the rest-frame luminosity in $u$ and $g$ as a function of redshift we
note a significant trend with higher redshift systems having higher
luminosity; the increase is roughly 30 to 50\% from $z\sim 0.4$ to
$z\sim 0.9$. It must be noted that such a luminosity evolution has a
negligible impact on our interpretation of colors within each
$z_{abs}$ bin; in fact, taking luminosity evolution into account
changes the observed-frame colors synthesized from the model SEDs by
0.02 mag at most for intermediate galaxy spectral types.
It is worth noting here that luminosities and colors are
\emph{scale-dependent}.  This effect is illustrated in Table
\ref{lum_col_tab} where we report the colors and luminosities for the
low-$z_{abs}$ sample measured between 10 and 50 kpc. With respect to
the total quantities the rest-frame luminosities are lowered by 0.45
($u$-band) to 0.83 mag ($i$-band), while the colors are systematically
bluer. This color gradient is primarily a consequence of the fact that
strong systems, which are also bluer, are more concentrated and
dominate the total light in the inner tens of kpc.

\subsection{The SED and luminosity of MgII absorbing galaxies in context}

At the beginning of this section we presented an
interpretation of the colors of the absorbing galaxies in terms of
local galaxy spectral templates. We showed that the average SED is
typical of intermediate type galaxies, with both old stellar
populations and a starburst component. This mix varies going from low-
to high-$W_0$ absorbers, with the latter having an enhanced starburst
component. It is interesting to note that, in a different context,
\cite{kacprzak+06} have recently obtained a result that goes in this
same direction, showing a positive correlation between $W_0$ and the
ratio between the asymmetry parameter, $A$, and impact parameter
for a sample of 24 intermediate redshift absorbing galaxies. By
interpreting $A$ as an index of galaxy ``activity'', the observed
correlation is fully consistent with our finding that higher $W_0$
absorbers are associated with galaxies having a smaller
impact parameter and a more ``active'' SED (in the sense of star
formation). The correlation between star formation and EW is further
supported by the observation, reported by \cite{nestor+05}, that
strong systems occur with a relatively higher frequency at high redshift
$\gtrsim 1$, thus paralleling the evolution of the cosmic star
formation rate.

No evidence is found for significant redshift evolution of the average
SED. This is an argument in favor of the thesis that MgII absorption
systems trace a well defined \emph{evolutionary phase} of
galaxies, rather than just a particular class of halos or galaxies in
a fixed mass range. 

In order to characterize this evolutionary phase in the cosmological
context, we compare the average colors and luminosities of the
absorbing galaxies with a well-understood sample of $I+B$-band
selected galaxies from the FORS Deep Field survey \citep[hereafter
FDF,][]{heidt+03}. In addition to multi-band photometry from the $U$
to the $K_S$-band, the FDF catalogs provide the photometric redshift
of each galaxy and the best fitting SED template
\citep[see][]{bender+01}. These can then be used to derive the rest-frame
photometric properties of the galaxy, and to convert the observed
magnitudes into the SDSS photometric system. We select all galaxies
with absolute rest-frame $i$-band magnitudes brighter than $-17.0$, and
split the sample into the same three redshift bins adopted for the
absorbers.
\begin{figure*}
\plotone{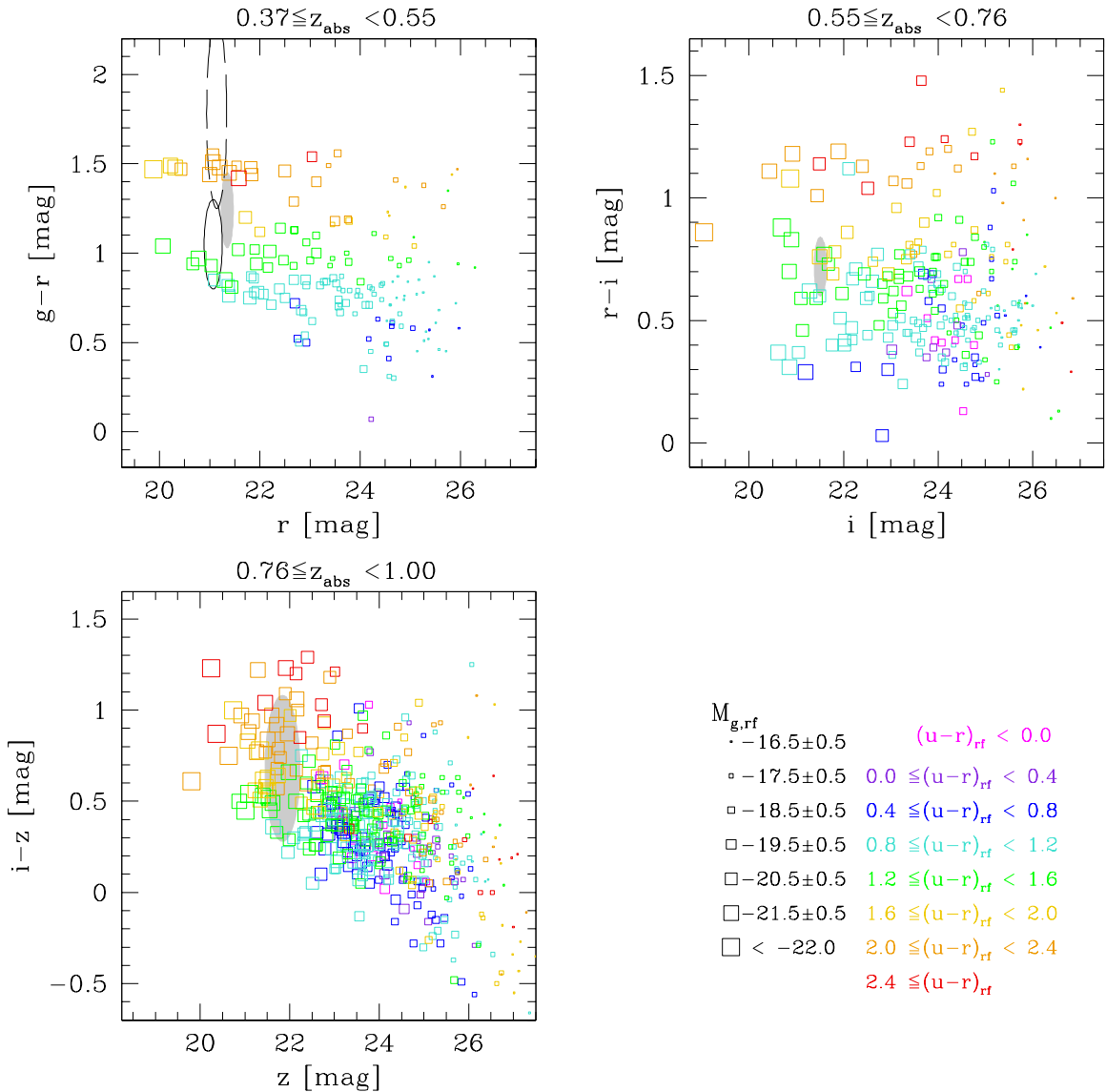} 
\caption{The \emph{observed frame} color-magnitude diagram (CMD) of
galaxies selected from the FORS Deep Field (FDF), compared to our
absorbers stack photometry in the three redshift bins.  Points are FDF
galaxies coded to indicate their rest-frame absolute $g$-band
magnitude (size) and rest-frame $u-r$ color (color), as shown in the
legend. The shaded ellipse is the 1-$\sigma$ confidence interval for
color/magnitude of our stack photometry for the ``All-'' EW sample. In the
low-$z_{abs}$ panel (\emph{upper left}) we also display the fiducial
estimates for the low- and high-$W_0$ absorber sub-samples with the
dashed and solid ellipses, respectively.\label{CMD_FDF}}
\end{figure*}
\begin{figure*}
\plotone{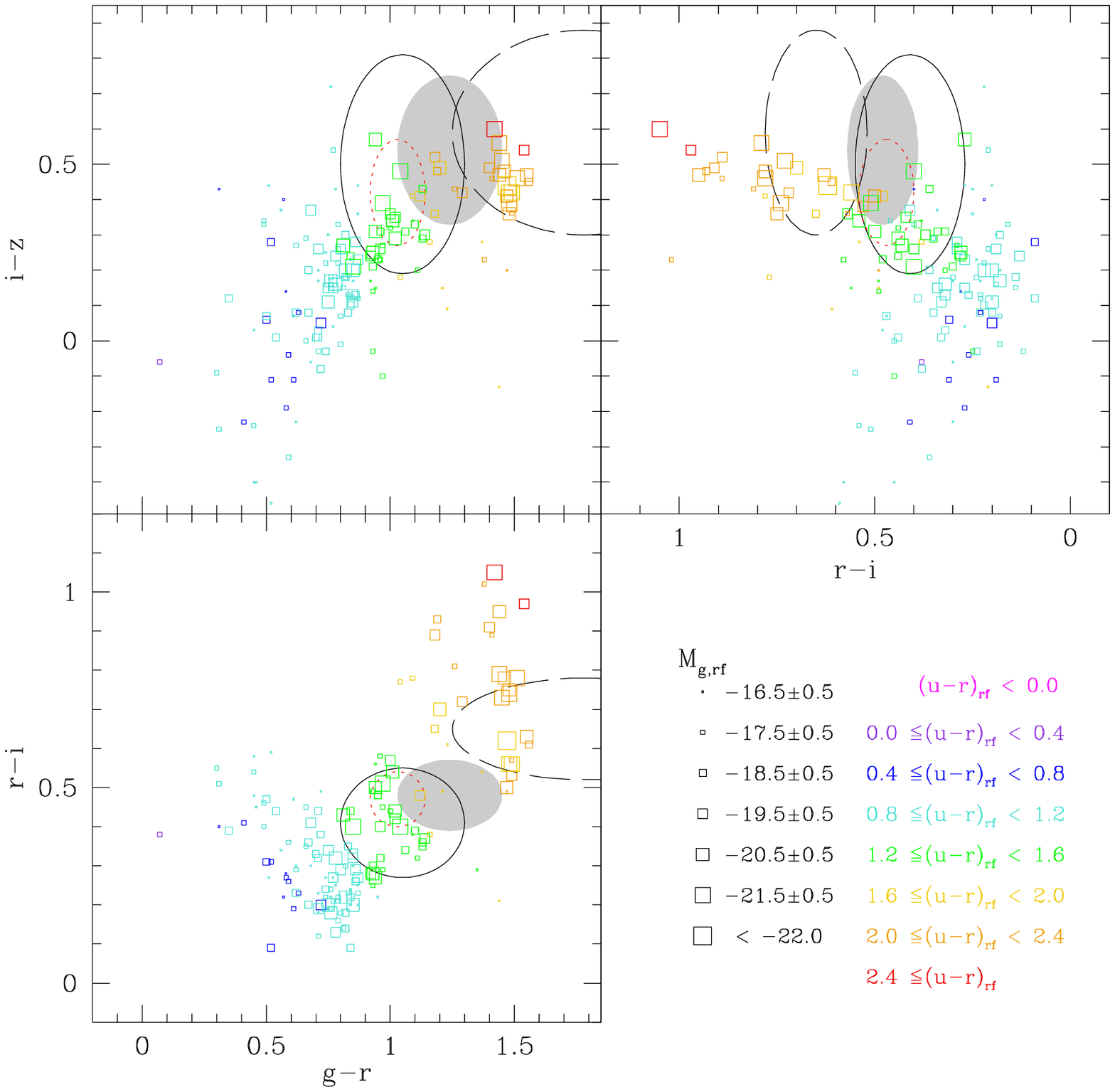}
\caption{The three projections of FDF galaxies in the \emph{observed
frame} $(g-r)$-$(r-i)$-$(i-z)$ observed color space, compared to our
absorbers stack photometry in the low-$z_{abs}$ bin ($0.37\leq
z_{abs}<0.55$).  As in Figure \ref{CMD_FDF}, points are FDF galaxies coded to
indicate their rest-frame absolute $g$-band magnitude (size) and
rest-frame $u-r$ color (color), as shown in the legend. The shaded
ellipse is the 1-$\sigma$ confidence interval for colors from our
stack photometry for the ``All'' EW sample. High- and low-$W_0$ absorber
sub-samples are shown with the solid and dashed ellipses,
respectively. The red dotted ellipse is used to display the colors in
the limited range 10-50 kpc.
\label{CC_FDF_lowz}}
\end{figure*}
Figure \ref{CMD_FDF} shows the \emph{observed-frame} color-magnitude
diagram (CMD) of the FDF galaxies in the three redshift
bins. Observed-frame bands and colors for different redshifts are
chosen so as to correspond to similar rest-frame spectral regions,
namely, the magnitude on the x-axis samples the flux at $\sim
4500$\AA, while the color on the y-axis roughly corresponds to 
rest-frame $u-g$ in all three panels. The size of the square
is indicative of the absolute $g$-band luminosity of each FDF galaxy
(as shown in the figure legend), while the color of the symbol codes
the rest-frame $u-r$, which is particularly suited to separate late-
from early-type galaxies \citep[e.g.][]{baldry_bimodality}. The
photometric measurements relative to our absorbers, including all $W_0$,
are indicated by the grey shaded ellipses in each panel, with the
shaded area representing the 1-$\sigma$ confidence limits. For the
low-$z_{abs}$ bin (upper left panel) we also overplot the two ellipses
relative to the high-$W_0$ systems (solid line) and the low-$W_0$ ones
(dashed line). Note that we always refer to photometric quantities
within the 100 kpc aperture, unless specified otherwise.

It appears that the average colors of the absorbing galaxies are
intermediate between the very red ones ($(u-r)_{rf}\geq 1.6$) of
passive galaxies on the red sequence (the ridge of yellow-red points
visible in the two top panels of Figure \ref{CMD_FDF}) and the bluer
colors, that characterize star forming spirals of intermediate
luminosity. This, again, is seen at all redshifts. We can thus conclude
that the average colors of the absorbing galaxies are intermediate not
only in absolute sense (as we already saw from the fit of local SED
templates), but also relative to the galaxies of similar luminosity at
the same redshift. Figure \ref{CMD_FDF} also supports the idea that MgII
absorbers reside in halos hosting a $L\lesssim L^\star$ galaxy. In
fact, if the total observed luminosity arose from a large number of
low-luminosity galaxies, then we would expect much bluer colors than
observed. Conversely, we note that the estimated absolute magnitude of
the absorbing galaxies ($M_{i,rf}\sim-21.7$) is perfectly consistent
with those of galaxies sharing the same region of the color-magnitude
diagram\footnote{For instance, this would have not been the case if
the observed $g-r$ of the low-$z_{abs}$ sample was 0.7 instead of 1.25; in
that case we would have been forced to conclude that the observed
total luminosity is produced by several blue low-luminosity
galaxies.}.

Figure \ref{CMD_FDF}, and the low-$z_{abs}$ panel in particular, at
first glance suggests that the absorbing galaxies could be those rare
galaxies that populate the so-called ``green valley'' between the red
sequence of passive galaxies and the blue sequence of the star-forming
ones. In turn, this would imply that the galaxies linked to MgII
absorbers are a kind of transition type that live in this phase for a
very limited amount of time (on the order of a few hundred Myr).
Alternatively, the intermediate color of the total sample may result
from a kind of ``morphological mix'' of passive and actively
star-forming galaxies. At the beginning of this \S \ref{SED_subsec} we
noted that high-$W_0$ absorbers are linked to bluer and more
star-forming galaxies than the low-$W_0$ systems. In Figure
\ref{CMD_FDF} the location of two ellipses for the high- and low-$W_0$
systems actually show that a systematic change in galaxy properties
occurs as a function of $W_0$: while high-$W_0$ systems occupy the
locus of luminous star-forming spirals, the low-$W_0$ ones overlap
with passive red-sequence objects.  In order to investigate whether
the two classes of absorbers reflect the color bimodality of the
overall galaxy population we can use the three projections of the
$(g-r)$-$(r-i)$-$(i-z)$ color space.  Figure \ref{CC_FDF_lowz} shows
these distributions for the FDF galaxies in the low $z_{abs}$ bin,
with the same symbols and conventions used in Figure \ref{CMD_FDF} and
reported in the legend. As in Figure \ref{CMD_FDF}, the shaded ellipse
represents the confidence region for the complete low-$z_{abs}$
sample, while the solid and dashed ellipses are the confidence
contours for the high- and low-$W_0$ subsamples, respectively. In
addition, we also report a red dotted ellipse to represent the colors
of the total low-$z_{abs}$ sample, but within the smaller aperture
$10<R<50$ kpc. The $(r-i)$-$(g-r)$ projection of the color space
appears particularly suited to divide between red, passive galaxies
and blue, active ones. The average colors of the total sample fall
exactly in between the two clouds, but now stronger systems are
clearly separated from the weaker ones: the stronger systems have
colors typical of star-forming spirals, while the weaker ones overlap
only with passive galaxies. It should be noted that intermediate-$W_0$
absorbers have the same colors as stronger ones, thus indicating that
the transition between active and passive types must be quite sharp
and occurs at $W_0\sim 1$\AA. This result indicates that absorbing
galaxies are likely to be representative of a large range of normal
galaxies with $L\lesssim L^\star$, whose color bimodality is also
reproduced. Also, absorbers weaker than $\sim 1$\AA~are most likely
associated with galaxies that have already moved to a phase of passive
evolution, while stronger systems mainly reside in halos where the
primary galaxy is still blue and actively star-forming. It is
interesting to note that the link between color bimodality and EW
holds also when so-called \emph{weak} systems ($W_0<0.3$\AA) are
considered. Using 7 absorbers with WFPC2/HST imaging and spectroscopic
spatial coverage \cite{churchill+06} have recently shown that passive
galaxies are also associated with such weak absorbers. Moreover, the
clustering analysis conducted by \cite{bouche+06} shows a similar
tendency, i.e., weaker systems appear to be more clustered than
stronger ones, which suggests that they are related to redder
galaxies. However, while the trend is reproduced, the large dark
matter halo masses inferred by their analysis for the low-EW systems
differ from the $\lesssim L^\star$ luminosities found in the present
study.

It should be noted that the average colors, and hence their
interpretation in terms of spectral types, is dependent on the choice
of the impact parameter limits. If one just considers the
luminosity-weighted average color of absorbing galaxies within 50 kpc,
instead of 100 kpc, bluer colors are obtained (as shown by the red
ellipses in Figure \ref{CC_FDF_lowz}, leading to the wrong conclusion
that passive galaxies are much less abundant among the
absorbers\footnote{The origin of this bias resides in the different
impact parameter distributions for low- and high-$W_0$ systems,
which results in the latter being the dominant source at small impact
parameters.}. Our extensive photometry represents a clear advantage in
this sense, with respect to most of the previous spectro-photometric
surveys that could cover only a limited range of impact parameters.

Finally, we have also observed a trend of increasing rest-frame
luminosity with redshift (30 to 50\% more luminous from $z\sim 0.4$ to
0.9). If we interpret the presence of MgII absorbers as indicative of
a given evolutionary phase for a galaxy, this luminosity evolution can
be seen as another phenomenon of \emph{downsizing}, i.e., the
occurrence of a given evolutionary phase is delayed in smaller
galaxies.

\section{Interpretation}\label{discussion_sect}

In this section we attempt to interpret the observational results
reported in this paper. We briefly discuss possible scenarios that may
reproduce the spatial and photometric measurements presented above,
encouraging detailed modeling in these directions.

In \S \ref{SB_impact_subsect} we presented measurements of a
spatial cross-correlation between MgII absorption and light,
i.e., galaxies, and we have shown that such a quantity is related to a
light weighted distribution of impact parameters.  As explained above
such observational results provide us with robust statistical
constraints as they do not make use of any assumption about the
nature, impact parameter and number of galaxies responsible for the
absorption. Moreover, the large number of absorption systems analyzed
(about 600 to 1250 in each redshift bin) allows us to measure the
gas-light correlation up to about 200 kpc. It is important to note
that the effects of galaxy clustering becomes increasingly important
as we consider larger scales. The gas detected in absorption can be
``related'' to a given galaxy which can be closely surrounded by a
number of neighbors. As recently shown by \cite{Churchill_Steidel+05}
this has led to several misidentifications in the past. 

Clustering effects are inherent to any statistical spatial studies and
must be taken into account for interpreting results.  Over the last
few years, galaxy-galaxy lensing studies have illustrated this
fact. This weak lensing technique provides us with a cross-correlation
between mass and light. Studies have shown that the galaxy masses
inferred from this technique have a significant contribution from
their environment \citep{mandelbaum+06}. Similar effects are expected
for gas-light cross-correlations. 

As we showed in \S \ref{SB_impact_subsect}, the slope of the
spatial cross-correlation between gas and light strongly depends on
the strength of the MgII absorbers. This change of slope can be due 
either to a clustering of the gas around galaxies being a function of
$W_0$ or to a change in the galactic environment.
Detailed modeling using, for example, the halo-model approach will be
required to accurately interpret these results.

Regarding the photometric properties of the light found in the
vicinity of MgII absorbers, we have shown that stronger absorbers are
related to bluer galaxies. We have found that systems with $W_0\gtrsim
1$\AA~ have colors matching those of star-forming galaxies with
$L\lesssim L^\star$. Similar results based on different techniques
were reported in previous studies
\citep[e.g.][]{guillemin_et_bergeron_91,churchill+99}. \cite{bond+01}
have also argued that the strong MgII systems can arise from the
``blow out'' of gaseous super-bubbles that are produced by supernovae
explosions in starbursting regions. Based on kinematic studies of the
MgII lines, \cite{prochter+06} have recently brought new evidence that
``the strong MgII phenomenon primarily arises from feedback processes
in relatively low mass galactic halos related to star formation.'' Our
results on systems with $W_0\gtrsim 1$\AA~ indeed provide a
statistically robust support to their conclusion, as far as the mass
and the star-formation activity are concerned. We have shown that
systems with $W_0\gtrsim 1.5$\AA~ are found to be in the direct
vicinity of a galaxy: 50\% of the associated light is within 25
kpc. Given typical super-wind expansion velocities of a few 100 km
sec$^{-1}\sim 1$~kpc Myr$^{-1}$ \citep{heckman+00}, the MgII clouds
have enough time to reach their observed location while the starburst
is still active (assuming that the duration of the burst is of the
order of 10 Myr). On the other hand we have shown that MgII systems
weaker than $\sim 1.1$\AA~are found at larger impact parameter (50\%
of the galaxy light is beyond 50 kpc from the MgII system), and are
predominantly associated with reddish and passive galaxies.  A possible
explanation may be that these weaker systems are also older,
i.e., observed a longer period of time after their outflow from the
parent galaxy. These MgII systems thus have had time to travel farther
away from the galaxy, and in the meanwhile the burst has finished and
the galaxy has turned into a passive red one. The interaction of the
clouds with the intergalactic medium might explain the lower $W_0$; on
long time-scales the MgII clouds can be eroded by hotter intergalactic
gas and the multiple components that give rise to large $W_0$ are
separated and diluted. Depending on the actual duration of the burst,
we can estimate the time-scale to turn the SED of the high-$W_0$
absorbing galaxies into those of the weaker ones as something ranging
from a few hundred Myr to 1 or 2 Gyr. Hence, the process of dilution
of the MgII clouds has to become effective on the same time-scale, and
the expansion speed at distances larger than 25 kpc must
progressively slow down to roughly $30~\mathrm{kpc}/
1~\mathrm{Gyr}\simeq 3~\mathrm{km}~\mathrm{sec}^{-1}$, in order to
explain the broader distribution of impact parameters for weaker
systems.

In contrast to outflow models, \cite{mo_miraldaescude_96} introduced a
model in which metal absorption lines are related to gas infalling
onto a galaxy which then triggers star formation. In such a scenario,
weak systems are associated with galaxies that still have to accrete gas
and trigger a (new) episode of star formation. In contrast, strong
systems can be associated with the gas that is being accreted onto a
galaxy and transformed into stars. The infall model therefore explains
the anti-correlation between EW and impact parameter and also the
correlation between blue colors and EW.  
In this model, weak absorbing galaxies must be interpreted as (the end
of) a quiescent phase that precedes a new substantial burst of star
formation, that will probably turn the galaxy back into a star-forming
spiral. The minimum duration of this quiescent phase can be estimated
to be 1-2 Gyr, i.e., the time to turn a galaxy red after a substantial
episode of star formation, consistent with the observed mean SED of
weak systems. However, such a kind of star formation history (SFH),
with intense bursts interleaved with long quiescent periods, appears
to be quite rare for $L\lesssim L^\star$ galaxies (corresponding to
the integrated light found within 100 kpc of MgII absorbers). Such
galaxies generally experience smooth continuous SFHs
\citep{kauffmann+03b}. If such an interpretation can be confirmed
quantitatively, it will show that the simple infall model does not
contain the necessary mechanisms to reproduce the observed trends
between the luminosity-weighted quantities and $W_0$.

\subsection{Implications for DLA systems}

Damped Lyman-$\alpha$ systems (DLAs) are believed to host most of the
hydrogen in the Universe and provide the reservoir for star formation.
As shown by \cite{Rao+2006}, strong MgII absorbers are often
associated with DLAs and the correlation between the two increases as
a function of $W_0(MgII)$. It is interesting to note that
\cite{Rao+2006} found that the mean $N_{HI}$ is roughly constant for
MgII absorbers with $W_0\gtrsim 0.6$ \AA~ but dramatically decreases
for weaker systems.  This result is similar to the color change found
between low- and high-$W_0$ systems: redder galaxies are expected to
have converted a significant fraction of the available neutral
hydrogen into stars. Moreover, while the
colors of MgII absorbing galaxies appear to be similar in the range
$W_0\gtrsim 1$\AA, we detected a transition scale at $W_0\sim 1$\AA,
below which the colors become significantly redder.  It would be of
great interest to investigate whether other properties of metal
absorbers or Lyman-$\alpha$ systems are seen to change around the
scale of $W_0(MgII)\sim 0.6-1$\AA. Understanding the origin of this
value would give us important insights into the nature and properties
of the gas in and around galaxies.

We can also note that, as the fraction of DLAs in MgII selected
systems increases as a function of $W_0(MgII)$, our results suggest
that stronger DLAs are expected to be associated with bluer galaxies, and
more specifically, with more star forming galaxies.

\section{Summary and conclusions}\label{summary}

In this paper we present a statistical analysis that allows us to
constrain the mean photometric properties of MgII absorbing galaxies
and the impact parameter distribution of the gas.  By stacking a large
number of PSF-subtracted images of quasars with absorbers and
comparing them with similar but non-absorbed quasars we isolate the
excess light associated with MgII absorber systems and investigate its
properties.
Contrary to a number of previous studies which attempted to identify
individual absorbing galaxies, our method measures a more general (and
well defined) cross-correlation between gas and light, and does not
use any assumption about the extent of the gas and/or a redshift
interval within which the galaxy can be found.  Moreover, it is
directly applicable to large samples, and can probe large scales.
In the present study we use a sample of about 2,800 quasar
sightlines with one strong MgII absorber ($W_0(\lambda2796)>0.8$~\AA) and
a control sample of more than 11,000 reference quasars extracted from
SDSS DR4.
The large number of systems allows us to study the properties of the
absorbing galaxies as a function of redshift and rest equivalent width
($W_0$), thus partly compensating the loss of information inherent to
any stacking procedure. Our results are as follows:

\newcounter{Scount}
\begin{list}{\arabic{Scount})}{}
{\usecounter{Scount}
\setcounter{Scount}{0}
\item We obtain significant detections of cross-correlations between
MgII gas and light out to $\sim 200$~kpc scales. We show that the
surface brightness (SB) distribution of such light is directly
proportional to the impact parameter distribution of MgII systems
around absorbing galaxies, weighted by galaxy luminosity.
\item Over the range 20-100 kpc, the SB profiles or impact parameter
distributions are well described by power laws with an index $\alpha$
that strongly depends on the rest equivalent width of the absorbers;
$\alpha\simeq-1$ for low-EW systems ($0.8\mathrm{\AA} \leq W_0 <
1.12\mathrm{\AA}$), while $\alpha\simeq-2$ for high-EW systems ($W_0
\geq 1.58\mathrm{\AA}$). More accurate fitting formulae are also
provided in \S \ref{impactpars_sec}.}
\end{list}
At low redshift, i.e. $0.37<z_{abs}\leq0.55$, where the S/N is high
enough, we find that
\begin{list}{\arabic{Scount})}{}
{\usecounter{Scount}
\setcounter{Scount}{2}
\item the mean luminosity enclosed within $10$ kpc $<r<100$ kpc is
$M_g=-20.65\pm0.11$ for the global sample, which is
$\sim0.5~L_g^\star$ of the field luminosity function in the same
redshift regime.
\item \emph{The spectral energy distribution of stronger systems
appears to be significantly bluer}. MgII systems with $W_0>1$\AA~are,
on average, characterized by an SED typical of star-forming/bursting
galaxies, while the light of weaker systems is dominated by passive
red galaxies (although they might not dominate the number counts).}
\end{list}
Moreover, considering the entire redshift range $0.37<z<1.0$, we find:

\begin{list}{\arabic{Scount})}{}
{\usecounter{Scount}
\setcounter{Scount}{4}
\item no detectable evolution with redshift of both spatial
distribution and \emph{global} average SED, which is always well
fitted by intermediate type spirals. This suggests that MgII systems
trace similar galaxies or a similar evolutionary phase of galaxies.
\item The rest-frame optical luminosity increases by $\sim 50\%$ from
$z\sim0.4$ to $z\lesssim1$. This phenomenon is reminiscent of the
so-called \emph{downsizing} which is observed in a number of
indicators of galaxy activity.
}
\end{list}

We have discussed possible interpretations of these observations
including inflowing/outflowing gas and clustering effects.  While
detailed modeling will be required to robustly test each model, we
point out that a scenario in which metal-enriched gas outflows from
star-forming/bursting galaxies is in qualitative agreement with
several types of trends seen in the data.

The observational results presented in this paper, namely the
cross-correlation between MgII gas and light and the mean colors of
absorbing galaxies, provide robust statistical constraints of great
interest for further modeling the gas distribution around
galaxies. They will help produce a comprehensive and consistent
picture of galaxy formation in both absorption and emission.

As a side-product of this study, we have shown that our stacking
technique is able to detect the light of QSO hosts and their
environment, thus opening new exciting perspectives for the study of
the relationships between QSOs and galaxy formation.

\acknowledgements{Our heartfelt thanks ought to go to Maurilio
Pannella for his precious help in using FDF data, to Dave Wilman, Eric
Bell and Jim Gunn for inspiring discussions, and to Houjun Mo, Simon
White, Jacqueline Bergeron, Roberto Saglia, Ralf Bender, and Lorenzo
Rimoldini for reading the manuscript and providing us with useful
comments. B.M. acknowledges the F. Gould foundation for its financial
support.

Funding for the SDSS and SDSS-II has been provided by the Alfred
P. Sloan Foundation, the Participating Institutions, the National
Science Foundation, the U.S. Department of Energy, the National
Aeronautics and Space Administration, the Japanese Monbukagakusho, the
Max Planck Society, and the Higher Education Funding Council for
England. The SDSS Web Site is http://www.sdss.org/.

The SDSS is managed by the Astrophysical Research Consortium for the
Participating Institutions. The Participating Institutions are the
American Museum of Natural History, Astrophysical Institute Potsdam,
University of Basel, Cambridge University, Case Western Reserve
University, University of Chicago, Drexel University, Fermilab, the
Institute for Advanced Study, the Japan Participation Group, Johns
Hopkins University, the Joint Institute for Nuclear Astrophysics, the
Kavli Institute for Particle Astrophysics and Cosmology, the Korean
Scientist Group, the Chinese Academy of Sciences (LAMOST), Los Alamos
National Laboratory, the Max-Planck-Institute for Astronomy (MPIA),
the Max-Planck-Institute for Astrophysics (MPA), New Mexico State
University, Ohio State University, University of Pittsburgh,
University of Portsmouth, Princeton University, the United States
Naval Observatory, and the University of Washington.

This research has made use of the NASA/IPAC Extragalactic
Database (NED) which is operated by the Jet Propulsion Laboratory,
California Institute of Technology, under contract with the National
Aeronautics and Space Administration.  }

\clearpage
\setcounter{table}{0}
\begin{landscape}
\begin{deluxetable}{lcccccccccc}
\tablewidth{0pt} \tablecolumns{11} \tabletypesize{\tiny}
\tablecaption{Shape parameters of the absorbers' SB profiles\label{shape_pars_tab}}
\tablehead{
\colhead{$W_0(2796)$ bin} & \multicolumn{5}{c}{Powerlaw slope $\alpha$} &
\multicolumn{5}{c}{First moment $R_1/\mathrm{kpc}$}\\
& \colhead{$g$} &\colhead{$r$} &\colhead{$i$} &\colhead{$z$} & \colhead{$<\alpha>$} &
\colhead{$g$} &\colhead{$r$} &\colhead{$i$} &\colhead{$z$} & \colhead{$<R_1>$}\\
\colhead{(1)} &\colhead{(2)} &\colhead{(3)} &\colhead{(4)} &\colhead{(5)} 
&\colhead{(6)} &\colhead{(7)} &\colhead{(8)} & \colhead{(9)}&\colhead{(10)} &\colhead{(11)} 
}
\startdata
\cutinhead{$0.37\leq z_{abs}<0.55$}
\textsc{All}          & $ -1.90^{+0.28}_{-0.37} $& $ -1.62^{+0.19}_{-0.21} $& $ -1.59^{+0.22}_{-0.23} $& $ -1.36^{+0.33}_{-0.44} $ & $-1.62 \pm 0.13$& $ 39.2\pm   6.1$ & $ 47.4 \pm  2.4$& $  47.8 \pm  2.7$ & $ 51.0 \pm  3.8$  & $47.6 \pm  1.6$\\
\textsc{low-$W_0$}         & $ -2.27^{+0.84}_{-1.73} $& $ -1.28^{+0.28}_{-0.32} $& $ -0.93^{+0.30}_{-0.30} $& $ -0.82^{+0.44}_{-0.48} $ & $-1.08 \pm 0.19$& $ 46.8\pm  11.8$ & $ 54.8 \pm  3.0$& $  56.5 \pm  2.8$ & $ 56.2 \pm  4.4$  & $55.6 \pm  1.8$\\
\textsc{Intermediate-$W_0$} & $ -1.59^{+0.44}_{-0.56} $& $ -1.36^{+0.36}_{-0.39} $& $ -1.88^{+0.51}_{-0.72} $& $ -0.31^{+0.49}_{-3.69} $ & $-1.51 \pm 0.27$& $ 33.9\pm  14.8$ & $ 43.4 \pm  6.4$& $  43.5 \pm  8.4$ & $ 57.5 \pm  9.1$  & $45.7 \pm  4.3$\\
\textsc{High-$W_0$}       & $ -2.14^{+0.44}_{-0.61} $& $ -2.33^{+0.43}_{-0.55} $& $ -2.25^{+0.40}_{-0.48} $& $ -2.33^{+0.68}_{-1.30} $ & $-2.25 \pm 0.27$& $ 38.9\pm   7.5$ & $ 43.8 \pm  3.9$& $  41.8 \pm  4.7$ & $ 42.8 \pm  7.5$  & $42.5 \pm  2.6$\\
\cutinhead{$0.55\leq z_{abs}<0.76$}\textsc{All}          & $ -1.32^{+0.30}_{-0.35} $& $ -1.61^{+0.21}_{-0.22} $& $ -1.26^{+0.16}_{-0.19} $& $ -1.34^{+0.36}_{-0.44} $ & $-1.38 \pm 0.12$& $ 49.9\pm   4.6$ & $ 47.2 \pm  3.2$& $  51.1 \pm  2.5$ & $ 55.0 \pm  4.5$  & $50.4 \pm  1.7$\\
\textsc{Low-$W_0$}         & $ -0.71^{+0.16}_{-0.67} $& $ -1.24^{+0.52}_{-0.63} $& $ -0.86^{+0.41}_{-0.39} $& $ -0.49^{+0.52}_{-0.66} $ & $-0.82 \pm 0.24$& $ 58.2\pm   9.0$ & $ 50.7 \pm  8.8$& $  56.8 \pm  5.0$ & $ 69.0 \pm  5.1$  & $60.7 \pm  3.1$\\
\textsc{Intermediate-$W_0$} & $ -1.40^{+0.78}_{-1.45} $& $ -2.46^{+0.51}_{-0.67} $& $ -1.70^{+0.37}_{-0.45} $& $ -2.07^{+0.79}_{-1.41} $ & $-1.92 \pm 0.31$& $ 50.9\pm  10.4$ & $ 45.1 \pm  7.2$& $  51.2 \pm  4.9$ & $ 49.8 \pm 11.7$  & $49.5 \pm  3.6$\\ 
\textsc{High-$W_0$}       & $ -1.60^{+0.33}_{-0.40} $& $ -1.43^{+0.21}_{-0.22} $& $ -1.31^{+0.26}_{-0.27} $& $ -1.56^{+0.61}_{-0.94} $ & $-1.43 \pm 0.15$& $ 45.5\pm   5.8$ & $ 46.7 \pm  3.4$& $  48.0 \pm  3.5$ & $ 40.7 \pm 13.5$  & $46.9 \pm  2.2$\\
\cutinhead{$0.76\leq z_{abs}<1.00$}\textsc{All}          & $ -1.92^{+0.30}_{-0.38} $& $ -1.42^{+0.27}_{-0.30} $& $ -1.42^{+0.27}_{-0.31} $& $ -1.45^{+0.57}_{-0.80} $ & $-1.55 \pm 0.17$& $ 43.6\pm   6.6$ & $ 44.8 \pm  6.4$& $  49.6 \pm  4.3$ & $ 53.3 \pm  6.3$  & $48.3 \pm  2.8$\\
\textsc{Low-$W_0$}         & $ -1.82^{+0.64}_{-1.06} $& $ -1.07^{+0.45}_{-1.38} $& $ -0.65^{+0.35}_{-0.53} $& $ -1.72^{+1.07}_{-2.28} $ & $-0.96 \pm 0.35$& $ 42.8\pm  12.7$ & $ 42.6 \pm 20.9$& $  56.7 \pm  5.5$ & $ 46.9 \pm 19.0$  & $53.4 \pm  4.8$\\
\textsc{Intermediate-$W_0$} & $ -2.50^{+0.68}_{-1.15} $& $ -1.42^{+0.34}_{-0.38} $& $ -1.58^{+0.53}_{-0.72} $& $ -1.23^{+0.93}_{-1.34} $ & $-1.55 \pm 0.29$& $ 33.0\pm  32.6$ & $ 47.7 \pm  8.2$& $  50.9 \pm  8.9$ & $ 64.5 \pm  5.8$  & $56.7 \pm  4.1$\\
\textsc{High-$W_0$}       & $ -1.63^{+0.33}_{-0.42} $& $ -1.77^{+0.50}_{-0.63} $& $ -2.03^{+0.43}_{-0.52} $& $ -1.58^{+1.06}_{-2.42} $ & $-1.78 \pm 0.26$& $ 47.2\pm   6.4$ & $ 41.2 \pm 11.0$& $  42.1 \pm  8.3$ & $ 39.5 \pm 20.4$  & $44.3 \pm  4.5$
\enddata
\end{deluxetable}
\clearpage
\end{landscape}

\appendix
\section{On the origin of the signal in reference QSOs}\label{refsys_appendix}

\begin{figure*}[h]
\plotone{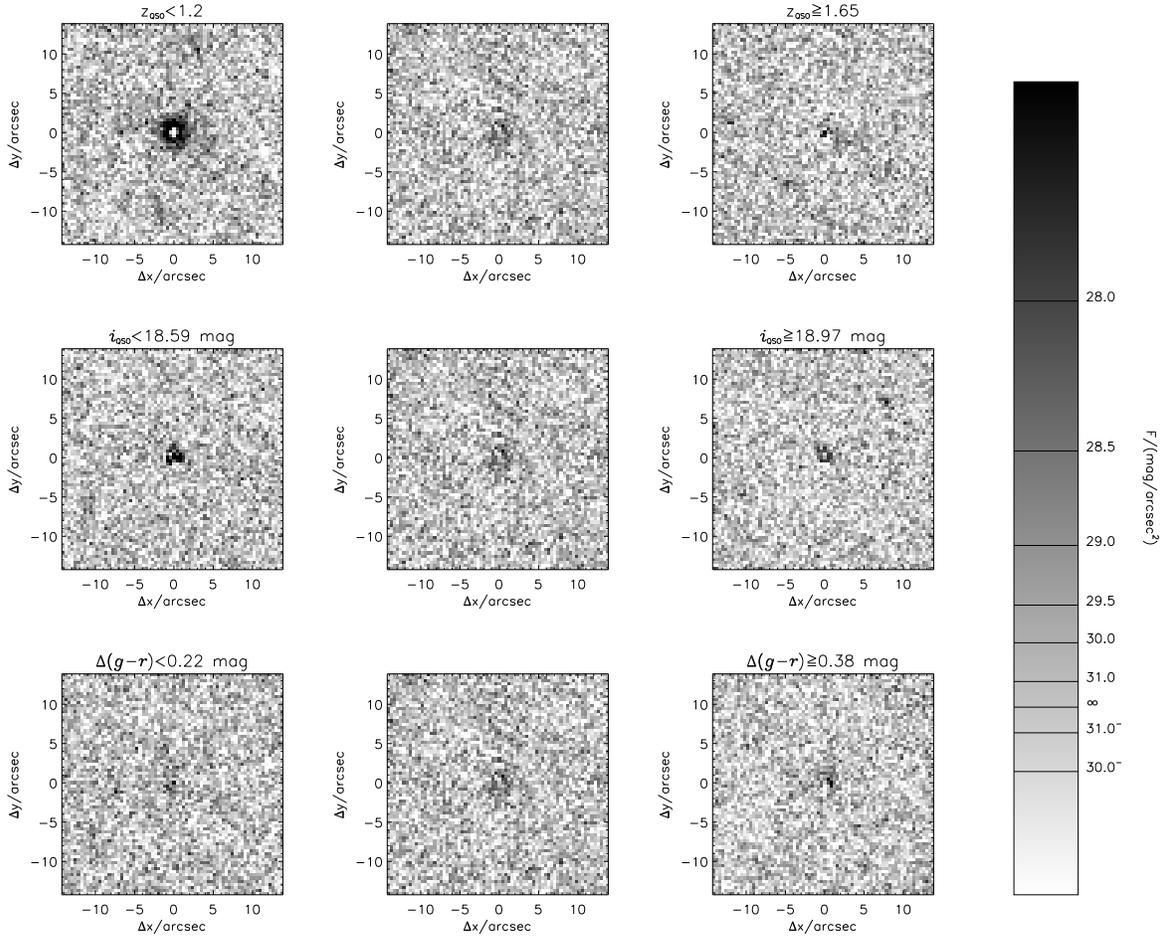}
\caption{The $i$-band stack images of reference QSOs in different
disjoint bins in the $z_{QSO}$-$i_{QSO}$-$\Delta(g-r)$ parameter
space. Only the central $30\times30~\mathrm{arcsec}^2$ are reproduced,
using a linear grey scale to map the intensity. The corresponding
levels in mag arcsec$^{-2}$ are reported in the key bar on the right
($\infty$ means background level).  The three images in the central
column represent the same reference central bin in the parameter
space, namely $1.2\le z_{QSO}<1.65$, $18.59\le i_{QSO}<18.97$,
$0.22~\mathrm{mag}\le\Delta(g-r)<0.38~\mathrm{mag}$. The left and
right panels in each rows are bins where only one of the three
parameters at time is varied with respect to the central reference
bin, as indicated by the labels.\label{maps_checkref}}
\end{figure*}

In \S\ref{stack_SB_subsec} we noted that the PSF subtraction leaves
significant SB residues even in the sample of reference QSOs. In this
appendix we further investigate the origin of this effect and
demonstrate that this is not due to a failure of the PSF subtraction
algorithm, rather to a genuine light excess found around QSOs with
respect to the actual PSF. To do this, we stack images of reference
QSOs only, without applying any geometrical transformation to the
images (apart from centering).
We consider as possible factors of mismatch with respect to the PSF:
\newcounter{L2count}
\begin{list}{\emph{(\roman{L2count})}}{}
{\usecounter{L2count}
\item the redshift of the QSO. This is the case when the light excess
is physically associated with the QSO, e.g. its host galaxy.
\item the magnitude of the QSO. This is the case when either the
excess is linked to intrinsic QSO properties, or when it arises from a
failure in deblending faint sources.
\item the color mismatch between the QSO and the star used to compute
the PSF. In fact, redder point sources are expected to have a narrower
PSF \citep[see e.g.][]{fried_65}.}
\end{list}
We split the three-dimensional parameter space given by $z_{QSO}$, the
QSO apparent magnitude in $i$-band $i_{QSO}$, and the color mismatch
$\Delta(g-r)\equiv (g-r)_{star}-(g-r)_{QSO}$ into $3^3$ cells
including roughly an equal number of reference QSOs ($\sim 400$).
Each of these $3^3$ subsamples is then processed and stacked
separately. We take the ``central'' cell in the parameter space as a
reference for the following comparisons. This is defined by $1.2\le
z_{QSO}<1.65$, $18.59\le i_{QSO}<18.97$, and
$0.22~\mathrm{mag}\le\Delta(g-r)<0.38~\mathrm{mag}$. The central
portion ($30\times30~\mathrm{arcsec}^2$) of the $i$-band stack image
for this reference subsample is shown in the central column of
fig. \ref{maps_checkref} and repeated in the three rows. The residuals
of the PSF subtraction are visible in these images and extend out to
$\sim 3~$arcsec with significant SB of $\sim 28.5~\mathrm{mag
arcsec}^{-2}$.
Let us now consider subsamples where we vary the range of each of the
three parameters separately, to investigate their effect on the
residuals of PSF subtraction. In the first upper row of
fig. \ref{maps_checkref} we show the images of the low-
($z_{QSO}<1.2$, to the left) and the high- ($z_{QSO}<1.2$, to the
right) QSO redshift subsamples, with the same $i_{QSO}$ and
$\Delta(g-r)$ cuts as in the reference. Similarly, in the second row
we make the $i_{QSO}$ vary (bright $i_{QSO}<18.59$ to the left and
faint $i_{QSO}\ge18.97$ to the right), while keeping fix the other two
parameters as in the reference subsample; and finally in the third row
the $\Delta(g-r)$ is varied (well matched star-QSO pairs to the left,
$\Delta(g-r)<0.22$~mag, and poorly matched pairs to the right,
$\Delta(g-r)\ge0.38$~mag).  The immediate conclusion that can be drawn
from the analysis of fig.\ref{maps_checkref} is that most of the SB
excess around unabsorbed QSOs is produced by low redshift QSOs. Also,
high redshift QSOs have no SB excess at all. The most logical
explanation for this is that at low redshift we do detect the light of
the QSO's host galaxy, which drops below our detection limits at
$z_{QSO}\gtrsim1.5$.  The lack of any significant dependence of the SB
excess on either $\Delta(g-r)$ or $i_{QSO}$ shows that the color
dependence of the PSF is well compensated by our algorithm (see
\S\ref{PSFsub_subsec}) and that there are no serious deblending
problems with fainter QSO being harder to deblend from neighboring
faint stars or galaxies.

\end{document}